\documentclass[12pt]{iopart}
\usepackage{iopams}
\expandafter\let\csname equation*\endcsname\relax

\expandafter\let\csname endequation*\endcsname\relax
\usepackage[svgnames]{xcolor}
\usepackage{graphicx}
\usepackage{amsmath}
\usepackage{amssymb}
\usepackage{times}
\usepackage{color}
\usepackage{url}
\usepackage{subfigure}
\usepackage{ulem}

\usepackage{etoolbox}

\makeatletter
\renewcommand\@appendixstar{\@@par
 \ifnumbysec
 \@addtoreset{table}{section}
 \@addtoreset{figure}{section}\fi
 \setcounter{section}{0}
 \setcounter{subsection}{0}
 \setcounter{subsubsection}{0}
 \setcounter{equation}{0}
 \setcounter{figure}{0}
 \setcounter{table}{0}
 \def\thesection{\Alph{section}} 
 \def\theequation{\ifnumbysec
      \Alph{section}\arabic{equation}\else
      \Alph{section}\arabic{equation}\fi}
 \def\thetable{\ifnumbysec
      \Alph{section}\arabic{table}\else
      A\arabic{table}\fi}
 \def\thefigure{\ifnumbysec
      \Alph{section}\arabic{figure}\else
      A\arabic{figure}\fi}}
\makeatother

\usepackage[pdftex,colorlinks=true,
pdfstartview=FitV,
linkcolor= linkcolor,
citecolor= linkcolor,
urlcolor= linkcolor,
hyperindex=true,
hyperfigures=false]
{hyperref}
\definecolor{linkcolor}{rgb}{0,0,0.6}

\newcommand{\cF}{\mathcal{F}}

\newcommand{\cG}{\mathcal{G}}

\newcommand\Ft{\mathcal{F}_{\rm topo}}

\newcommand{\cO}{\mathcal{O}}

\newcommand{\cN}{\mathcal{N}}

\newcommand{\dd}{{\rm d}}
\newcommand{\ee}{{\rm e}}

\newcommand{\bu}{\textbf{u}}

\newcommand{\bnab}{\boldsymbol{\nabla}}

\newcommand{\bq}{\textbf{q}}
\renewcommand{\br}{\textbf{r}}

\newcommand{\bx}{\textbf{x}}
\newcommand{\by}{\textbf{y}}
\newcommand{\bW}{\textbf{W}}

\newcommand{\ddd}{\mathcal{D}}
\renewcommand{\rmi}{{\rm i}}

\renewcommand{\eref}[1]{(\ref{#1})}
\renewcommand{\Fref}[1]{Fig.~\ref{#1}}

\DeclareMathOperator{\sinc}{sinc}

\definecolor{greenpyplot}{RGB}{0,128,0}

\usepackage{tikz}
\usetikzlibrary{arrows}
\usetikzlibrary{math}

\usepackage{pgfplots}

\if{
\usetikzlibrary{external}
\pgfrealjobname{Figure}
}\fi

\begin{document}

\title[Fluctuation-Induced First Order Transition to Collective Motion]{Fluctuation-Induced First Order Transition to Collective Motion}

\author{David Martin$^1$, Gianmarco Spera$^3$, Hugues Chat\'e$^4$, Charlie Duclut$^2$, Cesare Nardini$^4$, Julien Tailleur$^5$, Fr\'ed\'eric van Wijland$^3$}
\address{$^1$University of Chicago, Kadanoff Center for Theoretical Physics and Enrico Fermi Institute, 933 E 56th St, Chicago, IL 60637}

\address{$^2$ Laboratoire Physico-Chimie Curie, CNRS UMR 168, Institut Curie, Université PSL, Sorbonne Université, 75005, Paris, France
}
%
\address{$^3$ Universit\'e Paris Cité, Laboratoire Mati\`ere et Syst\`emes Complexes (MSC), UMR 7057 CNRS,F-75205 Paris,  France}
%
\address{$^4$ Service de Physique de l'Etat Condens\'e, CEA, CNRS Universit\'e Paris-Saclay, CEA-Saclay, 91191 Gif-sur-Yvette, France}
%
%
\address{$^5$ MIT Biophysics, 182 Memorial Drive, 77 Massachusetts Avenue, Cambridge, MA 02139, USA}
%

\date{\today\ -- \jobname}
\ead{dgmartin@uchicago.edu}

\begin{abstract}
The nature of the transition to collective motion in assemblies of
aligning self-propelled particles remains a long-standing matter of
debate. In this article, we focus on dry active matter and show that
weak fluctuations suffice to generically turn second-order
mean-field transitions into a `discontinuous' coexistence
scenario. Our theory shows how fluctuations induce a
density-dependence of the polar-field mass, even when this effect is
absent at mean-field level. In turn, this dependency on density
triggers a feedback loop between ordering and advection that
ultimately leads to an inhomogeneous transition to collective motion
and the emergence of inhomogeneous travelling bands. Importantly, we
show that such a fluctuation-induced first order transition is present
in both metric models, in which particles align with neighbors within
a finite distance, and in `topological' ones, in which alignment is
based on more complex constructions of neighbor sets. We compute analytically the noise-induced
renormalization of the polar-field mass using stochastic calculus,
which we further back up by a one-loop field-theoretical
analysis. Finally, we confirm our analytical predictions by numerical
simulations of fluctuating hydrodynamics as well as of topological
particle models with either $k$-nearest neighbors or Voronoi
alignment.  
\end{abstract}
\vspace{-1cm}
\submitto{\JSTAT}
\vspace{1cm}

Flocking is an emergent nonequilibrium phenomenon in which
interactions between agents produce collective motion at
large scales. This phenomenon is observed in a plethora of natural
systems such as flocks of starlings~\cite{ballerini2008empirical} or
human crowds~\cite{bain2019dynamic} as well as in artificial systems
ranging from self-propelled
colloids~\cite{bricard2013emergence,iwasawa2021algebraic} and shaken
grains~\cite{deseigne2010collective} to driven
filaments~\cite{schaller2010polar}. From a theoretical perspective,
flocking models typically include three ingredients: agents are
self-propelled, they align with each other, and their orientations
experience some form of
noise~\cite{vicsek1995novel,chate2020dry}. Over the years, such models
have proved relevant in fields as diverse as animal
behavior~\cite{buhl2006disorder,Ballerini2008,cavagna2017dynamic},
virtual entertainment~\cite{fathy2014flocking,reynolds1987flocks},
biology~\cite{schaller2010polar,zhang2010collective,sumino2012large,bi2016motility,liu2019self},
and swarm robotics~\cite{vasarhelyi2018optimized,fine2013unifying}.

Collective motion occurs when the aligning interactions overcome the
noise-induced randomization of the agent orientations. The question of
how this transition occurs, which is of both historical and
paradigmatic value, has led to a wealth of
theoretical~\cite{toner1995long,toner2005hydrodynamics,bertin2006boltzmann,mishra2010fluctuations,ihle2011kinetic,solon2013revisiting},
numerical~\cite{vicsek1995novel,gregoire2004onset,VicsekFirstOrder},
and experimental
works~\cite{narayan2007long,deseigne2010collective,schaller2010polar,bricard2013emergence}. It
is best understood in the context of \textit{metric} models, in which
particles align with neighbors within a finite distance.  At the
macroscopic level, the nature of the transition is now well
established~\cite{gregoire2004onset,VicsekFirstOrder,solon2013revisiting}:
It takes the form of a discontinuous `first-order' transition between
a disordered gas/paramagnetic phase and a polar-ordered
liquid/ferromagnetic phase, which are separated in the phase diagram
by a coexistence region~\cite{solon2015phase}. The origin of this
discontinuous transition has been traced back to a density-dependent
polar-field mass, which triggers a linear instability of the
homogeneous ordered state close to the transition and leads to the
formation of inhomogeneous travelling
structures~\cite{bertin2006boltzmann,bertin2009hydrodynamic,solon2013revisiting,caussin2014emergent,Solon2015PatternFI}. 
Notice, however, that the nature of the transition
is still the topic of ongoing research. For instance, metric alignment
interactions that are non-monotonic in the density were recently
suggested to induce a second-order transition at finite density using
active lattice gases with built-in diffusive scalings~\cite{agranov2024thermodynamically}.


\textit{Topological} or \textit{metric-free} systems, in which interactions between agents are
based on more complex constructions of neighbor sets such as Voronoi or $k$-nearest neighbors, form another important
class of flocking models.  These models play an important role due to their relevance
to studies of groups of
animals~\cite{Ballerini2008,niizato2011metric,gautrais2012deciphering,Camperi2012Interface,ginelli2015intermittent}
or pedestrians~\cite{moussaid2011simple}, where visual cues dominate
metric ones.  They are also the natural choice to model confluent
tissues where surrounding cells determine
interactions~\cite{honda2000differentiation,schaller2005multicellular,bock2010generalized,bi2016motility,barton2017active}.
In topological systems, mean-field descriptions of
Voronoi-based~\cite{peshkov2012continuous} and $k$-nearest-neighbor
models~\cite{chou2012kinetic} both lead to a density-independent
polar-field mass, hence predicting a continuous onset of order
where travelling bands remain absent.  This result is compatible with
the observation that doubling the distance between all particles, and
hence reducing the particle density, does not impact the aligning
dynamics. Topological models are thus expected to be much less
sensitive to density variations and they have been predicted to belong
to a universality class distinct from that of metric
models~\cite{ginelli2010relevance,peshkov2012continuous,chou2012kinetic,Camperi2012Interface}.

However, this conclusion has been recently
challenged~\cite{martin2021fluctuation}, thus triggering a debate
regarding the nature of the transition to collective motion in
topological models. The main point addressed in
Ref.~\cite{martin2021fluctuation} is the impact of fluctuations on the
mean-field descriptions of topological models. Dressing mean field
models with a weak noise has indeed been shown to induce a
density-dependent renormalization of the polar-field mass, which in
turn should generically trigger a discontinuous transition to collective
motion. In this article, we review, detail and extend the results of
Ref.~\cite{martin2021fluctuation}. In particular,
Ref.~\cite{martin2021fluctuation} reported the emergence of inhomogeneous
traveling bands for flocking models in which particles align with
their $k$-nearest neighbors, but it did not tackle the case of Voronoi
neighbors, leaving the door open to a different phenomenology. In this
article, we show that an active Ising model in which active particle
interact with their Voronoi neighbors also displays the
traveling bands that are typical of the first-order scenario. While
this does not close the debate on whether any topological model will
undergo such a discontinuous scenario, it further strengthens the idea
that the latter is the rule, independently of whether the alignment is
metric or topological. Furthermore, we complement the analytical
approach developed in~\cite{martin2021fluctuation} by a
field-theoretical one-loop perturbation theory, which yields
consistent results. We also perform a detailed study of finite size
effects and show that the critical size to observe bands is inversely
proportional to the derivative of the polar-field mass with respect to
the density: systems with a weakly density-dependent polar-field mass will feature a very large critical size. This explains the
difficulty in characterizing the nature of the transition to
collective motion in topological models, in which this
density-dependence is weak. In particular, while we report the direct
numerical observation of an ordered traveling domain for an active Ising model
with Voronoi-based alignment, its Vicsek-based counterpart resisted
our numerical efforts. Our evidence for the discontinuous nature of
the transition in this model thus remain circumstantial.

We first start by introducing in
Sec.~\ref{sec:intro} the models that will be studied throughout this
article, which are variations of the Vicsek Model (VM) and of the
Active Ising Model (AIM), including both their topological and metric
versions. We then review in Sec.~\ref{sec:MF_AIM} the mean-field
theory of the AIM. We present in Sec.~\ref{sec:hydro_AIM} a simple
method to build the mean-field equations and we show how their
predictions fail in the presence of noise in
Sec.~\ref{sec:linearStab_meanField}. To better understand the
emergence of the discontinuous scenario, we then show in
Sec.~\ref{sec:LinStab} that the minimal ingredients leading to the
linear instability of the homogeneous ordered phase are a
density-dependent polar-field mass and the advection of the density
field by the polar one. Next we demonstrate analytically in
Sec.~\ref{sec:FIFOT} that such a density-dependent polar-field mass is
generically generated by complementing the mean-field description of the AIM
with fluctuations. We present two complementary approaches to derive the
noise-induced density-dependent renormalization of the polar-field
mass in the one-dimensional mean-field description of the AIM
(Sec.~\ref{subsec:renorm_AIM} and~\ref{subsec:AIM_field_theory}), and
show these results to also hold in two dimensions
(Sec.~\ref{subsec:AIM_metric_2d}). We then extend our analysis to
topological AIMs in Sec.~\ref{part:FIFOT_topological}, and to
Vicsek-like models in Sec.~\ref{part:FIFOT_polar_case}. Throughout
this article, many numerical results are presented and the
corresponding numerical simulations are detailed
in~Appendix~\ref{sec:numerics}. We note that several of our
simulations are implemented in one dimension where it is well known
that flocks are metastable and ultimately undergo
reversals~\cite{o1999alternating,solon2013revisiting,benvegnen2022flocking}. All
our simulations were carried out using simulation times shorter than
the average time between reversals, to focus on the nature of the
transition and not on this otherwise interesting phenomenon. Finally,
we stress that part of our numerical and theoretical results have
already been announced in a shorter
account~\cite{martin2021fluctuation}. Everything which is not
explicitly described as such is new.

\section{Metric and topological flocking models}
\label{sec:intro}
\label{sec:meanField_AIM}

Historically, the emergence of collective motion was first studied as
a phase transition in the Vicsek Model~\cite{vicsek1995novel}.
In this two-dimensional model, one considers self-propelled point-like
particles carrying unit propulsion vectors $\bu_i$ and moving in an
$L_x\times L_y$ domain with periodic boundary conditions.  
In the following discrete-time version, particles align at each step according to
\begin{equation}
    \label{eq:Vicsek_micro}
    \text{arg}\Big[\bu_i\Big] \to \text{arg}\Big[\sum_{j\in \mathcal{N}_{i}} \bu_j \Big] + \sigma \eta\;,
  \end{equation}
where $\eta$ is a noise uniformly drawn in $[-\pi,\pi]$~\footnote{Note
that, alternatively, one may also consider adding a vectorial noise
inside the $\arg$ function.}.  In practice, the set $\mathcal{N}_{i}$
depends on the current configuration of the particles and can describe
any type of polar, ferromagnetic alignment, be it metric, in which case particles align with
neighbors within a finite distance, or topological, where interactions
are based on more complex constructions.
In this article, we consider three distinct cases: metric
alignment up to a distance $r_0$~\cite{vicsek1995novel}, topological
alignment between $k$-nearest neighbors~\cite{Ballerini2008}, and
topological alignment between Voronoi
neighbors~\cite{ginelli2010relevance}.  For the metric alignment, the set
$\mathcal{N}_{i}$ is defined as
\begin{align}
  \label{eq:OLAIM_dyn}
  \cN_i=\{j\ \text{such that}\ |\br_j-\br_i|<r_0\}\;,
\end{align}
where $r_0$ is the interaction range.
For the $k$-nearest-neighbor alignment, it is given by
\begin{align}
  \label{eq:OLAIM_dyn_knearest}
  \cN_i=\{\text{set of $k$-nearest neighbors of particle $i$}\}\;.
\end{align}
For the alignment between Voronoi neighbors it is given by
\begin{align}
  \label{eq:OLAIM_dyn_voronoi}
  \cN_i=\{\text{set of Voronoi neighbors of particle $i$}\}\;.
\end{align}
Historically, the metric version \eref{eq:OLAIM_dyn} of the VM has
been the first studied~\cite{vicsek1995novel}, while the topological
ones~\eref{eq:OLAIM_dyn_knearest}-\eref{eq:OLAIM_dyn_voronoi} were
introduced
later~\cite{peshkov2012continuous,chou2012kinetic,peshkov2014boltzmann}.
\if{ What originally attracted physicists' interest is the
  emergence of long range order within the polar flocking phase in 2D.
  This true long range order, which is prohibited for XY spins due to
  the Mermin-Wagner theorem~\cite{mermin1966absence}, relies on the
  self-propelled nature of the Vicsek Model's flying spins which
  drives the system out of equilibrium.  It has been captured in a
  generic theory for polar active fluid~\cite{toner1995long}, observed
  in numerical simulation~\cite{mahault2019quantitative} and also
  confirmed in experiments~\cite{iwasawa2021algebraic}.  }\fi The
nature of the phase transition to collective motion in these models
has been a topic of long-standing debate and interest. In the metric
case \eref{eq:OLAIM_dyn}, the transition was first described as a
continuous ferromagnetic-like transition taking the system from a
homogeneous disordered phase to a homogeneous ordered
one~\cite{vicsek1995novel}.  Later, the transition was instead shown
to be discontinuous, with the emergence of an inhomogeneous phase
separating the disordered and ordered phases~\cite{gregoire2004onset}.
This phase was missed in early studies because the transition is only
weakly first-order: one has to consider system sizes above a
large---but finite---critical length to observe the travelling
bands. For smaller systems, the transition appears critical. For the
topological cases, \eref{eq:OLAIM_dyn_knearest} and
\eref{eq:OLAIM_dyn_voronoi}, the flocking transition was also first
reported as continuous and no travelling bands were observed in early
studies, despite the use of large
systems~\cite{peshkov2012continuous,chou2012kinetic}. Therefore, it
seemed that metric and topological flocking models could belong to two
different universality classes, leading to different transitions to
collective motion.

To assess the genericity of these results and to allow for analytical
progress, another flocking model was introduced: the Active Ising
Model (AIM)~\cite{solon2013revisiting}.  In the (off-lattice version
of the) AIM, the continuous orientation vectors $\bu_i$
($\mathcal{O}_2$ symmetry) of the VM are replaced by discrete ones
($\mathcal{Z}_2$ symmetry): each particle carries an Ising spin
$\sigma_i=\pm 1$ indicating its orientation $\pm \bu_x$, with $\bu_x$
the unit vector in the $x$-direction. Particles move in an
$L_x\times L_y$ domain and the position $\br_i(t)$ of the $i$-th
particle at time $t$ evolves according to
\begin{align}
  \label{eq:flying_xy_spatial}
  \dot{\br}_i(t)= v \sigma_i(t) \bu_x + \sqrt{2D}\boldsymbol{\eta}_i(t)\;,
\end{align}
where $\boldsymbol{\eta}_i$ is a Gaussian white noise such that
$\langle\eta_{i,\alpha}(t)\eta_{j,\beta}(t')\rangle =
\delta_{ij}\delta_{\alpha\beta}\delta(t-t')$\footnote{Here, Greek
indices refer to spatial coordinates while Latin letters refer to
particle numbers.}, $v$ is the self-propulsion speed, and $D$ is a
translational diffusivity. The spin $\sigma_i$ of the $i$-th particle
then flips at a rate
\begin{align}
  \label{eq:flying_xy_alignment}
  W(\sigma_i \rightarrow -\sigma_i,t)=\Gamma e^{-\beta \sigma \bar{m}_i(\{\br_i(t)\})}\;,
\end{align}
where $\Gamma$ is a constant rate, $\beta$ is an `inverse temperature'
that controls the strength of the alignment, and the field
$\bar{m}_i(\{\br_i\})$ is given by
\begin{align}\label{eq:external_field}
  \bar{m}_i(\{\br_i\})= \sum_{j\in\cN_i}\sigma_j\bigg{/}\sum_{j\in\cN_i}1\;.
\end{align}
Note that the field $\bar{m}_i$ depends on the configuration of the
system through the set $\cN_i$, which characterizes the type of
alignment: metric for \eref{eq:OLAIM_dyn}, or topological for
$k$-nearest neighbors alignment \eref{eq:OLAIM_dyn_knearest} and
Voronoi alignment \eref{eq:OLAIM_dyn_voronoi}.  In this article, we
will show that, independently of the alignment, be it metric or
topological, the transition to collective motion generically remains
discontinuous in the AIM. This result had already been reported for
the metric~\cite{solon2013revisiting} and
$k$-nearest-neighbors~\cite{martin2021fluctuation} cases, but the
Voronoi case is new.

\section{The Mean-field description of the AIM}
\label{sec:MF_AIM}

Let us start by providing a simple derivation of the mean-field theory
of the AIM that can be used for any aligning field $\bar{m}_i$ and
thus any choice of $\cN_i$.

\subsection{A simple mean-field approach}
\label{sec:hydro_AIM}

To construct a mean-field description of aligning active particles, we
first consider the case of non-interacting particles whose
orientations experience alignment with an external field~$\bar
m(\br)$. Then, a mean-field description of the system is obtained by
replacing this external field by a functional of the particle
orientational and density field, $m(\br)$ and $\rho(\br)$,
respectively. The choice of $\bar m(\br,[\rho,m])$ then allows
describing a variety of aligning interactions, be they metric or
topological.

Let us show how to proceed in the case of the AIM.  In
Eq.~\eqref{eq:flying_xy_alignment}, we replace the many-body aligning
field $\bar{m}_i(\{\br_i\})$ with an external field $\bar{m}(\br)$ and
obtain
\begin{align}
  \label{eq:flying_xy_mf}
  W(\sigma_i \rightarrow -\sigma_i )= e^{-\beta \sigma \bar{m}(\br)}\;,
\end{align}
where we have set $\Gamma=1$ without loss of generality.  The
dynamics~\eqref{eq:flying_xy_spatial} then becomes non-interacting and
the one-body probability density $P(\br,\sigma,t)$ to find a particle
at position $\br$ at time $t$ with spin $\sigma$ evolves as:
\begin{equation}\label{eq:FP_flying_xy}
  \partial_t P(\br,\sigma,t)= D \nabla^2 P(\br,\sigma,t) -\partial_x[v \sigma P(\br,\sigma,t)]- e^{-\beta \sigma \bar{m}(\br)}P(\br,\sigma,t) + e^{\beta \sigma \bar{m}(\br)}P(\br,-\sigma,t)\;.
\end{equation}
\if{
  \begin{subequations}%
\begin{align}%
  \label{eq:FP_flying_xy_1}
  \partial_t P(\br,1,t)=& D \nabla P(\br,1,t) -\partial_x(v P(\br,1,t))- e^{-\beta \bar{m}(\br)}P(\br,1,t) +  e^{\beta \bar{m}(\br)}P(\br,-1,t)\;, \\
  \label{eq:FP_flying_xy_2}
  \partial_t P(\br,-1,t)=& D \nabla P(\br,\-1,t) +\partial_x(v P(\br,-1,t))+ e^{-\beta \bar{m}(\br)}P(\br,\-1,t) -  e^{\beta \bar{m}(\br)}P(\br,-1,t)\;.
\end{align}%
  \end{subequations}
  \fi One then introduces the one-body density and magnetization
  fields, $\rho(\br,t)=P(\br,-1,t)+P(\br,1,t)$ and
  $m(\br,t)=P(\br,1,t)-P(\br,-1,t)$, respectively. As frequently done
  in active matter, we refer below to $m$ as the polar
  field. Equation~\eref{eq:FP_flying_xy} is then equivalent to the
  joint evolution equations for $\rho(\br,t)$ and $m(\br,t)$ given by:
\begin{subequations}\label{eq:MF_active_spins}
\begin{align}
  \label{eq:MF_active_spins_rho}
  \partial_t \rho =& D\nabla^2 \rho-\partial_x(v m) \, ,\\
  \label{eq:MF_active_spins_m}
  \partial_t m =& D\nabla^2 m-\partial_x(v \rho) + \cF(\rho, m, \bar{m})\;,
\end{align}
\end{subequations}
where $\cF$ is given by
\begin{align}
\label{eq:f_bar}
  \cF(\rho, m, \bar{m})=2 \rho\sinh(\beta\bar{m})-2 m \cosh(\beta\bar{m})\;.
\end{align}
To turn Eq.~\eref{eq:MF_active_spins} into a mean-field description of
our microscopic model, we now replace the external field
$\bar{m}(\br)$ by a functional of the fields $\rho$ and $m$. In the
case of metric alignment~\eqref{eq:OLAIM_dyn}, $\bar{m}(\br)$
corresponds to
\begin{equation}
  \label{eq:mf_field_metric}
  \bar{m}(\br)=\frac{m(\br)}{\rho(\br)}\;,
\end{equation}
while its expression for topological models will be discussed in
Sec.~\ref{part:FIFOT_topological}.  We can now simplify the expression
of $\cF$ in~\eqref{eq:f_bar} at the onset of order by performing a
Taylor expansion up to third order in $\bar m$
\begin{align}
  \label{eq:F_onset}
  \cF(\rho,m)=2\rho \left(\beta\bar{m}+\frac{\beta^3\bar{m}^3}{6} \right)-2 m \left(1+\frac{\beta^2\bar{m}^2}{2} \right) \;.
\end{align}
Inserting \eref{eq:mf_field_metric} into \eref{eq:F_onset} allows
rewriting the mean-field dynamics~\eref{eq:MF_active_spins_m} at the
onset of order as
\begin{subequations}\label{eq:MF_AIM_final}
\begin{align}
  \label{eq:MF_AIM_rho}
  \partial_t \rho =& D\nabla^2 \rho-\partial_x(v m)\;,\\
  \label{eq:MF_AIM_m}
  \partial_t m =& D\nabla^2 m-\partial_x(v \rho) - \alpha m - \gamma\frac{m^3}{\rho^2} \, ,
\end{align}
\end{subequations}
where $\alpha=2(1-\beta)$ is the `mass' of the polar field $m$ and
$\gamma=2\left( \frac{\beta^2}{2}-\frac{\beta^3}{6}\right)$.

The similarity between Eq.~\eqref{eq:MF_AIM_m} and the Landau theory
of the equilibrium Ising model naturally leads us to identify $\alpha$
as $\alpha\propto T-T_c$. When $\alpha\geq 0$, the sole homogeneous
solution to Eq.~\eqref{eq:MF_AIM_final} is $\rho(\br)=\rho_0$ and
$m(\br)=0$, which we refer to as the ``disordered high-temperature
phase''. On the contrary, when $\alpha<0$, a second homogeneous
solution emerges: $\rho(\br)=\rho_0$ and
$m(\br)=m_0=\rho_0\sqrt{\frac{|\alpha|}\gamma}$. We refer to this
phase as the ``low-temperature ordered phase''. Importantly, we see
that the mean-field theory predicts a constant ``critical
temperature'' $\beta^{-1}=1$, independent of the average density
$\rho_0$ of the system. 

\subsection{Failure of the mean-field theory}
\label{sec:linearStab_meanField}
Let us now show that the mean-field
hydrodynamics~\eqref{eq:MF_AIM_final} fails at correctly describing the
transition to collective motion observed in the AIM. To do so, we
start by performing a linear stability analysis of the homogeneous
phases of~\eref{eq:MF_AIM_final} at the onset of order.

For simplicity, and without loss of generality, we perform the study
in one dimension and rewrite Eq.~\eref{eq:MF_AIM_final} in a
dimensionless form as:
\begin{subequations}\label{eq:active_Ising_mf_metric}%
\begin{align}%
  \partial_{\tilde{t}}\rho &=  \partial_{\tilde{x}\tilde{x}}\rho-\partial_{\tilde{x}} m \\
  \partial_{\tilde{t}} m &= \partial_{\tilde{x}\tilde{x}} m - \partial_{\tilde{x}} \rho - \tilde{\alpha}m-\tilde{\gamma} \frac{m^{3}}{\rho^{2}}\;,
\end{align}%
\end{subequations}%
where we used the dimensionless variables $\tilde{x}=x v/D$, $\tilde{t}=tv^{2}/D$, and we defined $\tilde{\alpha}=D\alpha/v^{2}$ and $\tilde{\gamma}=D\gamma/v^{2}$.
In the following, we drop the tilde to lighten the notation.

Considering a perturbation around the homogeneous ordered profile, $\rho(x,t) = \rho_0 + \delta\rho(x,t)$ and $m(x,t)=m_0 + \delta m(x,t)$, the linearized dynamics of the perturbations $\delta m$ and $\delta \rho$ in Fourier space then reads
\begin{equation}
  \label{eq:dynamic_alpha_rho_metric}
  \partial_t
\begin{pmatrix}
\delta {\rho_q} \\
\delta {m_q}
\end{pmatrix} = \begin{pmatrix}
- q^{2} & -\rmi q \\
-\rmi q  - 2\alpha\sqrt{|\alpha|/\gamma} & -q^{2} + 2\alpha
\end{pmatrix}
\begin{pmatrix}
\delta \rho_q \\
\delta m_q
\end{pmatrix}\;,
\end{equation}
where we have used the relation
$m_0=\rho_0\sqrt{|\alpha|/\gamma}$. The system is linearly unstable
whenever the growth rate of the perturbation is positive,
\textit{i.e.} when the matrix in
Eq.~\eqref{eq:dynamic_alpha_rho_metric} has an eigenvalue with a
positive real part. Painful but straightforward algebra detailed in
App.~\ref{app:linear_stab} shows that this never happens when $\alpha$
is a constant, independent of the value of the density field $\rho_0$.
\if{
One can show (see App.~\ref{app:linear_stab} for details) that this is equivalent to having $b(q)>0$, where
\begin{align}
  \label{eq:instability_criterion_metric}
  b(q) \!=& 16 q^2 \left(\frac{2|\alpha_0|\alpha_0^2}{\gamma}+4 (2 \alpha_0 -1) \alpha_0 ^2\!\right)-64 \alpha_0  (5 \alpha_0 -2) q^4+64 (4 \alpha_0 -1) q^6-64 q^8 \, .
\end{align}
Note that, because we are linearizing around the ordered phase, we must have $\alpha_0<0$.
This implies that only the term of order $q^2$ in $b(q)$ can change sign and become positive to trigger an instability.
Finally, since we are interested in the behavior of the system at the onset of order, $|\alpha_0|\ll 1$, and the necessary condition for the emergence of an instability at lowest order in  $|\alpha_0|$ then reads
\begin{equation}
  - 4 \alpha_0^2 >0 \, ,
\end{equation}
which is never satisfied.}\fi The homogeneous ordered solution is thus
always linearly stable close to the onset of order.

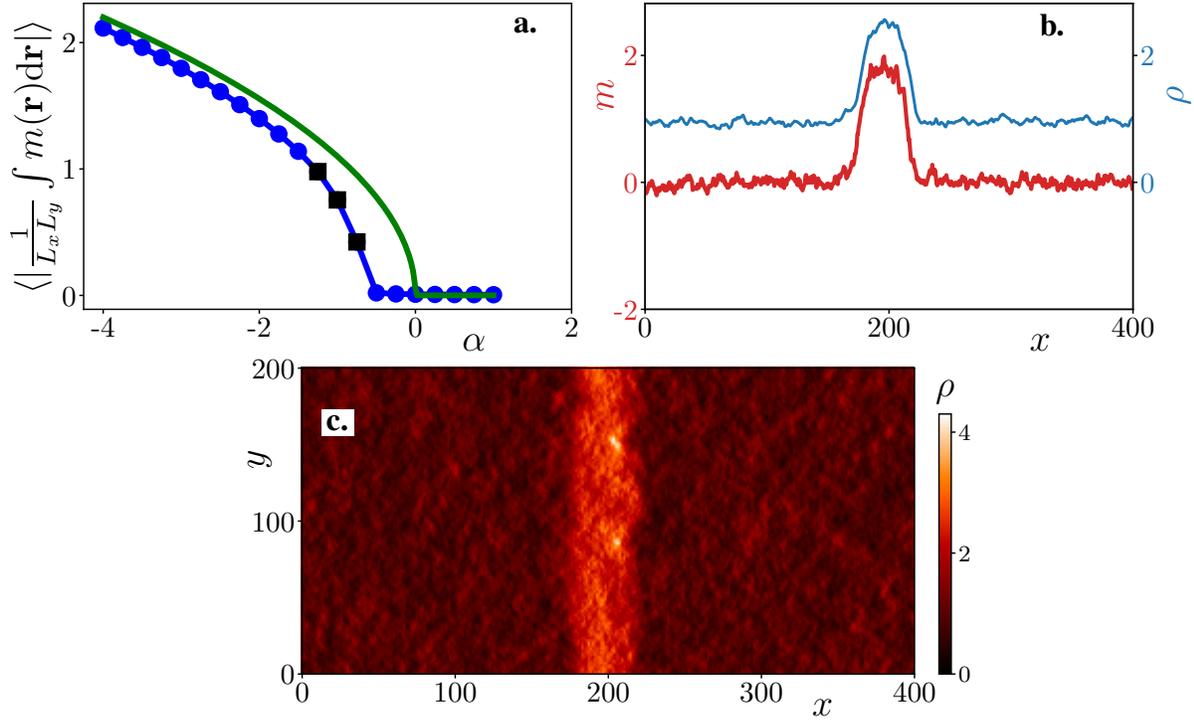
\begin{figure}[t]
  \flushright
  \begin{tikzpicture}
    \def\y{4.5}
    \node at (0,0) {\includegraphics[width=\textwidth]{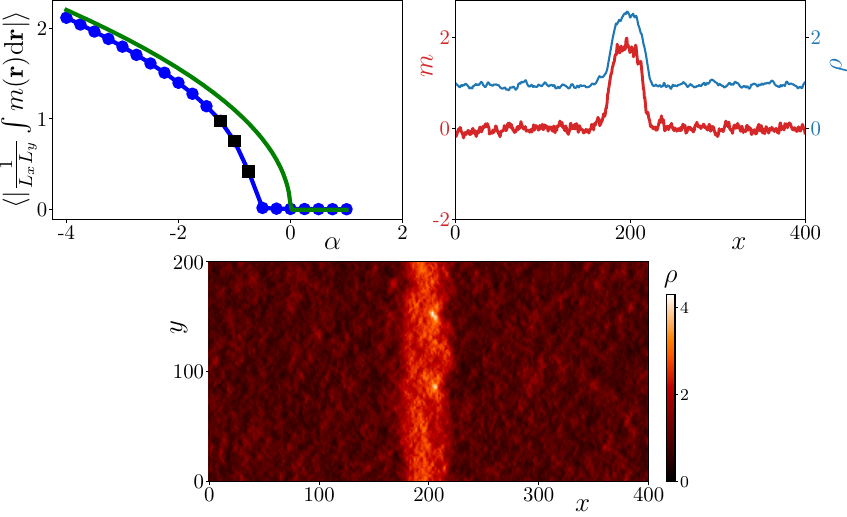}};
    \node[font=\bf] at (-1,\y) {a.};
    \node[font=\bf] at (6,\y) {b.};
    \node[font=\bf,fill=white,inner sep=2pt] at (-3.5,-0.8) {c.};
  \end{tikzpicture}
  \vspace{-0.3cm}
  \caption{ Numerical simulations of
    \eqref{eq:hydro_AIM_noise_2D_1}-\eqref{eq:hydro_AIM_noise_2D_2}. {\bf
      a.} Average magnetization as $\alpha$ is varied.  The
    transition occurs at a value $\alpha_c<0$, shifted from the
    mean-field prediction $\alpha_c=0$ (green line).  At the onset of
    order, inhomogeneous profiles (black squares) separate homogeneous
    ordered and disordered phases (blue dots).  Parameters:
    $D=v=\gamma=\sigma=1$, $dx=0.5$, $dt=0.01$, $L_x=400$, $L_y=40$,
    $\rho_0\equiv N/(L_x L_y)=1.1$.  {\bf c.} A snapshot close to
    the transition shows an ordered travelling band in a disordered
    background.  {\bf b.} The corresponding density and
    magnetization fields averaged along $y$.  Parameters: same as
    before up to $L_y =200$, $dx=0.5$, $\alpha = -0.9$. 
    }
    \label{fig:SDEmetric}
\end{figure}

However, particle-level simulations using Eq.~\eqref{eq:OLAIM_dyn}
contradict this prediction. They instead show that the ordered phase is
unstable close to the transition, where ordered travelling domains are
observed~\cite{solon2015flocking}. The mean-field description
Eq.~\eqref{eq:MF_AIM_final} of the metric AIM thus fails to capture
the first-order nature of the transition. A natural culprit for this
failure is the effect of fluctuations, which are hitherto neglected at
mean-field level. To confirm this hypothesis, we carried out
simulations of the mean-field evolution~\eref{eq:MF_AIM_final}
supplemented by fluctuations according to
\begin{subequations} \label{eq:fluctuating_mf_AIM}
\begin{align}
  \label{eq:hydro_AIM_noise_2D_1}
  \partial_t \rho &= D \nabla^2\rho -v\partial_x m \, , \\
  \label{eq:hydro_AIM_noise_2D_2}
  \partial_t m &= D \nabla^2 m - v \partial_x \rho -\mathcal{F}(\rho,m) + \sqrt{2 \sigma \rho} \, \eta
  \;,
\end{align}
\end{subequations}
where $\eta$ is a Gaussian white noise of variance
$\langle\eta(\bx,t)\eta(\by,t')\rangle=\delta(\bx-\by)\delta(t-t')$
and $\sigma$ controls the strength of the fluctuations.  As shown in
Fig.~\ref{fig:SDEmetric}, the continuous transition predicted in the
mean-field case ($\sigma=0$) is instead replaced by the standard first
order scenario~\cite{solon2013revisiting,solon2015flocking} and the
emergence of travelling-band solutions.  Therefore, including
fluctuations to the mean-field evolution \eref{eq:MF_AIM_final} allows
recovering the phenomenology observed in the particle-level simulations.

Before we quantify how fluctuations dress the mean-field dynamics to
yield a discontinuous transition in Sec.~\ref{sec:FIFOT}, we first
investigate in Sec.~\ref{sec:LinStab} the minimal ingredients of a
hydrodynamic theory that suffice to make the ordered phase unstable near the
onset of order.

\section{First-order transition to collective motion: mechanism and finite-size effects}
\label{sec:LinStab}
As mentioned in the introduction, the linear instability of
homogeneous ordered profiles close to the onset of order can be
connected to the density-dependence of the linear term entering the
dynamics of the average polar field (denoted by $\alpha$ in
Eq.~\eqref{eq:F_onset})~\cite{bertin2006boltzmann,bertin2009hydrodynamic,solon2013revisiting,caussin2014emergent,Solon2015PatternFI}. As
we show in Sec.~\ref{subsec:minimal}, the presence of a
density-dependent $\alpha$ and of the advection of the density by the
polar field are indeed sufficient to generate an instability of the
ordered phase at its onset, independently of the sign of
$\alpha'(\rho_0)$. 
We then restrict our analysis to the hydrodynamic theory of the AIM in
Sec.~\ref{subsec:linstap_AIM_renorm} and show that, indeed, a
density-dependent $\alpha$ suffices to generate a linear
instability. This allows us to address in Sec.~\ref{sec:finite_size}
the importance of finite-size effects, by showing that the critical
size above which this instability can be seen diverges as $L_c \sim
1/|\alpha'(\rho_0)|$ when $\alpha'(\rho_0)\to 0$: the weaker the
density-dependency of $\alpha$ with $\rho$, the larger the system
needs to be for the discontinuous nature of the transition to be seen.

\subsection{The origin of the ordered-phase instability}
\label{subsec:minimal}
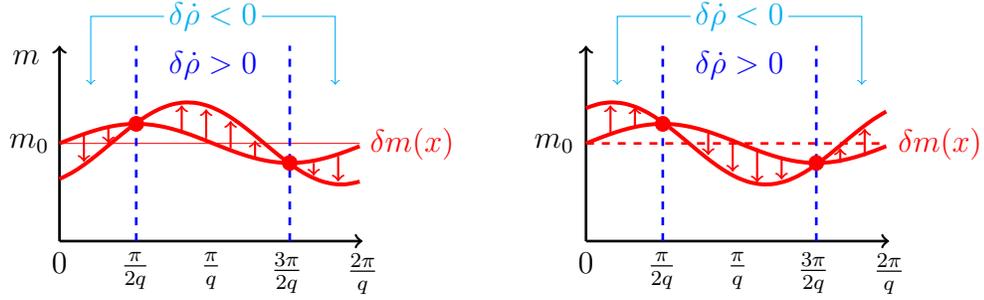
\begin{figure}
  \flushright
  \includegraphics{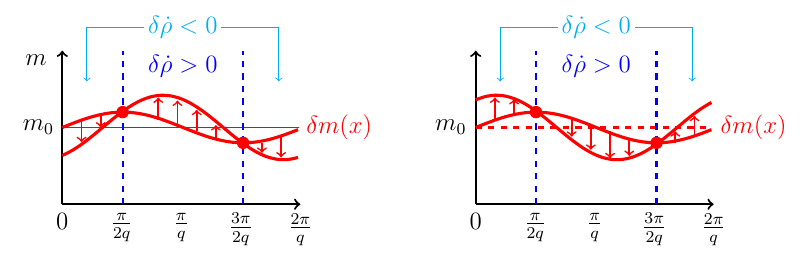}
  \if{
  \beginpgfgraphicnamed{Fig2}
  \begin{tikzpicture}
    \node at (0,0) {
      \begin{tikzpicture}[domain=0:3.07,scale=1.3]
        \draw[line width=1pt,->] (0,0) node[anchor=north] {$0$} -- (3.1,0) node[anchor=north] {$\frac{2\pi}{q}$};
        \draw (3.1*.5,0) node[anchor=north] {$\frac{\pi}{q}$};
        \draw (3.1*.25,0) node[anchor=north] {$\frac{\pi}{2q}$};
        \draw (3.1*.75,0) node[anchor=north] {$\frac{3\pi}{2q}$};
        \draw[line width=1pt,->] (0,0) -- (0,2) node[pos=0.94,anchor=east,xshift=-3] {$m$};
        \draw[red,line width=0pt] (0,1) node[anchor=east] {${\color{black} m_0}$} -- (3.07,1) node[anchor=west] {$\delta m(x)$};
        \draw[red,line width=1.5pt] plot[smooth] function{0.2*sin(2*x)+1};
        \draw[line width=1pt,dashed,blue] (3.14*.25,0) -- (3.14*.25,2);
        \draw[line width=1pt,dashed,blue] (3.14*.75,0) -- (3.14*.75,2);
        \draw[blue] (3.14*.5,1.8) node {$ \delta\dot \rho>0$};
        \draw[cyan] (3.14*.5,2.3) node {$ \delta \dot \rho<0$};
        \draw[cyan,->] (3.14*.5-.5,2.3) -- + (-.75,0) -- +(-.75,-0.7);
        \draw[cyan,->] (3.14*.5+.5,2.3) -- + (.75,0) -- +(.75,-0.7);

        \def\x{.25}
        
        \draw[red,->,thick] (\x,1.1) -- + (0,-.29);
        \draw[red,->,thick] (.5,1.15) -- + (0,-.15);
        
        \draw[red,->,thick] (1.25,1.1) -- + (0,+.29);
        \draw[red,->,thick] (1.5,1.05) -- + (0,+.30);
        \draw[red,->,thick] (1.75,.95) -- + (0,+.28);
        \draw[red,->,thick] (2,.85) -- + (0,+.18);
        
        \draw[red,->,thick] (2.6,.8) -- + (0,-.13);
        \draw[red,->,thick] (2.85,.87) -- + (0,-.26);
        
        \draw[red,line width=1.5pt] plot[smooth] function{0.42*sin(2*x-1.05)+1};
        
        \filldraw[red] (3.14*.25,1.2) circle (.075);
        \filldraw[red] (3.14*.75,.8) circle (.075);
        
      \end{tikzpicture}
    };
    
  \node at (7,0) {
  \begin{tikzpicture}[domain=0:3.07,scale=1.3,shift={(0,0)},anchor=center]

  \draw[line width=1pt,->] (0,0) node[anchor=north] {$0$} -- (3.1,0) node[anchor=north] {$\frac{2\pi}{q}$};
  \draw (3.1*.5,0) node[anchor=north] {$\frac{\pi}{q}$};
  \draw (3.1*.25,0) node[anchor=north] {$\frac{\pi}{2q}$};
  \draw (3.1*.75,0) node[anchor=north] {$\frac{3\pi}{2q}$};
  \draw[line width=1pt,->] (0,0) -- (0,2);
  \draw[red,line width=1pt,dashed] (0,1) node[anchor=east] {${\color{black} m_0}$} -- (3.07,1) node[anchor=west] {$\delta m(x)$};
  \draw[red,line width=1.5pt] plot[smooth] function{0.2*sin(2*x)+1};
  \draw[line width=1pt,dashed,blue] (3.14*.25,0) -- (3.14*.25,2);
  \draw[line width=1pt,dashed,blue] (3.14*.75,0) -- (3.14*.75,2);
  \draw[blue] (3.14*.5,1.8) node {$\delta \dot \rho>0$};
  \draw[cyan] (3.14*.5,2.3) node {$\delta \dot \rho<0$};
  \draw[cyan,->] (3.14*.5-.5,2.3) -- + (-.75,0) -- +(-.75,-0.7);
  \draw[cyan,->] (3.14*.5+.5,2.3) -- + (.75,0) -- +(.75,-0.7);

  \draw[red,->,thick] (.25,1.1) -- + (0,+.29);
  \draw[red,->,thick] (.5,1.15) -- + (0,+.22);

  \draw[red,->,thick] (1.25,1.1) -- + (0,-.22);
  \draw[red,->,thick] (1.5,1.05) -- + (0,-.34);
  \draw[red,->,thick] (1.75,.95) -- + (0,-.35);
  \draw[red,->,thick] (2,.85) -- + (0,-.22);

  \draw[red,->,thick] (2.6,.8) -- + (0,.15);
  \draw[red,->,thick] (2.85,.87) -- + (0,.28);

  \draw[red,line width=1.5pt] plot[smooth] function{0.42*sin(2*x-5.25)+1};

  \filldraw[red] (3.14*.25,1.2) circle (.075);
  \filldraw[red] (3.14*.75,.8) circle (.075);

  \end{tikzpicture}
  };
  \end{tikzpicture}
  \endpgfgraphicnamed{Fig2}
  }\fi

\caption{Sketch of the fate of an initial fluctuation $\delta m$ of
  the magnetization field, when $\alpha'(\rho)<0$ ({\bf left}) or
  $\alpha'(\rho)>0$ ({\bf right}). The initial fluctuation triggers an
  increase of the density field ahead of its maxima and a decrease in
  its wake. Close to the onset of order $\delta m \simeq
  -\alpha'(\rho_0) \delta \rho$ so that $\delta m$ remains constant at
  its maxima, but it is amplified either ahead of them ($\alpha'<0$,
  left panel) or behind them ($\alpha'>0$, right panel). In both
  cases, the initial fluctuation is amplified and the sign of
  $\alpha'(\rho_0)$ only determines the direction in which it
  propagates.}
\label{fig:instab_lin}
\end{figure}

We consider a minimal model in which, to model self-propulsion, the
density field is advected by the magnetization field and we choose an
ordering dynamics with a density-dependent linear term:
\begin{align}
  \label{eq:minimal_hydro}
  \dot{\rho}=-v\partial_x m\;, \qquad \dot{m}=-\alpha(\rho)m-\Gamma m^3 \;.
\end{align}
Let us consider the impact of a small perturbation $\delta m= \epsilon
\sin(qx)$ on a homogeneous ordered profile $\rho=\rho_0$,
$m=m_0=\sqrt{-\alpha/\Gamma}$ (See Fig.~\ref{fig:instab_lin}). A
positive fluctuation of $\delta m$ enhances the rightward motion of the
particle while a negative fluctuation favors leftward motion so that
$\delta \rho$ is enhanced ahead of the perturbation and decreased
behind its maxima. Mathematically, since $\dot{\rho}=-v\epsilon q
\cos(qx)$, $\rho(x)$ develops an out-of-phase response to $\delta m$
that vanishes only at the extrema of $\delta m$. To linear order,
$\partial_t \delta m=2 \alpha \delta m - \alpha'(\rho_0) \delta \rho$
so that, close to the onset of order, $\partial_t \delta m \simeq -
\alpha'(\rho_0) \delta \rho$. If $\alpha'(\rho_0)<0$, $\delta m$
follows $\delta \rho$ and increases ahead of its maxima---where it
remains constant since $\delta\rho$ vanishes there---leading to the
rightward propagation and to the amplification of $\delta m$
(\Fref{fig:instab_lin}, left panel). If $\alpha'(\rho_0)<0$, $\delta
m$ increases behind its maxima, leading to a leftward
propagation of the fluctuation, which is again amplified
(\Fref{fig:instab_lin}, right panel). In both cases, an initial
perturbation is necessarily amplified, leading to a linear
instability.

Equation~\eqref{eq:minimal_hydro} is of course too simple to capture
the full transition of flocking models, but it captures the main
ingredient of the linear instability of the ordered phase: a
fluctuation of the polar field leads to an out-of-phase fluctuation of
the critical temperature. In turn, this leads to an amplification of
the initial fluctuation of order, either ahead or behind of its
initial maxima. Advection of the density by the polar field and a
density-dependent critical temperature are thus sufficient ingredients to
prevent the occurrence of a continuous transition.

\subsection{Linear stability analysis of the renormalized AIM}
\label{subsec:linstap_AIM_renorm}


We now discuss in more detail the instability leading to the
emergence of travelling domains at the onset of order for the metric
Active Ising Model.  As highlighted in Sec.~\ref{subsec:minimal}, the
only missing ingredient preventing the formation of ordered domains in the
mean-field description~\eqref{eq:MF_AIM_final} is a density-dependent
linear Landau term $\alpha(\rho)$. As suggested
in~\cite{solon2013revisiting} and shown
in~\cite{martin2021fluctuation}, fluctuations actually dress the
mean-field value of $\alpha$ by density-dependent corrections.  The
derivation of such corrections is detailed in
Sec.~\ref{sec:FIFOT}. For now, we simply postulate that the main
effect of fluctuations on the mean-field Eq.~\eref{eq:MF_AIM_final} is
the addition of an unspecified density-dependent polar-field mass:
\begin{subequations}\label{eq:MF_AIM_renormalized_generic}
\begin{align}
  \label{eq:MF_AIM_rho_ren_generic}
  \partial_t \rho =& D\nabla^2 \rho-\partial_x(v m)\;,\\
  \label{eq:MF_AIM_m_ren_generic}
  \partial_t m =& D\nabla^2 m-\partial_x(v \rho) - \alpha(\rho) m - \gamma\frac{m^3}{\rho^2}\;.
\end{align}
\end{subequations}
To study the emergence of traveling domains, we then perform a linear stability
analysis of the homogeneous ordered fixed points of
Eq.~\eref{eq:MF_AIM_renormalized_generic}, detailed in
App.~\ref{app:linear_stab}. As expected from
Sec.~\ref{subsec:minimal}, the necessary and sufficient condition for
a linear instability of the ordered phase at onset reads
\begin{equation}
\label{eq:stab_cond}
\alpha'(\rho_0)^2 >0 \, .
\end{equation}
The homogeneous ordered solution is thus unstable as soon
$\alpha'(\rho_{0})\neq 0$. Let us now discuss the critical system size
beyond which this instability can be observed.

\subsection{Finite-size effects}
\label{sec:finite_size}

\begin{figure}[t]
\flushright
\begin{minipage}[c]{0.65\linewidth}
  \includegraphics{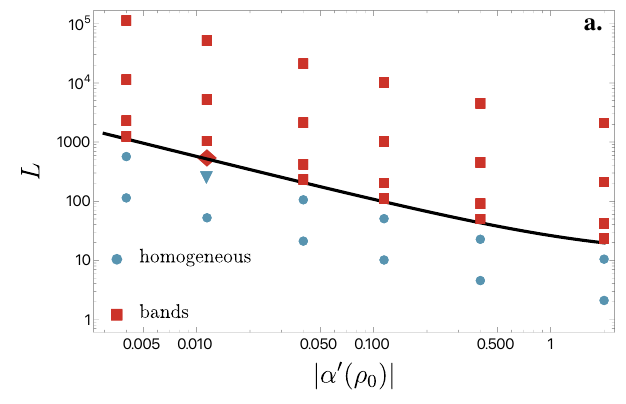}
  \if{
    \beginpgfgraphicnamed{Fig3a}
    \begin{tikzpicture}
      \def\y{0}
      \path (0,\y) node {\includegraphics[width=1.\textwidth]{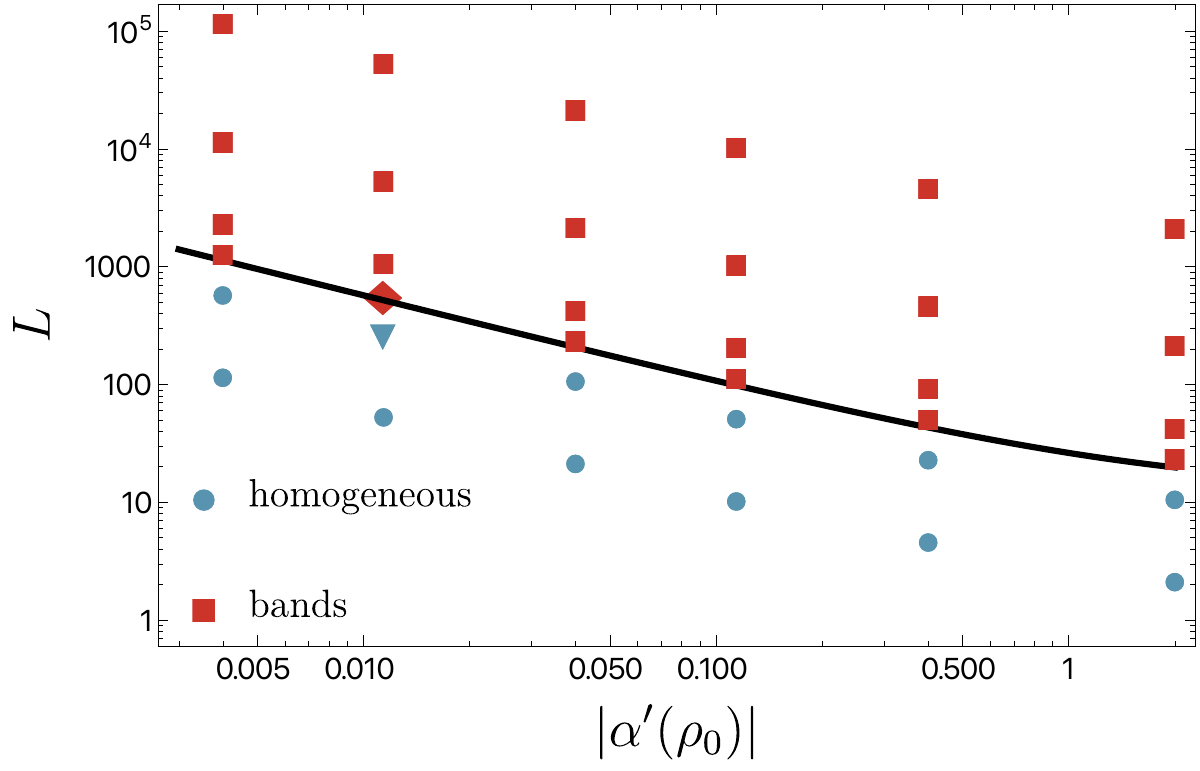}};

      \def\xlabel{4.7}
      \def\ylabel{3}
      \path (\xlabel, \ylabel) node {\fontsize{12}{2}\selectfont \textbf{a.}};

    \end{tikzpicture}
    \endpgfgraphicnamed{Fig3a}
  }\fi
  \end{minipage}
  \hfill
  \begin{minipage}[c]{0.3\linewidth}
    \includegraphics{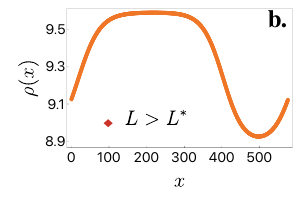}
    \if{
  \beginpgfgraphicnamed{Fig3b}    
    \begin{tikzpicture}
      \def\y{0}
      \path (0,\y) node {\includegraphics[width=1.\textwidth]{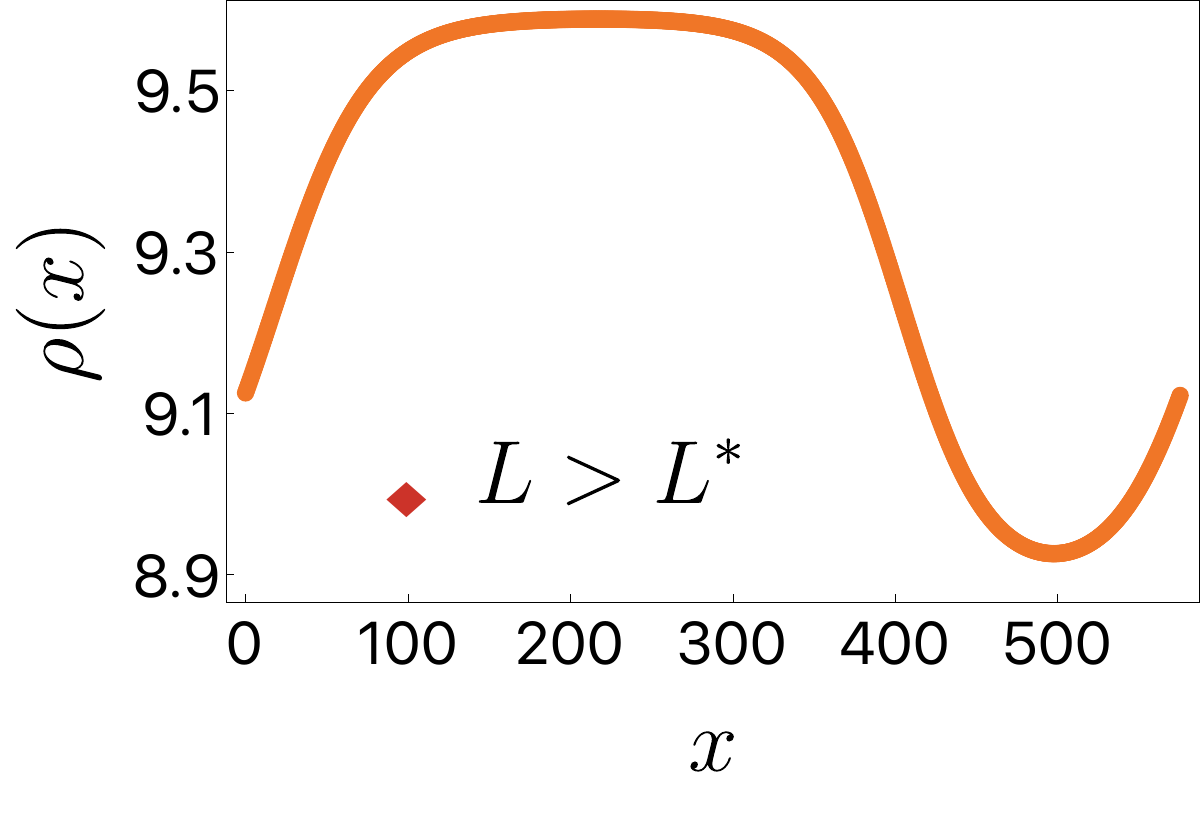}};

      \def\xlabel{2.1}
      \def\ylabel{1.4}
      \path (\xlabel, \ylabel) node {\fontsize{12}{2}\selectfont \textbf{b.}};

    \end{tikzpicture}
    \endpgfgraphicnamed{Fig3b}    
    }\fi
    \includegraphics{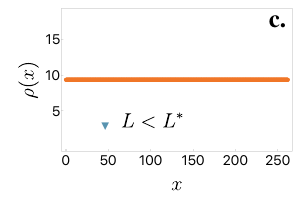}
    \if{
  \beginpgfgraphicnamed{Fig3c}
    \begin{tikzpicture}
      \def\y{0}
      \path (0,\y) node {\includegraphics[width=1.\textwidth]{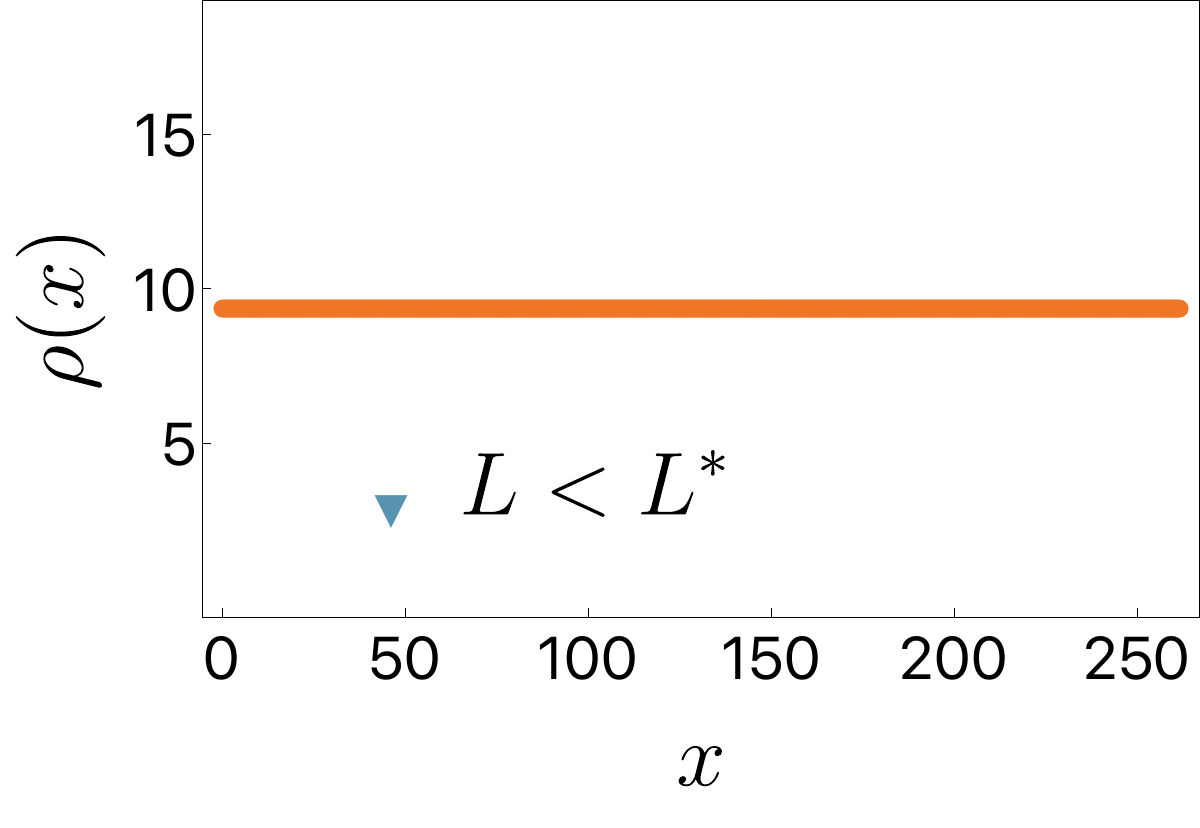}};

      \def\xlabel{2.1}
      \def\ylabel{1.4}
      \path (\xlabel, \ylabel) node {\fontsize{12}{2}\selectfont \textbf{c.}};

    \end{tikzpicture}
    \endpgfgraphicnamed{Fig3c}
    }\fi
  \end{minipage}

    \caption{Linear instability and finite-size effects in the
      evolution of the metric AIM,
      Eq.~\eqref{eq:MF_AIM_renormalized_generic}, in the presence of a
      density-dependent polar-field mass modeled as
      $\alpha(\rho)=a_0+a_1/\rho$. \textbf{a.} Phase diagram obtained
      by simulating Eq.~\eqref{eq:MF_AIM_renormalized_generic} for
      different system sizes $L$ and values of $|\alpha'(\rho_0)|$.
      Blue circles (respectively red squares) indicate a steady state
      with homogeneous density (resp. bands). Solid black line shows
      the estimation of the critical length $L^*$ obtained from
      Eq.~\eqref{eq:Lstar}. 
      Initial condition is $\rho(x)=\rho_0+a\rho_0 \cos(2\pi x/L)$ and $m(x)=m_0+ b m_0 \cos(2\pi x/L)$ with $a=b=2 \times 10^{-5}$. 
      \textbf{b.} Snapshot of the
      phase-separated density field $\rho(x)$, in the steady state for
      $L>L^*$ (red diamond in panel a).  \textbf{c.} Snapshot of the
      homogeneous density field $\rho(x)$, in the steady state for
      $L<L^*$ (blue triangle in panel a). Parameters: $\gamma=1$,
      $a_0=-\varepsilon-2\sqrt{\varepsilon}$, $a_1=1$,
      $\rho_0=1/(2\sqrt{\varepsilon})$ and $\varepsilon=[1/1000,
        1/350, 1/100, 1/35, 1/10, 1/2]$.  }
      \label{fig:system_size}
  \end{figure}
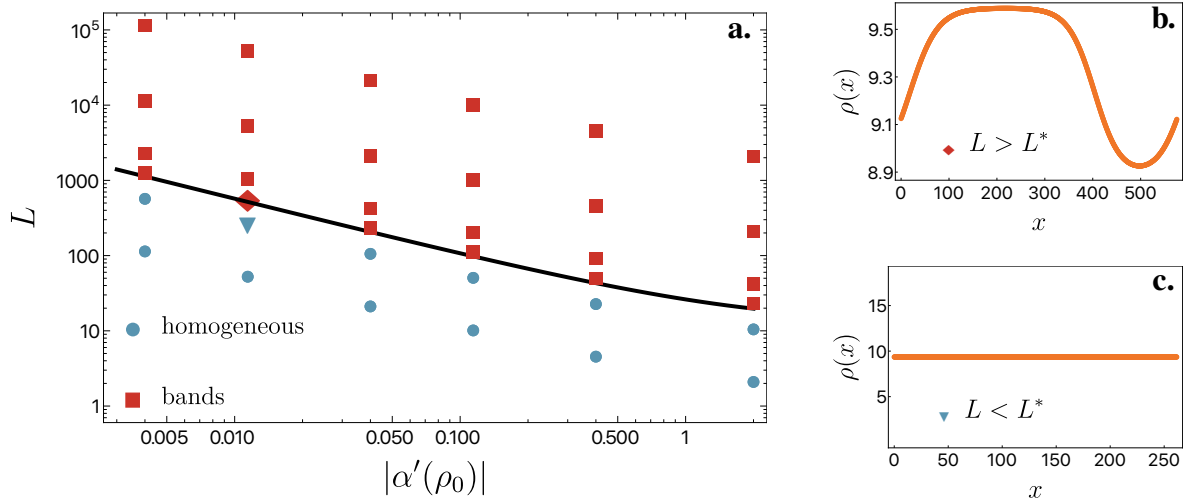

We have shown in Sec.~\ref{subsec:linstap_AIM_renorm} that, in the
presence of a density-dependent polar-field mass, the evolution of the
hydrodynamic Eq.~\eref{eq:MF_AIM_renormalized_generic} describing the
metric AIM exhibits an instability heralding the formation of
travelling domains. The linear instability detailed in
App.~\ref{app:linear_stab} however tells us that this instability
happens at a finite wavenumber $q$ and is therefore system-size
dependent: there is room for it to be prevented in small systems. Indeed, the first
unstable mode $q^*$ is such that the instability can be observed
provided the system size $L$ verifies
\begin{equation}
  \label{eq:Lstar}
  L > L^* = \frac{2\pi}{q^*} \simeq \frac{4 \pi \sqrt{\gamma (2+5 |\alpha_0|)}}{\sqrt{|\alpha_0|^2(4-8\gamma)-4|\alpha_0|(\gamma+\alpha_0' \rho_0) + (\alpha_0' \rho_0)^2  }} \, ,
\end{equation}
which simplifies to $L^*\simeq 4\pi
\sqrt{2\gamma}/(\rho_0|\alpha_0'|)$ at the onset of order where
$|\alpha_0|\ll 1$.  $L^*$ is larger for weaker density dependence of
$\alpha$ and eventually diverges in the limit where $\alpha_0'\to0$.

Even though nonlinear effects play an important role in the formation
of the traveling domains~\cite{caussin2014emergent,Solon2015PatternFI},
it is natural to expect that $L^*$ could provide a good estimate of
the typical system size above which they may be
observed. To test this idea, we simulated
Eq.~\eqref{eq:MF_AIM_renormalized_generic} close to the onset of
order, using a density-dependent linear term given by
$\alpha(\rho)=a_0+a_1/\rho$.  The results are displayed in
Fig.~\ref{fig:system_size}, which compares $L^*$ with the minimal
system size required to observe a traveling domain. We note
that $L^*$, as given in Eq.~\eqref{eq:Lstar}, correctly captures the
scaling of the critical system size above which traveling domains are observed
over several orders of magnitude\footnote{We stress that $L^*$ is,
however, only an order-of-magnitude estimate and we do not expect an
exact agreement.}.

\section{Fluctuation-Induced First-Order Transition in the metric AIM}
\label{sec:FIFOT}

In this section, we study how fluctuations renormalize the mean-field
hydrodynamics of the metric AIM~\eref{eq:MF_AIM_final}. In
Sec.~\ref{subsec:renorm_AIM}, we first
follow~\cite{martin2021fluctuation} and use stochastic calculus to
compute the leading-order corrections in the noise strength to the
mean-field dynamics. Then, in Sec.~\ref{subsec:AIM_field_theory}, we
propose an alternative derivation using a path integral perturbative
approach. Both methods coherently lead to the same result: the mass of
the polar field is renormalized by fluctuations in a density-dependent
way. Together with the results of Sec.~\ref{sec:LinStab}, we thus
conclude that the metric AIM undergoes a fluctuation-induced first
order transition (FIFOT) to collective motion. Finally, we show that
these results extend to two dimensions in
Sec.~\ref{subsec:AIM_metric_2d}.

\subsection{Renormalization of the mean-field AIM}
\label{subsec:renorm_AIM}

In this section, we compute to leading order in the noise strength the
dynamics of the average density and magnetization fields starting from
the mean-field dynamics for the metric AIM complemented by a Gaussian
noise, \textit{i.e.} Eq.~\eqref{eq:fluctuating_mf_AIM}. To facilitate
our analysis, we first detail the derivation in $1$D before discussing
the $2$D case in Sec.~\ref{subsec:AIM_metric_2d}.  Our starting point
is thus the It\=o-stochastic differential equations
\begin{subequations}\label{eq:AIM_noise_1d}%
\begin{align}%
  \label{eq:hydro_AIM_noise_1}
  \partial_t \rho &= D \partial_{xx}\rho -v\partial_x m \, , \\
  \label{eq:hydro_AIM_noise_2}
  \partial_t m &= D \partial_{xx} m - v \partial_x \rho -\mathcal{F}(\rho,m) + \sqrt{2 \sigma \rho} \, \eta
  \;,
\end{align}%
\end{subequations}%
where $\eta(x,t)$ is a zero-mean delta-correlated Gaussian white noise
field and $\cF(\rho,m)=\alpha m + \gamma m^3/\rho^2$. Note that,
hereafter, $\rho(x,t)$ and $m(x,t)$ represent fluctuating fields.  The
noise acting on $m(x,t)$ is multiplicative; it describes the
fluctuations of a sum over $\propto \rho$ particles and we take it
proportional to $\sqrt{\rho}$. Note that we could also complement
Eq.~\eref{eq:hydro_AIM_noise_1} by a conserved noise. In
App.~\ref{app:renorm_conserved_noise}, we show that including such a
noise does not alter our results, in agreement with the intuition that
this conserved noise should be subdominant at large scales.

To assess the role of fluctuations, we construct the dynamics of the
average fields $\tilde{\rho}(x,t)=\langle \rho(x,t)\rangle$ and
$\tilde{m}(x,t)=\langle m(x,t) \rangle$ to leading order in the noise
strength $\sigma$. Denoting by $\rho_0(x,t)$ and $m_0(x,t)$ the
solution of Eq.~\eqref{eq:AIM_noise_1d} in the absence of noise
(\textit{i.e.} when $\sigma=0$), we introduce the deviations
$\Delta\rho$ and $\Delta m$ from these mean-field solutions and expand
them in powers of $\sigma^{1/2}$, which will prove below to be a correct
scaling for the series:
\begin{subequations}\label{eq:expans_Delta}%
\begin{align}%
  \label{eq:expans_Delta_rho}
  \Delta \rho &= \rho - \rho_0 = \sigma^{\frac{1}{2}}\delta\rho_1 + \sigma\delta\rho_2 + \cdots \; ,\\
  \label{eq:expans_Delta_m}
  \Delta m &= m - m_0 = \sigma^{\frac{1}{2}}\delta m_1 + \sigma\delta m_2 + \cdots \, .
\end{align}%
\end{subequations}%
Note that for $k\geq 1$ the $\delta \rho_k$ and $\delta m_k$ are stochastic fields
while $\rho_0$ and $m_0$ are deterministic ones.

The dynamics of $\delta\rho_k$ and $\delta m_k$ can then be obtained
at arbitrary order in $k$ by inserting the
expansion~\eqref{eq:expans_Delta} into Eq.~\eqref{eq:AIM_noise_1d} and
equating terms of order $\sigma^{k/2}$.  Using this systematic
expansion, we show in App.~\ref{app:AIM_FIFOT} that the dynamics of
the average fields $\tilde{\rho}(x,t)$ and $\tilde{m}(x,t)$ is
given, to first order in $\sigma$, by:
\begin{subequations}%
\begin{align}%
  \label{eq:bar_rho}
  \partial_t \tilde{\rho} =& D \partial_{xx}\tilde{\rho} -v\partial_x \tilde{m}\;, \\
  \begin{split}
    \label{eq:bar_m}
    \partial_t \tilde{m} =& D \partial_{xx} \tilde{m} - v \partial_x \tilde{\rho} -\cF(\tilde{\rho},\tilde{m})-\sigma \frac{\partial^{2} \mathcal{F}}{\partial m^{2}} \left(\frac{\langle\delta m_1^{2}\rangle -\langle\delta m_1\rangle^{2}}{2}\right) \\
    &-  \sigma\frac{\partial^{2} \mathcal{F}}{\partial \rho^{2}} \left(\frac{\langle\delta \rho_1^{2}\rangle-\langle\delta \rho_1\rangle^{2}}{2}\right)
    - \sigma \frac{\partial^{2} \mathcal{F}}{\partial m \partial \rho} \left(\langle\delta m_1\delta \rho_1\rangle-\langle\delta m_1\rangle \langle\delta \rho_1\rangle\right) \, ,
  \end{split}
\end{align}%
\end{subequations}%
where the derivatives of $\cF$ are evaluated at
$\tilde{\rho},\tilde{m}$. In Eq.~\eqref{eq:bar_m}, we see that the
fluctuations modify the mean-field expression of $\cF$.  In order to
compute this fluctuation-induced renormalization, we need to evaluate
the correlators involving $\delta\rho_1$ and $\delta m_1$.  To perform
this computation, we assume that we are far enough from the linear
instability so that $\delta \rho_1$ and $\delta m_1$ are fast modes which
relax on length-scales and timescales much smaller than the ones
relevant for $\rho_0$ and $m_0$. Under this assumption, $\rho_0(x,t)$
and $m_0(x,t)$ are considered as constant in time and space when
computing the correlators in terms of $\rho_0$ and $m_0$ and their
dependency on $x$ and $t$ is re-established a posteriori in the final
expression. Averages computed using these assumptions are denoted by $\langle\ldots\rangle_0$. We show in
App.~\ref{app:correlators_AIM} that the correlators rapidly relax to
\begin{subequations}%
\label{eq:corr_delta_tot}
\begin{align}%
\label{eq:corr_deltam2}
    \langle\delta m_1^2 \rangle_0 =& \frac{\rho_0}{2v}\frac{\sqrt{\frac{2}{u}}+\sqrt{1+u}}{2+u}  - \frac{3 \gamma D}{v^3}\frac{\frac{u}{\sqrt{1+u}}+\frac{\sqrt{2}(2+3u)}{u^{3/2}}}{4(u+2)^2} \frac{m_0^2}{\rho_0}+ \cO(m_0^3) \, , \\
  \label{eq:corr_deltarho2}
  \langle\delta \rho_1^2 \rangle_0 =& \frac{\rho_0}{2v}\frac{\sqrt{\frac{2}{u}}-\frac{1}{\sqrt{1+u}}}{2+u} + \cO(m_0^2) \;,\\
  \langle\delta \rho_1\delta m_1 \rangle_0 =& 0 + \cO(m_0^2) \, ,
\end{align}%
\end{subequations}%
where $u=\alpha D/v^2$. 
Note that in \eqref{eq:corr_delta_tot} we discarded higher-order terms in $m_0$ since we are interested at the onset of collective motion where $m_0 \ll 1$.

Inserting the correlators~\eref{eq:corr_deltam2}-\eref{eq:corr_deltarho2} into Eq.~\eref{eq:bar_m} and discarding terms beyond order $\tilde{m}^3$, we finally obtain the renormalized dynamics for $\tilde{m}$, valid up to order $\sigma$, which reads
\begin{subequations}%
\begin{align}
  \label{eq:bar_m_closed}
  \partial_t \tilde{m} =& D \partial_{xx} \tilde{m} - v \partial_x \tilde{\rho} -\hat{\alpha}(\tilde{\rho})\tilde{m}-\hat{\gamma}(\tilde{\rho})\frac{\tilde{m}^3}{\tilde{\rho}^2}\ ,
\end{align}
with the renormalized Landau coefficients $\hat{\alpha}(\tilde{\rho})$ and $\hat{\gamma}(\tilde{\rho})$ given by 
\begin{align}
  \label{eq::dressed_gamma_metric}
  \hat{\gamma}(\tilde{\rho}) =& \gamma + \frac{3\sigma\gamma}{2v\tilde{\rho}}\frac{ \sqrt{\frac{2}{u}}-\frac{1 }{\sqrt{1+u}}}{u+2 } - \frac{9\sigma\gamma^2 D}{4v^3\tilde{\rho}}\frac{ \left(\frac{u}{\sqrt{1+u}}+\frac{\sqrt{2} \left(2+3 u\right)}{u^{3/2}}\right)}{ \left(u+2\right)^2} \, , \\
\hat{\alpha}(\tilde{\rho}) =& \alpha + \frac{3 \sigma \gamma}{ 2 \tilde{\rho} v} \frac{\sqrt{2/u}+\sqrt{1+u}}{2+u}\ ,
\label{eq:RenAlpha}
\end{align}%
\end{subequations}%
with $u=\alpha D/v^2$.
Fluctuations have thus, to order $\sigma$, dressed $\alpha$ and
$\gamma$ into $\hat{\alpha}$ and $\hat{\gamma}$ respectively.
Crucially, we remark that the linear mass $\hat\alpha$ now depends
explicitly on the density in Eq.~\eref{eq:RenAlpha}.  According to the
stability analysis performed in Sec.~\ref{sec:LinStab}, a
density-dependent $\alpha$ turns the continuous transition predicted
by the mean-field evolution~\eref{eq:MF_AIM_final} into a first-order
phase transition exhibiting travelling domains near the onset of order. The derivation
above thus explains why the numerical simulations of
Eq.~\eqref{eq:fluctuating_mf_AIM} led to the first order scenario
reported in \Fref{fig:SDEmetric}.

\begin{figure}[t]
  \flushright
  \includegraphics{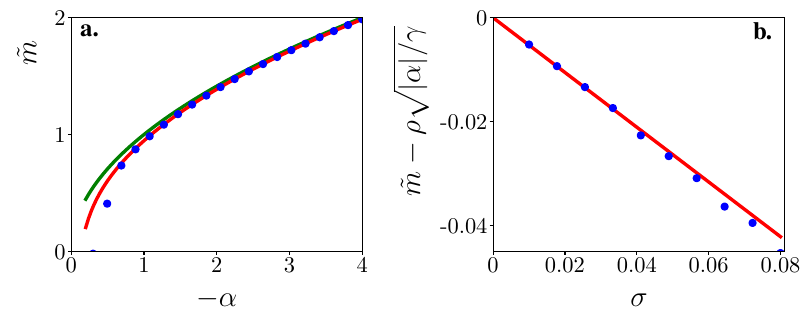}
  \if{
  \beginpgfgraphicnamed{Fig4}
  \begin{tikzpicture}
  \node at (0,0) {\includegraphics[width=0.85\textwidth]{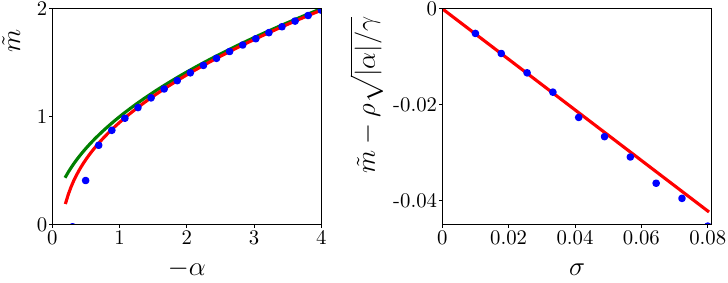}};
  \node[font=\bf] at (-5.4,2.2) {a.};
  \node[font=\bf] at (6,2.2) {b.};
  \end{tikzpicture}
  \endpgfgraphicnamed{Fig4}
  }\fi
\caption{{\bf a.} Averaged magnetization in the low temperature phase as $\alpha$ is varied.
      The dots are obtained from numerical simulations of \eqref{eq:AIM_noise_1d} while the green line is the mean-field prediction $m=\sqrt{|\alpha|/\gamma}\rho$ and the red line is the renormalized prediction \eref{eq:prediction_m_low_temp}.
      Parameters: $D=1$, $v=1$, $\gamma=1$, $\rho=1$.
      {\bf b.} Renormalized correction to mean-field $\tilde{m}-\sqrt{|\alpha|/\gamma}\rho$ as a function of the noise strength $\sigma$.
      Blue dots are obtained from numerical simulations of \eqref{eq:AIM_noise_1d}, while the red line is the renormalized prediction \eref{eq:prediction_m_low_temp}.
      Parameters: $D=1$, $v=1$, $\rho=1$, $\gamma=1$, $\alpha=4$.}
    \label{fig:magn_low_temp}
\end{figure}

In addition to accounting for the nature of the transition, our
perturbation theory in the noise strength also allows us to predict
how the fluctuations in Eq.~\eref{eq:AIM_noise_1d} renormalizes
the mean-field average magnetization in the ordered phase. As detailed
in App.~\ref{app:AIM_FIFOT_ordered}, the leading-order correction in
$\sigma$ to the average magnetization is given by
\begin{subequations}%
\begin{align}%
  \label{eq:prediction_m_low_temp}
  \tilde{m} = \sqrt{\frac{|\alpha|}{\gamma}}\rho_0 - \sigma \frac{3\sqrt{\gamma}}{2\sqrt{2D}|\alpha|}f_w\left(\frac{v^2(\gamma-|\alpha|)}{2D|\alpha|\gamma},\frac{v^2}{2D\gamma}\right)\; ,
\end{align}
where $f_w(s,u)$ is defined as
\begin{align}
  f_w(s,u)=& \int_{-\infty}^{+\infty} \frac{d\tilde{q}}{2\pi}\frac{ s\left(1 + 2\tilde{q}^2\right) + (1+2\tilde{q}^2)^2+4u\tilde{q}^2}{s\left(1+2  \tilde{q}^2\right)^2+\left(1+\tilde{q}^2 \right) \left[\left(1+2 \tilde{q}^2 \right)^2+4 u \tilde{q}^2 \right]} \, .
\end{align}%
\end{subequations}%
As shown in Fig.~\ref{fig:magn_low_temp}, our prediction
\eref{eq:prediction_m_low_temp} quantitatively match---without any
fitting parameter---the numerical measurement of the magnetization in
the low temperature phase, as long as $\alpha$ is large enough that
the system is far away from the transition. 

In the next section, we show that the results derived here are
consistent with a perturbation theory based on the dynamical action
corresponding to our stochastic dynamics. While the framework
presented here is appealing due to its simplicity, field-theoretical
methods are more straightforwardly extended, for instance to higher
order in the noise strength.



\subsection{AIM: Field-theoretic version}
\label{subsec:AIM_field_theory}

In this section, we present a field-theoretical derivation of the
fluctuation-induced renormalization of $\alpha$ in
Eq.~\eqref{eq:AIM_noise_1d}.  A path-integral representation of the
coupled Langevin equations~\eqref{eq:hydro_AIM_noise_1}
and~\eqref{eq:hydro_AIM_noise_2} can be obtained using the
Martin--Siggia--Rose--De Dominicis--Janssen (MSRDJ)
formalism~\cite{tauber2014critical,oerding2000fluctuation}, such that the average of an
observable $\mathcal{A}[\rho,m]$ is obtained as
\begin{subequations}%
\begin{align}%
    \langle \mathcal{A}[\rho,m] \rangle = \int \ddd \rho \ddd \hat{\rho} \ddd m \ddd \hat{m} \,
    \mathcal{A}[\rho,m] \ee^{-S[\rho,\hat{\rho},m,\hat{m}]} \, ,
\end{align}%
where we have introduced the response fields $\hat{m}$ and $\hat{\rho}$.
The action $S$ reads 
\begin{align}%
\nonumber
    S[\rho,\hat{\rho},m,\hat{m}] =&
    \int_{x,t} \hat{m}  \left[ \partial_t m - D \partial_{xx} m + v \partial_{x} \rho +\alpha m +\frac{\gamma}{\rho^2}m^3 - \sigma \rho \hat{m} \right] \\ 
    & + \int_{x,t} \hat{\rho}\left[ \partial_t\rho - D \partial_{xx} \rho + v \partial_{x} m \right]\, ,
\end{align}%
\end{subequations}%
where the space and time dependence of the fields $\rho,\hat{\rho},m,\hat{m}$ is implicit.
The action can be simplified by considering an expansion around the homogeneous density, $\rho (x,t)= \rho_0+\delta\rho(x,t)$ with $\rho_0 \neq0$, and by performing explicitly the integral over~$\hat{\rho}$. This is best done in Fourier space, where we find:
\begin{align}
    \delta\rho (q,\omega) = \frac{-\rmi q v}{D q^2 - \rmi \omega} m(q,\omega) \, .
\end{align}
The action $S$ can then be expressed as $S = S_0 + S_{\rm int}$, where $S_{\rm int}$ is the interaction part while $S_0$ is the Gaussian part.
The later is given by
\begin{align}
  \label{eq:gaussian_part_action}
    S_0 &= \frac{1}{2} \int_Q M_i(-Q) H_{ij} M_j(Q) \, .
\end{align}
In \eref{eq:gaussian_part_action}, we have introduced the short-hand notation $Q=(q,\omega)$, $\int_Q = (2\pi)^{-2} \int \dd q \dd \omega$, $M_i(Q) = (m(Q), \hat{m}(Q))_i$, and adopted the implicit summation over repeated indices. The inverse of the Gaussian kernel $H_{ij}$ is the $2\times2$ matrix of the bare correlators:

\begin{align}%
    (H^{-1})_{ij} =
    \begin{pmatrix}
    \langle m(Q) m(-Q) \rangle_0 & \langle m(Q) \hat{m}(-Q) \rangle_0 \\
    \langle \hat{m}(Q) m(-Q) \rangle_0 &   \langle \hat{m}(Q) \hat{m}(-Q) \rangle_0
    \end{pmatrix}_{ij} \, .
\end{align}%
It will prove convenient to define the bare propagator $G_0$ and the bare correlation function $C_0$ as:
\begin{subequations}%
\begin{align}%
    \langle m(Q_1) \hat{m}(Q_2) \rangle_0 = (2\pi)^2 \delta (Q_1+Q_2) G_0(Q_1) , \quad G_0(Q) &= \left(D q^2 -\rmi \omega + \alpha + V(Q) \right)^{-1} , \\
    \langle m(Q_1) m(Q_2) \rangle_0 = (2\pi)^2 \delta (Q_1+Q_2) C_0(Q_1) , \quad C_0(Q) &= 2\sigma\rho_0 G_0(Q) G_0(-Q) \, ,
\end{align}%
\end{subequations}%
with $V(Q)=q^2 v/(D q^2-\rmi \omega)$. We also introduce a diagrammatic notation:
\begin{align}
    G_0(Q) &=
    \begin{tikzpicture}[baseline=-.1cm]
        \coordinate (V1) at  (-1,0);
        \coordinate (V2) at  (-1/2,0);
        \coordinate (V3) at  (0,0);
        \draw[->,>=latex]   (V1)   -- (V2);
        \draw[dashed]   (V2)   -- (V3);
        \path (V2) node [above] {{\small $Q$}};
        \end{tikzpicture} \, , \quad
    C_0(Q) =
    \begin{tikzpicture}[baseline=-.1cm]
        \coordinate (V1) at  (-1,0);
        \coordinate (V2) at  (-1/2,0);
        \coordinate (V3) at  (0,0);
        \coordinate (V4) at  (1/2,0);
        \coordinate (V5) at  (1,0);
        \draw[->,>=latex]   (V1)   -- (V2);
        \draw[dashed]   (V2)   -- (V3);
        \draw[dashed]   (V3)   -- (V4);
        \draw[->,>=latex]   (V5)   -- (V4);
        \filldraw [black] (V3) circle (2pt);
        \filldraw [white] (V3) circle (1.5pt);
        \path (V2) node [above] {{\small $Q$}};
        \path (V4) node [above] {{\small $-Q$}};
        \end{tikzpicture}
\end{align}

The interaction part of the action $S_{\rm int}$ is given by
\begin{subequations}%
\begin{align}%
    S_{\rm int} &= S^{(\sigma)}+ \sum_{k=0}^\infty S^{(\gamma)}_k \, , \\
    S^{(\sigma)} &= - \sigma \int_{Q_1,Q_2,Q_3} \hat{m}(Q_1)\hat{m}(Q_2) \delta\rho (Q_3) (2\pi)^2 \delta\left(\sum_i Q_i \right) \, , \\
\begin{split}
    S^{(\gamma)}_k &= (k+1) (-1)^k \frac{\gamma}{\rho_0^2} \int_{Q_1, \cdots, Q_{4+k}} (2\pi)^2\delta\left(\sum_{i=1}^{4+k} Q_i \right) \\
    & \hspace{3cm}  \times\hat{m}(Q_1) m(Q_2) m(Q_3) m(Q_4) \prod_{j=1}^k \frac{\delta\rho (Q_{4+j})}{\rho_0} \, ,
\end{split}
\end{align}%
\end{subequations}%
from which we deduce the (amputated) interaction vertices
\begin{subequations}%
\begin{align}%
    W^{(\sigma)}(Q_1,Q_2,Q_3) &=
    \begin{tikzpicture}[baseline=-.1cm]
        \coordinate (V0) at  (0,0);
        \coordinate (V1) at  (-1.41/2,0);
        \coordinate (V2) at  (1/2,1/2);
        \coordinate (V3) at  (1/2,-1/2);
        \draw[dashed,->,>=latex]   (V0)   -- (V1);
        \draw[dashed,->,>=latex]   (V0)   -- (V2);
        \draw[->,>=latex]   (V0)   -- (V3);
        \filldraw [black] (V0) circle (2pt);
        \path (V1) node [above] {{\small $Q_1$}};
        \path (V2) node [right] {{\small $Q_2$}};
        \path (V3) node [right] {{\small $Q_3$}};
        \end{tikzpicture}
    = \sigma \frac{\rmi q_3 v}{D q_3^2 - \rmi \omega_3} (2\pi)^2 \delta(Q_1+Q_2+Q_3) \, , \\
 W^{(\gamma)}_0(Q_1,\cdots,Q_{4}) &=
 \begin{tikzpicture}[baseline=-.1cm]
     \coordinate (V0) at  (0,0);
     \coordinate (V1) at  (-1.41/2,0);
     \coordinate (V2) at  (1/4,0.65);
     \coordinate (V3) at  (1.41/2,0);
     \coordinate (V4) at  (1/4,-0.65);
     \draw[dashed,->,>=latex]   (V0)   -- (V1);
     \draw[->,>=latex]   (V0)   -- (V2);
     \draw[->,>=latex]   (V0)   -- (V3);
     \draw[->,>=latex]   (V0)   -- (V4);
     \filldraw [black] (V0) circle (2pt);
     \path (V1) node [above] {{\small $Q_1$}};
     \path (V2) node [above] {{\small $Q_2$}};
     \path (V3) node [above] {{\small $Q_3$}};
     \path (V4) node [right] {{\small $Q_4$}};
     \end{tikzpicture}
 = \frac{\gamma}{\rho_0^2} (2\pi)^2 \delta(Q_1+Q_2+Q_3+Q_4)   \, , \\
 \begin{split}
 \label{eq:third_amputated_vertex}
 W^{(\gamma)}_k(Q_1,\cdots,Q_{4+k}) &=
 \begin{tikzpicture}[baseline=-.1cm]
     \coordinate (V0) at  (0,0);
     \coordinate (V1) at  (-1.41/2,0);
     \coordinate (V2) at  (1/4,0.65);
     \coordinate (V4) at  (1/2,1/2);
     \coordinate (V3) at  (1/2,-1/2);
     \draw[dashed,->,>=latex]   (V0)   -- (V1);
     \draw[->,>=latex]   (V0)   -- (V2);
     \draw[->,>=latex]   (V0)   -- (V3);
     \draw[->,>=latex]   (V0)   -- (V4);
     \filldraw [black] (V0) circle (2pt);
     \path (V1) node [above] {{\small $Q_1$}};
     \path (V2) node [above] {{\small $Q_2$}};
     \path (V4) node [right] {{\small $Q_3$}};
     \path (V3) node [right] {{\small $Q_{k+4}$}};
     \path (3/4,0.1) node {{\vdots}};
     \end{tikzpicture} \\
 &= (k+1) (-1)^k \frac{\gamma}{\rho_0^{2+k}} (2\pi)^2 \delta\left(\sum_{i=1}^{4+k} Q_i \right) \left[ \prod_{j=0}^k \frac{-\rmi q_j v}{D q_j^2 - \rmi \omega_j}  \right]^{\rm sym} \, .
\end{split}
\end{align}%
\end{subequations}%
In \eref{eq:third_amputated_vertex}, $[\cdot]^{\rm sym}$ indicates that the vertex must be symmetrized with respect to the momenta $Q_2,\cdots,Q_{4+k}$, since the corresponding legs are all represented by a solid line while only $k$ of them bear a nontrivial momentum structure.

In the limit of small noise $\sigma\to 0$, a perturbative expansion close to the Gaussian action can be performed. The dressed response function can therefore be obtained perturbatively as
\begin{align}
\begin{split}
    \langle m(Q_1) \hat{m}(Q_2) \rangle &= (2\pi)^2 \delta(Q_1+Q_2) G(Q_1) \, , \\
    &= \int \ddd m \ddd \hat{m} \,
    m(Q) \hat{m}(-Q) \ee^{-S_0-S_{\rm int}} \, ,\\
    &= \langle m(Q_1) \hat{m}(Q_2) \rangle_0 - \langle \int_Q m(Q) \hat{m}(-Q) S_{\rm int} \rangle_0 + \cdots \, .
\end{split}
\end{align}
The mass renormalization is then deduced from the inverse propagator $G^{-1}$ at vanishing external momentum $\hat{\alpha}-\alpha = G^{-1}(P=0)$, where $G^{-1}$ is obtained from Dyson's equation as $G^{-1}(P)=G_0^{-1}(P)-\Sigma(P)$.
In the later, the so-called self-energy $\Sigma(P)$ can be computed from the diagrammatic expansion
\begin{align}
    \Sigma(P) =    \quad
    \begin{tikzpicture}[baseline=-.1cm]
        \coordinate (V0) at  (0,0);
        \coordinate (Vc) at  (0,1/2);
        \coordinate (V11) at  (-1,-1/2);
        \coordinate (V12) at  (-1/2,-1/4);
        \coordinate (V21) at  (1/2,-1/4);
        \coordinate (V22) at  (1,-1/2);
        \coordinate (C1) at  (-1/2,1/2);
        \coordinate (C2) at  (1/2,1/2);
        \draw[->,>=latex]   (V11)   -- (V12);
        \draw[dashed]   (V12)   -- (V0);
        \filldraw [black] (V0) circle (2pt);
        \draw[->,>=latex]   (V0)   -- (V21);
        \draw[dashed]   (V21)   -- (V22);
        \draw[<-,>=latex] (C2) arc (0:-90:1/2);
        \draw[dashed,-] (C2) arc (0:90:1/2);
        \draw[<-,>=latex] (C1) arc (180:270:1/2);
        \draw[dashed] (C1) arc (180:90:1/2);
        \filldraw [black] (0,1) circle (2pt);
        \filldraw [white] (0,1) circle (1.5pt);
        %
        %
        \path (V12) node [above = 2pt, left=3pt] {{\small $P$}};
        \path (V21) node [above = 2pt, right=3pt] {{\small $P$}};
        \path (C1) node [left] {{\small $Q$}};
        \path (C2) node [right] {{\small $-Q$}};
    \end{tikzpicture}
    \quad + \quad O\left( \sigma^2 \right) \, .
\end{align}
Importantly, the only vertex that can contribute at first order in $\sigma$ is $W^{(\gamma)}_0$, hence simplifying tremendously the computation despite the infinitely many vertices $W^{(\gamma)}_k$ that could have contributed.
All in all, we only have to compute a single integral to obtain the mass renormalization
\begin{align}
\begin{split}
    \hat{\alpha}-\alpha=\Sigma (0) &= 3 \int_Q  W^{(\gamma)}_0 (0,0,Q,-Q) C_0(Q) + O\left( \sigma^2 \right) \, ,\\
    &= \frac{3\sigma\gamma}{\rho_0} \frac{\sqrt{2} v^2 + \sqrt{D\alpha (v^2+D\alpha)}}{2\sqrt{D\alpha} (2v^2+D \alpha)} + O\left( \sigma^2 \right)\, ,
\end{split}
\end{align}
which gives a result identical to the one obtained in
Sec.~\ref{subsec:renorm_AIM}.

Note that the perturbative expansion
presented in this section is systematic and can be extended to higher
(loop) order. Our two perturbation theories thus consistently show how
fluctuations make $\alpha$ density-dependent and lead to the same
renormalized evolution of the metric AIM given by
Eq.~\eqref{eq:MF_AIM_renormalized_generic}.

\subsection{AIM: The 2d case}
\label{subsec:AIM_metric_2d}

In this section, we show that the results obtained in
Secs.~\ref{subsec:renorm_AIM} and~\ref{subsec:AIM_field_theory} are
also valid in dimension $2$: the renormalized linear mass
$\hat{\alpha}$ is also density-dependent in this case.  Starting from
Eq.~\eqref{eq:fluctuating_mf_AIM}, the perturbative method employed in
\ref{subsec:renorm_AIM} can be conducted in two dimensions as well, and
Eqs.~\eqref{eq:bar_rho}-\eref{eq:bar_m} remain valid (up to replacing
$\partial_{xx}$ by $\bnab^2$).  However, due to the additional
dimension, the stochastic evolution of the fields $\delta m_1$ and
$\delta \rho_1$ is changed, as well as the correlators $\langle \delta
m_1^2 \rangle_0$, $\langle \delta \rho_1^2 \rangle_0$ and $\langle \delta
m_1 \delta \rho_1 \rangle_0$.  As shown in App.~\ref{sec:AIM_2D}, they
now read
\begin{subequations}\label{eq:corr_aim_2D_m_m}
\begin{align}%
  \langle\delta m_1^2 \rangle_0
  =& \frac{\rho_0 \alpha}{v^2}h_1\left(\frac{\alpha D}{v^2}\right) + \mathcal{O}(m_0^2)\; , \quad \langle\delta \rho_1^2 \rangle_0
  =  \frac{\rho_0\alpha}{v^2}h_2\left(\frac{\alpha D}{v^2}\right) + \mathcal{O}(m_0^2)\;,\\ \langle\delta \rho_1\delta m_1 \rangle_0 =& 0 + \mathcal{O}(m_0^2)\; ,
\end{align}
\end{subequations}
where the functions $h_1$ and $h_2$ are given by
\begin{subequations}\label{eq:func_corr_AIM_2d}
\begin{align}  
  h_1(u)=&\frac{1}{4\pi^2}\int_{V} \frac{\tilde{\bq}^2(2u^2\tilde{\bq}^2+u)+\tilde{q}_x^2}{\left(2\tilde{\bq}^2+1\right)\left(\tilde{\bq}^2(u^2\tilde{\bq}^2+u)+\tilde{q}_x^2\right)}\, {\rm d}^{2} \tilde{\bq}\, , \\ 
  h_2(u)=&\frac{1}{4\pi^2}\int_{V} \frac{\tilde{q}_x^2}{\left(2\tilde{\bq}^2+1\right)\left(\tilde{\bq}^2(u^2\tilde{\bq}^2+u)+\tilde{q}_x^2\right)}\, {\rm d}^{2}\tilde{\bq}\, .
\end{align}%
\end{subequations}%
Using \eref{eq:corr_aim_2D_m_m}, we obtain the following renormalized dynamics for $\tilde{m}$
\begin{align}
  \label{eq:dyn_m_2D_renorm}
  \partial_t \tilde{m} =& D \bnab^2 \tilde{m} - v \partial_x \tilde{\rho} -\cF(\tilde{\rho},\tilde{m})-\sigma \frac{3\gamma \tilde{m}}{\tilde{\rho} \alpha}h_1\left(\frac{\alpha D}{v^2}\right) + \mathcal{O}(\sigma \tilde{m}^2)\, ,
\end{align}
from which we deduce the expression of the renormalized mass $\hat{\alpha}$ at first order in $\sigma$
\begin{align}\label{eq:dyn_renorm_2d}
  \hat{\alpha} = \alpha -\sigma \frac{3\gamma\alpha}{\tilde{\rho} v^2}h_1\left(\frac{\alpha D}{v^2}\right)\, .
\end{align}
We observe that the renormalized mass $\hat{\alpha}$ has become density dependent due to the effect of fluctuations, hence extending our one-dimensional result to dimension $2$. Since the stability analysis performed in Sec.~\ref{sec:LinStab} remains valid in 2D, the density-dependence of $\hat{\alpha}(\tilde{\rho})$ turns the continuous transition predicted by the mean-field evolution~\eqref{eq:fluctuating_mf_AIM} into a first-order, liquid-gas-like phase separation exhibiting ordered traveling domains.

However, an interesting difference between the one- and
two-dimensional cases can be noted by computing explicitly $h_1$ in
Eq.~\eqref{eq:dyn_renorm_2d}: the renormalization is infinite in
dimension~$2$ due to a UV divergence. The latter can be traced back to
the approximation of the microscopic AIM by a continuous
partial-differential equation at mean-field level. This approximation
is valid only for scales much larger than the correlation length and
produces quantitative errors when describing our microscopic models on
short scales. Indeed, the correlators $\langle O_q O_{-q} \rangle$
entering the renormalization of the mass scale as $q^{-2}$ when
$q\to\infty$.  In dimension $2$, taking into account the measure ${\rm
  d}^{2}{\bf q}$, the integral of the correlators thus behave as $\int
q^{-1}dq$, which leads to logarithmically diverging corrections of the
mass. The scales at which our mean-field partial-differential equation
fails at describing the microscopic model are precisely the scales
that dominate the integral. While this spurious divergence prevents us
from predicting quantitatively the prefactor of the mass
renormalization, the dimensional analysis of
Eq.~\eqref{eq:dyn_renorm_2d} nevertheless guarantees a
density-dependent renormalization. It would be interesting to check
whether the difference between 1D and 2D implies different
``universal'' behaviours of the renormalized mass $\hat\alpha$: Our
computations indeed suggest that different microscopic models leading to the
same mean-field equations in 1D should have the same renormalized mass
while this form of universality would not hold in 2D.

\section{Fluctuation-Induced First Order Transition in topological AIMs}
\label{part:FIFOT_topological}

In this section, we show that the topological version of the AIM also
experiences a FIFOT. We start in
Sec.~\ref{sec:generic_alignment_topol} by
following~\cite{martin2021fluctuation} to argue that topological
alignment should generically induce a density-dependent polar-field
mass once fluctuations are taken into account. We then consider
explicitly the case of the $k$-nearest-neighbour alignment in
Sec.~\ref{sec:k_nearest_renorm}, and confirm the existence of a
discontinuous transition using particle-level simulations. We finally
discuss the case of Voronoi alignment in
Sec.~\ref{sec:voronoi_alignment_renorm} where we show that the onset
of order indeed depends on density and is accompanied with the
traveling domains that are typical of first-order
scenarios. Sections~\ref{sec:generic_alignment_topol}
and~\ref{sec:k_nearest_renorm} are extended accounts of arguments and
results presented in~\cite{martin2021fluctuation} whereas
Sect.~\ref{sec:voronoi_alignment_renorm} is entirely new.

\subsection{Generic aligning field $\bar m(x,[\rho,m])$ and dimensional analysis}
\label{sec:generic_alignment_topol}

Before we turn to specific models, let us consider the case in which
the aligning field $\bar{m}$ is an intensive field taking the form of
an unspecified functional of $\rho$ and $m$:
\begin{equation}
  \label{eq:functional_topol_m}
  \bar m (x) = {\cal G}(x,[\rho,m]) \,.
\end{equation}
Note that, ${\cal G}_{\rm metric}(x,[\rho,m])=m(x)/\rho(x)$
corresponds to the case of the metric AIM studied so far. For
simplicity, we work in 1D and consider the mean-field evolution of
\begin{subequations}\label{eq:AIM_generic_noisy}
\begin{align}%
  \partial_t \rho &= D \partial_{xx}\rho -v\partial_x m \, ,\\
\label{eq:AIM_generictopological_noisy}
  \partial_t m &= D \partial_{xx} m - v \partial_x \rho -\mathcal{F}(\rho,m,\bar{m}) + \sqrt{2\sigma\rho}\,\eta
  \, ,
\end{align}\label{eq:topological_general}%
\end{subequations}%
where $\bar m (x)$ is given by Eq.~\eqref{eq:functional_topol_m} and
$\mathcal{F}(\rho,m,\bar{m})$ by Eq.~\eref{eq:F_onset}. Note that, as
before, $m(x)$ in Eq.~\eqref{eq:AIM_generictopological_noisy} is the
sum of the orientations of $\sim \rho(x)$ particles so that the noise
is multiplicative and proportional to $\sqrt{\rho(x,t)}$. The possible topological nature of $\bar m$ has no reason to change this scaling.
Because of the fluctuations and according to Sec.~\ref{sec:FIFOT}, we expect that the mean-field approximation for the dynamics of $\tilde{\rho}=\langle\rho\rangle$ and $\tilde{m}=\langle m\rangle$ must be corrected by an additional term $\Delta\mathcal{F}$
\begin{subequations}%
\begin{align}%
  \partial_{t}\tilde{\rho} &= D\partial_{xx} \tilde{\rho} -v\partial_x \tilde{m} \,, \\
  \label{eq:rho_topological_general_ren}
  \partial_{t} \tilde{m} &= D\partial_{xx} \tilde {m} -v\partial_x \tilde{\rho} - \mathcal{F}(\tilde{m},\tilde{\rho},\tilde{\bar{m}}) -\sigma \Delta\mathcal{F} \, .
\end{align}%
\end{subequations}%
Using scaling arguments and dimensional analysis in the renormalization procedure, we show in App.~\ref{app:generic_alignment} that the generic dependency of $\Delta\mathcal{F}$ should take the following form (to leading order in $\sigma$)
\begin{align}
  \label{eq:delta_Ft_scaling_final_main}
  \Delta\mathcal{F} &=  \tilde{m}\; \bar{\cF}\left(\frac{\Gamma D}{v^2}, \frac{\Gamma}{v\tilde{\rho}},\beta\right) + \cO(\tilde{m}^2)\ ,
\end{align}
where $\bar{\cF}$ is a dimensionless function that depends on the specific choice of the aligning functional $\cG$.
If the aligning field is generic, there is no reason that the
dependency of $\bar{\cF}$ on its second argument vanish, and we thus
expect a correction to $\alpha$ that depends on the local density
$\tilde{\rho}$. This result suggests that flocking models should
generically exhibit a phase separation with travelling bands near the
onset of order, irrespective of whether metric or topological alignment is
involved. In the following sections, we confirm this statement and the
scaling form of~\eqref{eq:delta_Ft_scaling_final_main} in the case of
the alignment with the $k$ nearest neighbors.

Before we do so, it is nevertheless interesting to ask whether one can
find a case for which the dependence on density suggested by
Eq.~\eqref{eq:delta_Ft_scaling_final_main} vanishes. So far, the only
strategy that we have found to achieve this is rather brutal: to
disconnect the positional degrees of freedom and the aligning
dynamics. For instance, consider a fully-connected version of the AIM
for which
\begin{equation}
  \label{eq:bar_m_FC_main}
  \bar m = \frac{\int_0^L  m(z) dz}{N};\qquad\text{where}\qquad N=\int_0^L \rho(z) dz\;.
\end{equation}
In this case, to leading order in $\sigma$, $\Delta\mathcal{F}$ reads (see App.~\ref{app:generic_alignment} for details):
\begin{align}
  \label{eq:delta_ft_FC_main}
  \Delta\mathcal{F} =\frac{\tilde{m}}{N}\left(\frac{\beta^2}{2}+\frac{\beta^2}{1-\beta}\right) + \cO(\tilde{m}^2) \ .
\end{align}
Comparing Eq.~\eqref{eq:delta_ft_FC_main} with
Eq.~\eqref{eq:delta_Ft_scaling_final_main}, we remark that $\bar{\cF}$
does not depend on $\Gamma D/v^2$ nor on $\Gamma/(v\tilde{\rho})$ for
the fully connected AIM. For this specific kind of alignment,
fluctuations leads to a density-{\it independent} correction of the linear
Landau term and, consistent with the results of
Ref.~\cite{chepizhko2021revisiting}, the continuous mean-field transition to
collective motion resists fluctuations.

\subsection{Alignment with the $k$-nearest neighbors}
\label{sec:k_nearest_renorm}

In this section, we consider an AIM in which particles align with
their $k$-nearest neighbours, as in \eref{eq:OLAIM_dyn_knearest}. At
the field-theoretical level and in one space dimension, particles at
position $x$ thus align with a ``topological'' field $\bar m(x,t)$
computed as
\begin{equation}
  \label{eq:def_topol_m}
\bar{m}(x)=\frac 1 k \int^{x+y(x)}_{x-y(x)}m(z){\rm d}z\; ,
\end{equation}
where the field $y(x)$ measures the interaction range of a particle
located at $x$. It is constructed so that alignment occurs with the
$k$-nearest neighbours by requiring that
\begin{equation}
  \label{eq:definitiony}
  \int^{x+y(x)}_{x-y(x)}\rho(z){\rm d}z=k\;.
\end{equation}
Note that doubling the distance between the particles does not alter
the values of $\bar m(x)$, as expected in topological
models~\cite{Ballerini2008,ginelli2010relevance}.  For the
$k$-nearest-neighbor alignment \eref{eq:def_topol_m}, the homogeneous
solution of \eref{eq:AIM_generic_noisy} in the absence of noise is
given by $\rho=\rho_0$, $m=m_0$, entailing that $y=y_0=k/(2\rho_0)$
and $\bar m= m_0/\rho_0$.  The linear term in $\cF$ then reduces to $2
\Gamma (1-\beta) m_0$, leading to a density-independent mean-field
critical temperature $\beta_m=1$.  Following the same steps as in
Sec.~\ref{sec:linearStab_meanField}, we show in
App.~\ref{app:lin_stab_topo} that the homogeneous disordered and
ordered solutions remain linearly stable for $\beta<\beta_m$ and
$\beta>\beta_m$ respectively.  The topological mean-field theory of
Eq.~\eref{eq:AIM_generic_noisy} with $k$-nearest alignment
\eref{eq:def_topol_m} thus predicts a continuous transition in the
absence of noise.

\begin{figure}[t]
\hspace*{2cm}
  \flushright
  \begin{tikzpicture}
  \node at (0,0) {\includegraphics[width=0.9\textwidth]{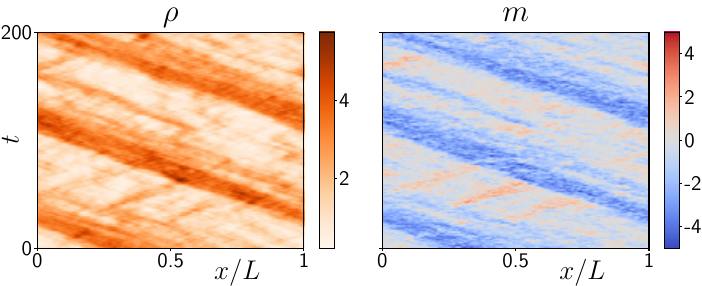}};
  \end{tikzpicture}
  \caption{Kymograph of the density (left) and magnetization profiles (right) resulting from a numerical integration of the stochastic PDE~\eqref{eq:topological_general} with $k$-nearest alignment \eref{eq:def_topol_m}.
  It shows a dense polar band propagating in a disordered gas.
  Parameters: $D=\Gamma=v=1$, $k=0.5$, $\rho_0=2$, $L = 100$, $\beta =1.1$, $dx=0.05$, $dt=0.001$, $\sigma =0.3$. 
  }
  \label{fig:TAIM_band}
\end{figure}

We now discuss the role of fluctuations. Starting from
Eq.~\eqref{eq:topological_general}, we construct the dynamics of the
average fields $\tilde{\rho}=\langle\rho(x,t)\rangle$ and
$\tilde{m}=\langle m(x,t)\rangle$ to order $\sigma$. We stress that
Eq.~\eqref{eq:definitiony} directly yields the field $y(x,t)$ in terms
$\rho(x,t)$. As in the metric case, there are thus only two
independent hydrodynamic fields, $\rho(x,t)$ and $m(x,t)$, and the
renormalization process presented in Sec.~\ref{sec:FIFOT} is still
valid. Note that the definition of $y(x)$ suggests that, to leading
order in $\sigma$, it varies over lengthscales comparable to those
over which $\rho_0(x)$ and $m_0(x)$ varies. The same thus holds for
$\bar m(x)$ and we thus assume that $\delta\rho_1$ and $\delta m_1$
vary on lengthscales much smaller than the lengthscales over which
$\rho_0(x)$, $m_0(x)$, $\bar m(x)$ and $y(x)$ vary.  To first order in
the noise strength $\sigma$, we then find (see
App.~\ref{sec:renormalization_generic} for details):
\begin{subequations}%
\begin{align}%
  \label{eq:tilde_rho_topo_main}
  \partial_{t}\tilde{\rho} &= D\partial_{xx} \tilde{\rho} -v\partial_x \tilde {m} \, ,\\
  \label{eq:tilde_m_topo_main}
  \partial_{t} \tilde{m} &= D\partial_{xx} \tilde {m} -v\partial_x \delta \tilde{\rho} - \cF\left(\tilde{m},\tilde{\rho},\tilde{\bar{m}}\right)-\sigma\Delta\cF \ ,
\end{align}%
\end{subequations}%
where $\tilde{\rho}=\langle\rho\rangle$, $\tilde{m}=\langle m\rangle$, $\tilde{\bar{m}}=\bar{m}(\tilde{\rho},\tilde{m})$ and where the correction $\Delta\cF$ due to renormalization is given by
\begin{equation}
  \label{eq:delta_Ft_renorm_topo}
  \Delta\cF= \tilde{m}\;\frac{2}{k}\;g\left(\beta,\frac{\Gamma k}{v\tilde{\rho}},\frac{\Gamma D}{v^2}\right) + \cO(\tilde{m}^2\sigma)+\cO(\sigma^\frac{3}{2}) \, ,
\end{equation}
with $g$ a positive function whose expression is given in
Eq.~\eqref{eq:g_app} of
App.~\ref{sec:renormalization_generic}. Importantly, $g$ depends
explicitly on the density, and the linear term of the aligning
dynamics thus depends on the density.  As discussed in
Section~\ref{sec:LinStab}, this entails the occurrence of a phase
separation with travelling bands near the onset of
order. Figure~\ref{fig:TAIM_band} shows the result of numerical
simulations of Eq.~\eqref{eq:topological_general} with the $k$-nearest
alignment \eref{eq:def_topol_m}. As predicted, we indeed observe the
formation of inhomogeneous propagating bands close to the onset of
order. (Note that the figure show some small flocks reversing their directions of motion and back-propagating; it would be interesting to investigate whether this is a one-dimensional effect~\cite{benvegnen2022flocking} or the signature of a greater fragility of topological flocks~\cite{benvegnen2023metastability}.)   

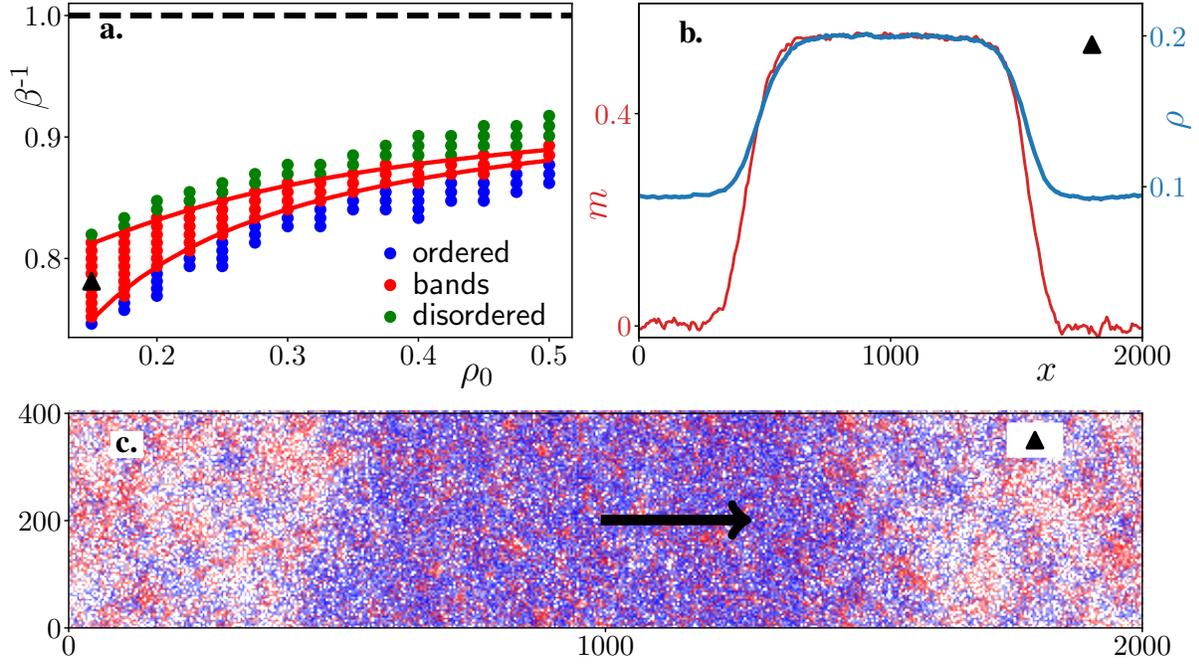
\begin{figure}[h]
  \flushright

    \begin{tikzpicture}
    \node at (0,0) {\includegraphics[width=\textwidth]{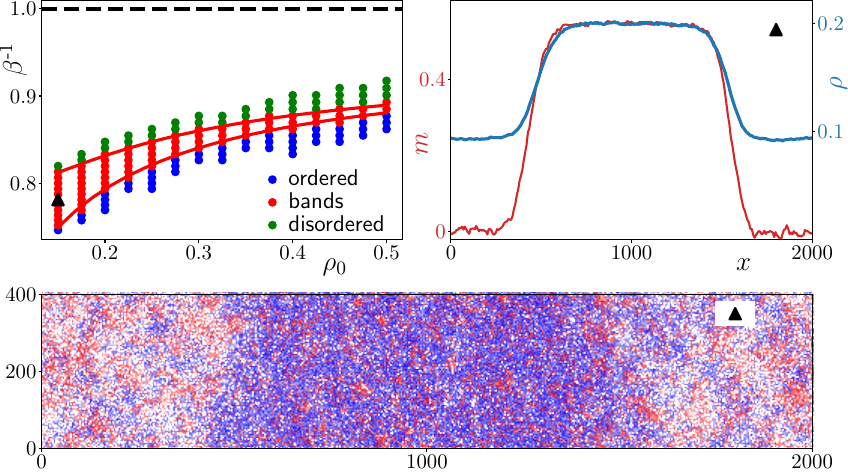}};
    \def\dylabel{4}
      \node[font=\bf] at (-6.5,\dylabel) {a.};
    \node[font=\bf] at (1.2,\dylabel) {b.};
      \node[font=\bf,fill=white,inner sep=2pt] at (-6.3,-1.5) {c.};
      \draw[->,black,line width=4] (0,-2.5) --+ (2,0); 
    \end{tikzpicture}
    \caption{Particle-level simulations of the AIM
      \eqref{eq:flying_xy_spatial}-\eqref{eq:flying_xy_alignment} with
      $k$-nearest neighbors alignment \eref{eq:OLAIM_dyn_knearest}.
            {\bf a.} Phase diagram in the $(\rho_0,\beta^{-1})$
            plane. The homogeneous ordered (blue) and disordered
            (green) regions are separated by a coexistence phase
            (red). 
            The red lines are guide to the eyes which
            show how the transition shifts as the mean density varies.
            {\bf c.} Snapshot of a propagating band corresponding to
            the black triangle in the phase diagram. Blue and red
            particles correspond to positive and negative spins. The
            corresponding density and magnetization fields, averaged
            over $y$ and time, are shown in panel {\bf b.}. The black arrow indicates the direction of motion of the band. 
            Parameters: $D=8$, $L_{y}=400$, $L_{x}=2000$, $k=3$,
            $\Gamma =0.5$, $v=0.9$. Figure adapted from
            Ref.~\cite{martin2021fluctuation}.}
    \label{fig:TAIM_micro}
\end{figure}

The analysis of our topological field-theory thus predicts both a
density-dependent onset of order as well as the emergence of
ordered traveling domains. We now probe these two predictions in
particle-level simulations of the AIM using
Eq.~\eref{eq:flying_xy_spatial} with the $k$-nearest alignment defined
in \eref{eq:OLAIM_dyn_knearest}. Figure~\ref{fig:TAIM_micro} shows the
phase diagram measured numerically in the density-temperature plane.
In this diagram each point $(\rho, \beta^{-1} )$ is colored according
to its steady state: homogeneously ordered, homogeneously disordered,
or travelling domains.  In agreement with our predictions, the onset of
order occurs at noise values that depend on the mean density $\rho_0$.
Furthermore, simulations reveal a discontinuous transition to
collective motion and the emergence of ordered travelling bands
similar to the one described in~\Fref{fig:SDEmetric}
and~\Fref{fig:TAIM_band}.

\subsection{Voronoi alignment}
\label{sec:voronoi_alignment_renorm}

Finally, we consider the case of a topological Voronoi AIM where the aligning field $\bar{m}$ is computed for each particle as the averaged magnetization of its Voronoi neighbors \eref{eq:OLAIM_dyn_voronoi}.
In this case, the lack of a proper field-theoretical expression for
$\bar m$ prevents an explicit derivation of the functionals
$\mathcal{G}$ and $\mathcal{F}$, as well as an analytical proof that
the onset of order becomes density-dependent due to fluctuations.  On
the other hand, direct simulations of the particle-level
dynamics~\eqref{eq:flying_xy_spatial}-\eqref{eq:flying_xy_alignment}
using the CGAL libray allow us to probe whether such a
dependency on the density exists.  We now report the results of these
simulations, while the details of their implementation can be found in
App.~\ref{app:numerics_Voronoi}.

\begin{figure}[t]
  \flushright
  \includegraphics{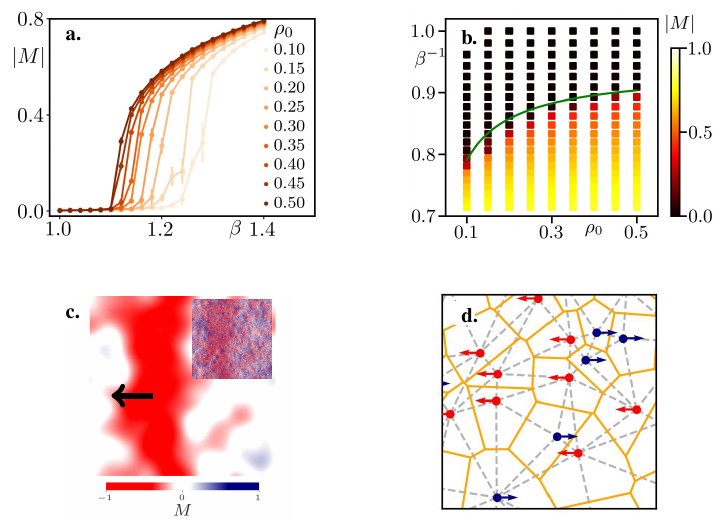}
  \if{
    \beginpgfgraphicnamed{Fig7}
\begin{tikzpicture}










  \def\x{-9.2}

  \path (\x,0) node {\includegraphics[]{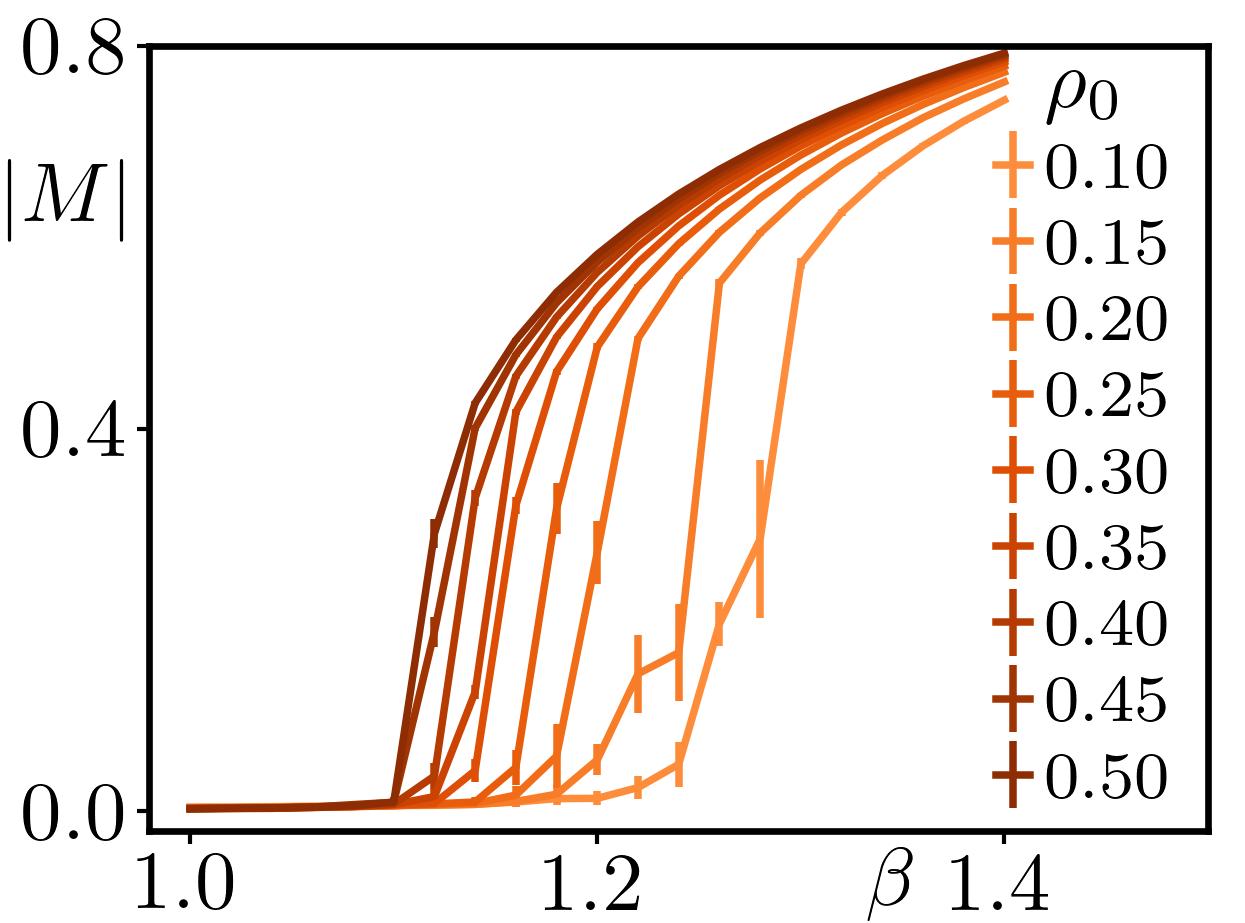}};

  \def\xlabel{-1.65}
  \def\ylabel{1.42}
  \path ( \x + \xlabel, \ylabel) node {\color{black}\fontsize{12}{2}\selectfont \textbf{a.}};

  \def\x{-3.5}

  \path (\x,0) node {\includegraphics[]{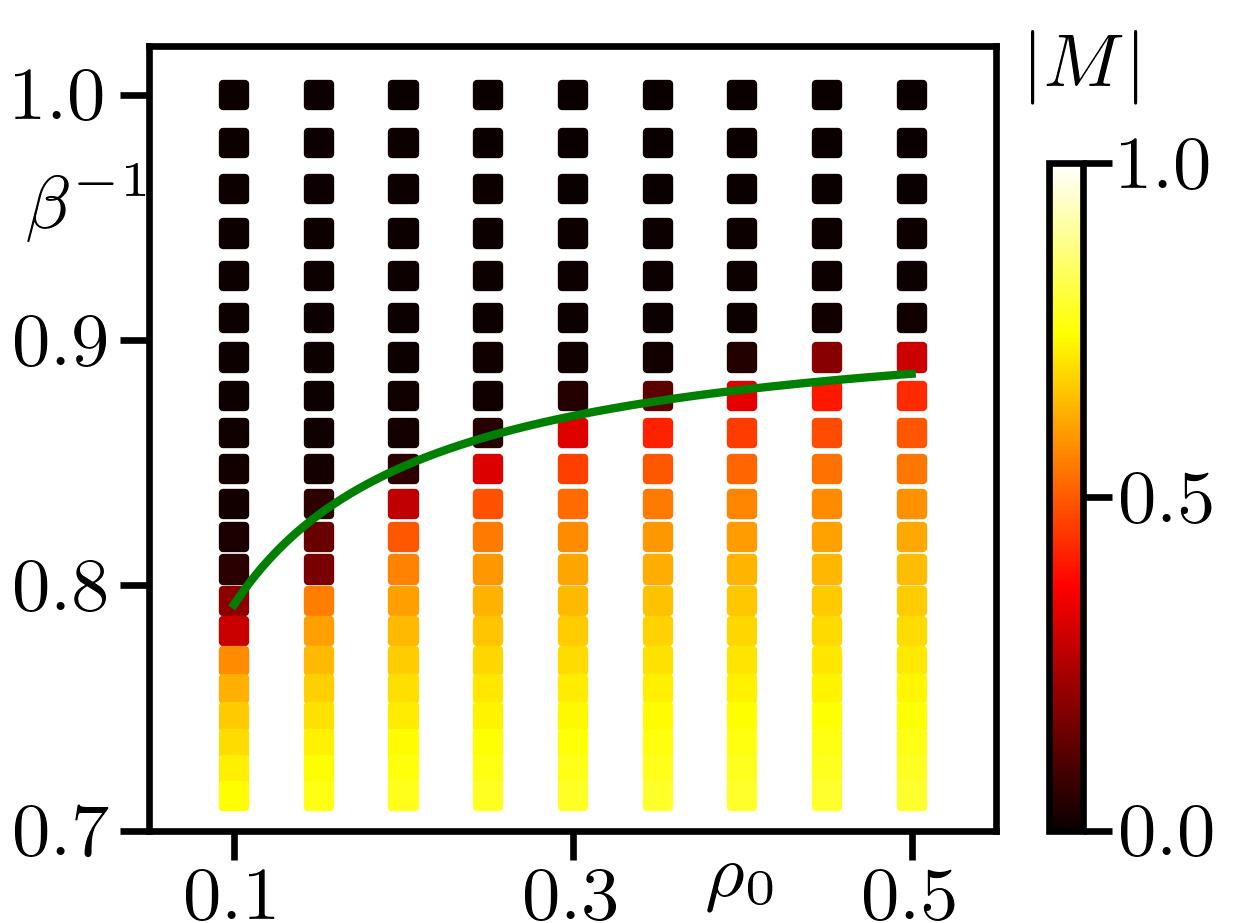}};

  \def\dwx{0.17}
  \def\dwy{0.2}
  \filldraw[white] (\x+\xlabel-\dwx,  + \ylabel-\dwy) rectangle (\x+\xlabel+\dwx, \ylabel+\dwy);
  \path (+\x+ \xlabel, \ylabel) node {\color{black}\fontsize{12}{2}\selectfont \textbf{b.}};

  \def\x{2}
  \def\xinset{3}
  \def\yinset{-1}

    \path (\x,0) node {\includegraphics[]{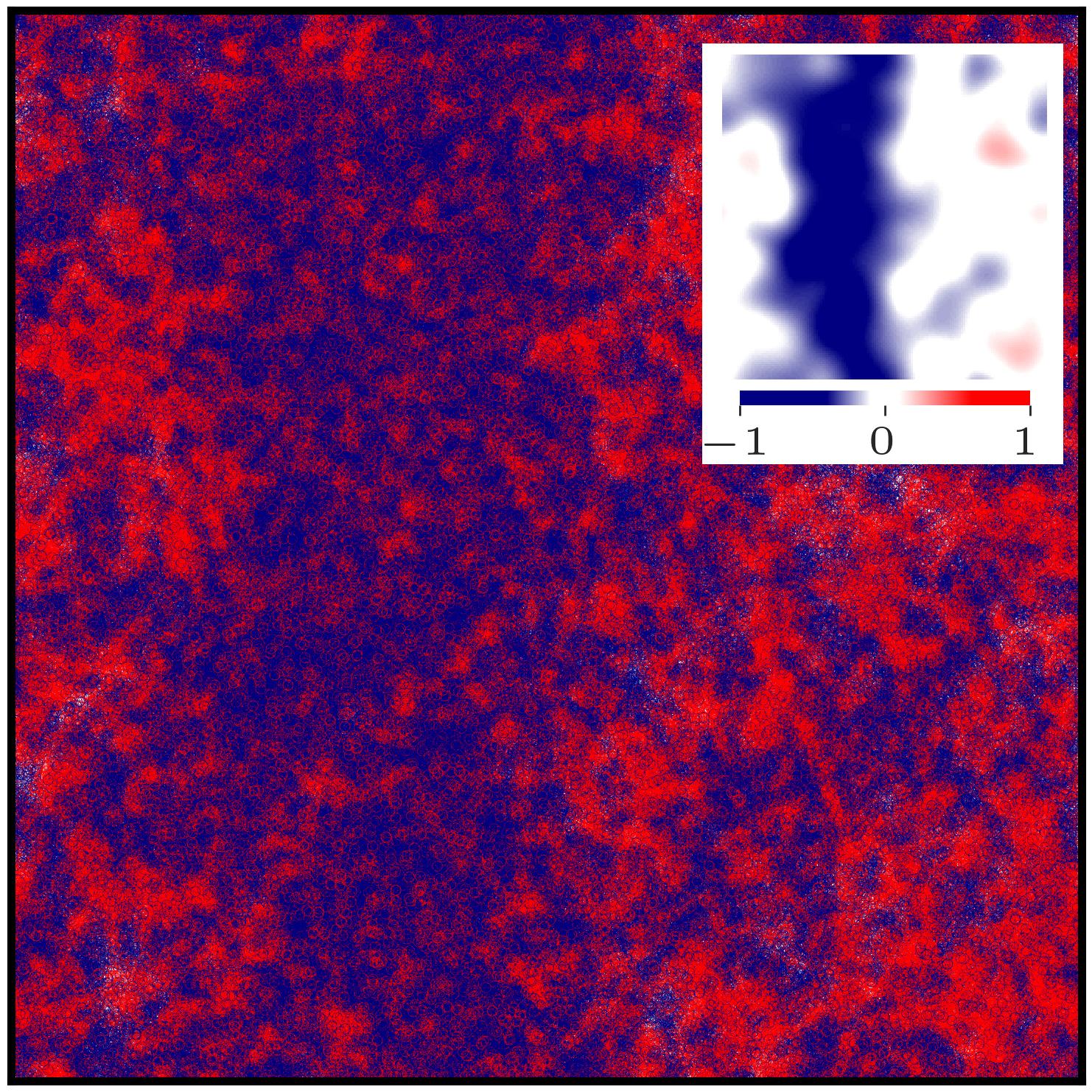}};
  
  \filldraw[white] (\x+0.98*\xlabel-\dwx,  + \ylabel-\dwy) rectangle (\x+0.98*\xlabel+\dwx, \ylabel+\dwy);
  \path (+\x+ 0.98*\xlabel, \ylabel) node {\color{black}\fontsize{12}{2}\selectfont \textbf{c.}};

  \def\xlabel{4}
  \def\ylabel{2.3}
  \path (\xlabel, \ylabel) node {\color{white}\fontsize{12}{2}\selectfont \textbf{c}};

\end{tikzpicture}
\endpgfgraphicnamed{Fig7}
}\fi

    \caption{Particle-level simulations of the AIM
      \eqref{eq:flying_xy_spatial}-\eqref{eq:flying_xy_alignment} with
      Voronoi alignment \eref{eq:OLAIM_dyn_voronoi}.  \textbf{a.}
      Average global magnetization magnitude $|M|$ as a function of
      the inverse temperature $\beta$. Different colors correspond to
      different values of the density $\rho_0$ (decreasing from left
      to right). \textbf{b.} Phase diagram of the flocking transition
      in the temperature $(\beta^{-1},\rho_0)$ plane.  The
      color-coding indicates the magnitude of the steady-state global
      magnetization. The solid green line is a fit of the
      line separating the ordered ($|M|\neq0$) phase from the
      disordered ($M=0$) one to the function $\beta^*( \rho_0)
      = \beta_0 + \beta_1 / \rho_0$, with $\beta_0 \simeq 1.067$ and
      $\beta_1 \simeq 0.020$.  \textbf{c.} Heatmap of the local magnetization $M$ of a travelling band. The inset shows the corresponding snapshot with right-moving (blue circles) and left-moving (red circles) particles. The black arrow indicates the direction of motion of the band. {\bf d.} Close-up showing the Voronoi tessellation (yellow lines) extracted from the inset of panel c. Interacting spins are connected by gray dashed lines. Parameters used in c. and d.: $\rho_0 = 0.2$, $\beta = 1.18$,
      $v=0.5$, $D=5$, $\Gamma=0.5$; box size: $L_x=L_y=1500$. }
      \label{fig:aim_voronoi}
  \end{figure}
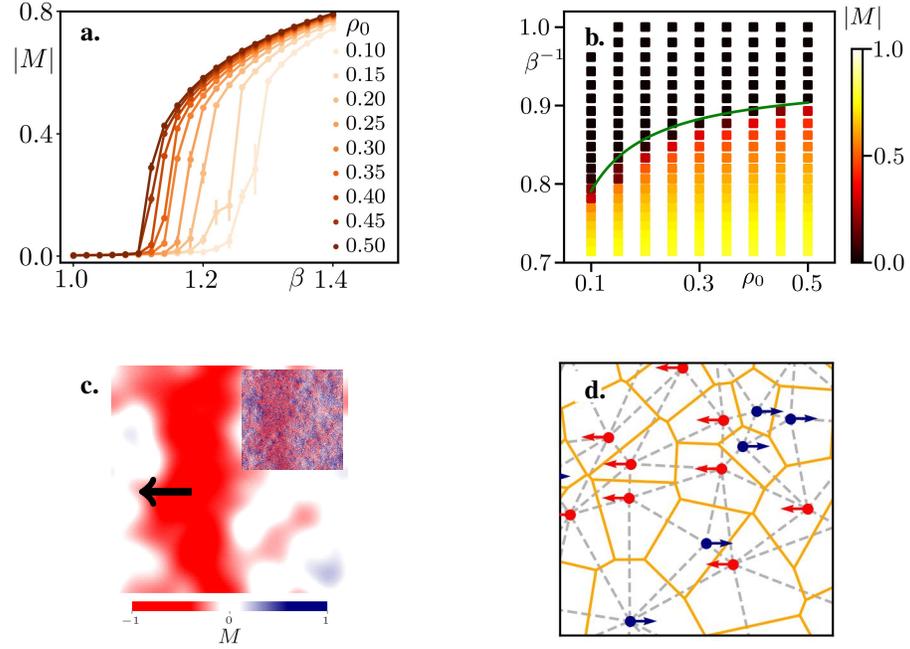

In Fig.~\ref{fig:aim_voronoi}a, we measured the global magnetization
magnitude $|M|=\langle| m |\rangle$ as a function of the inverse
temperature $\beta$ for different densities $\rho_0$.  We observe that
the critical temperature $\beta^*$ at which a global magnetization
develops depends on the average density. Figure~\ref{fig:aim_voronoi}b
displays the phase diagram of the Voronoi AIM model in the
density-temperature plane. Each point in the $(\rho_0, \beta^{-1})$
plane corresponds to an equilibrated simulation colored according to
its steady-state global magnetization magnitude. The green curve is a
fit to the line separating the disordered and ordered regions, with
$\beta^*( \rho_0) = \beta_0 + \beta_1 / \rho$, (with $\beta_0 \simeq
1.067$ and $\beta_1 \simeq 0.020$), hence confirming the
density-dependence of the critical temperature. 

Given the shape of the phase diagram, we expect from the linear
stability analysis of Sec.~\ref{sec:LinStab} that the transition is
characterized by non-homogeneous travelling solutions. This is
confirmed in simulations of the particle-level dynamics performed close
to the onset of order for which a painstaking numerical effort allowed
us to report the existence of stable ordered bands. Let us note that,
due to the weak density-dependence of the critical temperature on
$\rho$, very large system sizes are required to observed traveling
bands, in accordance with Eq.~\eqref{eq:Lstar}. As a consequence,
despite the large system sizes used, we are completely unable to
measure binodal densities and properly delimit the band
region. Figure~\ref{fig:aim_voronoi}c nevertheless displays a snapshot
of such a band made of left-moving particles.

\section{Fluctuation-Induced First Order Transition in Toner--Tu models}
\label{part:FIFOT_polar_case}

In this final part, we discuss the case of vectorial flocking models.
We begin in Sec.~\ref{subsec:lin_stab_toner_tu} by considering the Toner-Tu hydrodynamic equations and showing that, in
this case as well~\cite{bertin2009hydrodynamic,mishra2010fluctuations}, a density-dependent polar-field mass generically
entails a discontinuous flocking transition. Then, by complementing the
mean-field description of a topological version of Toner--Tu equations
with noise, we show analytically in
Sec.~\ref{subsec:vicsek_voronoi_renorm} how a renormalized
density-dependent polar-field mass emerges in this model. 
Finally, in Sec.~\ref{subsec:numerical_topol_vicsek}, we perform particle-level simulations and confirm in
\Fref{fig:vicsek_topological_neigh_microscopic}a and
\Fref{fig:vicsek_topological_microscopic} a density-dependent onset of
order in the Vicsek model for both $k$-nearest neighbors (using
results of~\cite{martin2021fluctuation}) and for Voronoi alignment
(using new simulation
results). 
In addition, Figure~\ref{fig:vicsek_topological_neigh_microscopic} also
shows the emergence of travelling bands for
$k$-nearest-neighbour alignment. In the Voronoi case, the very weak
dependency of the onset to order suggests that enormous sizes would be
required to see bands, and the latter have indeed so far resisted our
numerical efforts.

\subsection{Linear stability analysis of a vectorial polar active fluid}
  \label{subsec:lin_stab_toner_tu}

Our starting point is the Toner--Tu hydrodynamic equations
describing the evolution of density $\rho(\br,t)$ and velocity fields
$\bW(\br,t)$ as~\cite{toner2012reanalysis}
\begin{subequations}%
\begin{align}%
  \partial_t\rho + \bnab\cdot \bW=& 0 \, .
\label{eq:Caussin_W_full_2}\\
    \partial_{t} \bW=& -\lambda_1(\bW\cdot\bnab)\bW- \lambda_2(\bnab\cdot\bW)\bW -\lambda_3 \bnab(|\bW|^2) - \alpha\bW-a_4 |\bW|^{2}\bW -\bnab P_1\nonumber \\
    & -\bW \left( \bW \cdot \bnab  P_2 \right) + D_{B} \bnab (\bnab \cdot \bW)+ D_{T}\nabla^{2}\bW + D_{2}(\bW\cdot\bnab)^{2}\bW \, ,
 \label{eq:Caussin_W_full_1} 
\end{align}\label{eq:Caussin_W_full}%
\end{subequations}%
where the coefficients $\lambda_i$, $\alpha$, $a_4$, $D_B$ , $D_T$,
$D_2$, as well as the scalar functions $P_1$ and $P_2$, depend, in
general, on the local density $\rho(\br,t)$ and on the amplitude
$|\bW(\br,t)|$ of the velocity field.  Since we are interested in the
onset of flocking, we focus on a simplified one-dimensional version of
Eq.~\eqref{eq:Caussin_W_full}, where this single dimension corresponds
to the main direction of motion in the ordered phase.  In that case,
${\bf W}$ becomes a scalar and Eq.~\eqref{eq:Caussin_W_full}
simplifies into
\begin{subequations}%
\begin{align}%
\partial_t\rho + \partial_x W =&\, 0 \, ,
\label{eq:Caussin_W_final_2}
\\
\partial_{t}
W + \lambda W \partial_x W =&\, -\alpha W - a_4 W^3 - \partial_x P_1 -W^2\partial_x P_2  + D \partial_{xx}W + D_2 W(\partial_x W)^2\;.
\label{eq:Caussin_W_final_1} 
\end{align}%
\end{subequations}
In order to perform a linear stability analysis of the homogeneous profiles, we introduce a characteristic velocity scale $v$ such that $W=v\hat{W}$. Introducing similarly the variables $\hat \lambda=v\lambda$, $\hat{a}_4=v^2 a_4$, $\hat{P}_1=v^{-1}P_{1}$, $\hat{P}_2=v P_{2}$ and $\hat{D}_2=v^2 D_2$, the dynamics can be rewritten as
\begin{subequations}%
\begin{align}%
  \partial_t\rho + v \partial_x W =&\, 0 \, ,
  \label{eq:Caussin_W_adim_2}
\\  \partial_{t}
  W + \lambda W \partial_x W =&\, -\alpha W - a_4 W^3 - \partial_x P_1 -W^2\partial_x P_2  + D \partial_{xx}W + D_2 W(\partial_x W)^2 \, , \label{eq:Caussin_W_adim_1}
\end{align}\label{eq:Caussin_W_adim}%
\end{subequations}
where we have dropped the hat for simplicity and where $\rho$ and $W$
in Eq.~\eqref{eq:Caussin_W_adim} now have the same dimension.  We then
consider perturbations $\delta \rho $ and $\delta W$ around the
homogeneous solution $\rho_0$, $W_0=\sqrt{|\alpha|/a_4}$. The
linearized dynamics of $\delta\rho$ and $\delta W$ in Fourier space
read
\begin{align}
  \label{eq:Caussin_stabMatrix}
  \partial_t \begin{pmatrix}
  \delta {\rho_q} \\
  \delta {W_q}
  \end{pmatrix} = \begin{pmatrix}
  0 & -i v q \\
  \xi_1 -iq \xi_2  & \xi_3 -iq \xi_4 - \xi_5 q^2
  \end{pmatrix} \cdot
  \begin{pmatrix}
  \delta \rho_q \\
  \delta W_q
  \end{pmatrix}\, ,
\end{align}
where the $\xi_i$ are functions of $\rho_0$ and $W_0$.  Their exact
expressions can be directly deduced from a Taylor expansion around
$\rho_0$ and $W_0$ of the parameters and scalar functions appearing in
\eref{eq:Caussin_W_adim_1}.  Here, we are only interested in their
scaling with $W_0$ near the transition, when $|W_0|\ll 1$, where they
satisfy
\begin{subequations}\label{eq:xi_scalings}
\begin{align}%
  \xi_1 =& W_0 \alpha^{\prime} + \cO(W_0^2)\, , \quad
  \xi_2 = \xi_2^0 + \cO(W_0)\, , \quad
  \xi_3 = -\gamma W_0^2 + \cO(W_0^3)\, , \\
  \xi_4 =& \xi_4^0 W_0 + \cO(W_0^2)\, , \quad
  \xi_5 = \xi_5^0 + \cO(W_0)\, ,
\end{align}
\end{subequations}
with $\alpha^{\prime}$ and $\gamma$ defined as
\begin{align}
  \alpha^{\prime} = \frac{\partial\alpha}{\partial \rho}\bigg{|}_{\rho=\rho_0,\,|W|=0}\, \quad, \quad \gamma = 2 a_4(\rho=\rho_0,|W|=0)\, .
\end{align}%
From the analysis of the eigenvalues of the stability matrix in Eq.~\eqref{eq:Caussin_stabMatrix}, we show in App.~\ref{app:Caussin_stab} that the homogeneous solution is unstable as soon as $b(q)>0$ where
\begin{align}
  b(q) =& 16 q^2 v \left(\xi_1^2 v+\xi_3 \xi_4\xi_1-\xi_2 \xi_3^2\right)+16 q^4 v \xi_5\; (2 \xi_2\; \xi_3-\xi_1\; \xi_4)-16 q^6 \left(\xi_2\; \xi_5^2\; v\right) \ .
\end{align}
Close to the transition, $|W_0| \ll 1$, and Eq.~\eqref{eq:xi_scalings}
allows us to rewrite $b(q)$ as
\begin{align}
  \label{eq:bq_transition_TT}
  b(q) = 16 v^2 q^2\left[(W_0 \alpha')^2 + \cO(W_0^3)\right] \, .
\end{align}
The homogeneous solution is therefore unstable provided that the polar-field mass depends on density, \textit{i.e.} $\alpha'\neq0$. 

To find the system size needed to observe the instability when $\alpha'\to0$, we study the scaling of the largest root $q^{\star}$ of  $b$ in this limit. Since $q^\star$ is expected to vanish, we can truncate $b(q)$ at the lowest nontrivial order in q (order 4) to estimate $q^{\star}$ as 
\begin{align}
q^{\star}= \sqrt{\frac{\xi_1^2 v+\xi_3\xi_4\xi_1-\xi_2\xi_3^2}{2(\xi_1\xi_4-2\xi_2\xi_3)\xi_5}}\;,
\end{align}
from which we deduce the critical system size $L^{\star}$ close to the transition ($|W_0|\ll 1$) as 
\begin{align}\label{eq:LstartVM}
L^{\star}\equiv \frac{2\pi}{q^\star}\simeq 2\pi\sqrt{\frac{2\xi_5^0(\alpha^{\prime} \xi_4^0+\gamma \xi_2^0)}{v\alpha^{\prime \: 2}}}\; .
\end{align}
When $\alpha$ weakly depends on the density, $\alpha^{\prime} \ll 1$ and $L^{\star}\sim 1/|\alpha^{\prime}|$: the critical system size follows the same scaling as computed in Section \ref{sec:finite_size} in the case of the Active Ising Model.


The finite-wavelength instability of the homogeneous ordered phase discussed above is of the same nature as that reported for other hydrodynamics description of polar active matter~\cite{bertin2009hydrodynamic,mishra2010fluctuations}. It confirms that the dependence of the polar-field mass on the density is a generic mechanism making the homogeneous ordered phase unstable at onset. Finally, Eq.~\eqref{eq:bq_transition_TT} shows that this instability is independent of the sign of $\alpha'$ and
holds for any active polar fluid described by a Toner--Tu
hydrodynamics.


\subsection{Renormalization of a topological Toner--Tu model: first order}
\label{subsec:vicsek_voronoi_renorm}

To make progress, we now turn to an explicit model for which the
values of the coefficients appearing in Eq.~\eqref{eq:Caussin_W_full}
are known.  For this explicit model, we derive the fluctuation-induced
renormalization of the mass term and show that, similarly to our
discussion in Sec~\ref{sec:FIFOT} for the AIM, it also exhibits a
dependency on the density.
  
We consider the mean-field dynamics derived in  Ref.~\cite{peshkov2012continuous} for the case of the Vicsek model with Voronoi alignment:
  \begin{subequations}%
  \begin{align}%
    \label{eq:hydro_Voronoi_1}
    \partial_t \rho + \nabla \cdot \bW &= 0 \, , \\
    \label{eq:hydro_Voronoi_2}
    \partial_t \bW + \frac{\lambda}{\rho} (\bW \cdot \nabla) \bW &= -\frac{v^2}{2} \nabla \rho + \frac{\kappa}{2\rho}\nabla\bW^2 -\left(\alpha+\frac{\gamma}{\rho^2} \bW^2\right) \bW +D \nabla^2 \bW -\frac{\kappa}{\rho} (\nabla \cdot \bW)\bW \, ,
\end{align}\label{eq:hydro_Voronoi}%
\end{subequations}%
where the parameters $D$, $\alpha$, $\lambda$, $\gamma$, $\kappa$, and $v$ are density-independent.
Note that in Eq.~\eref{eq:hydro_Voronoi_2}, we have introduced the
self-propulsion speed $v$, which was set equal to unity in
Ref.~\cite{peshkov2012continuous} by rescaling time and space.

As in Sec.~\ref{subsec:lin_stab_toner_tu}, we consider the one-dimensional version of Eq.~\eqref{eq:hydro_Voronoi}, which reads
\begin{subequations}\label{eq:hydro_voronoi_1D}
  \begin{align}%
    \label{eq:hydro_Voronoi_1D_1}
    \partial_t \rho + \partial_x W &= 0 \, , \\
    \label{eq:hydro_Voronoi_1D_2}
    \partial_t W + \frac{\lambda}{\rho} W \partial_x W &= -\frac{v^2}{2} \partial_x \rho + D \partial_{xx}W - \alpha W -\frac{\gamma}{\rho^2} W^3 \, .
  \end{align}%
\end{subequations}%
Introducing the variables $\bar{W}=v^{-1}W$, $\bar \lambda=v\lambda$, $\bar{\gamma}=v^2 \gamma$ and $\bar{D}=v^2 D$, Eqs~\eref{eq:hydro_voronoi_1D} can then be expressed as
\begin{subequations}%
\begin{align}%
  \label{eq:hydro_Voronoi_1D_hat_1}
  \partial_t \rho + v \partial_x \bar{W} &= 0 \, , \\
  \label{eq:hydro_Voronoi_1D_hat_2}
  \partial_t \bar{W} + \frac{\bar{\lambda}}{\rho} \bar{W} \partial_x \bar{W} &= -\frac{v}{2} \partial_x \rho + D \partial_{xx}\bar{W} - \alpha \bar{W} -\frac{\bar{\gamma}}{\rho^2} \bar{W}^3 \, ,
\end{align}%
\end{subequations}
Dropping the bar notation for clarity and dressing the dynamics with noises, we obtain a stochastic evolution describing an active polar fluid with Voronoi alignment as
\begin{subequations}\label{eq:hydro_Voronoi_fluct}
\begin{align}%
  \label{eq:hydro_Voronoi_fluct_1}
  \partial_t \rho + v\partial_x W &= \partial_x\left(\sqrt{2\sigma\epsilon\rho}\;\eta_1\right) \\
  \label{eq:hydro_Voronoi_fluct_2}
  \partial_t W + \frac{\lambda}{\rho} W \partial_x W &= -\frac{v}{2} \nabla \rho + D \partial_{xx}W - \alpha W -\frac{\gamma}{\rho^2} W^3 + \sqrt{2\sigma\rho}\;\eta_2 \; ,
\end{align}%
\end{subequations}%
where $\eta_{1,2}$ are Gaussian noises with unit variance while
$\sigma$ and $\epsilon$ are constants controlling their strengths (We
consider here the role of the conserved noise for the sake of
generality).  Equation~\eref{eq:hydro_Voronoi_fluct} is the starting
point for our renormalization procedure.  Following the method
developed in Sec.~\ref{sec:FIFOT}, we show in App.~\ref{app:ren_TT}
that the average fields $\tilde{\rho}=\langle \rho \rangle$ and
$\tilde{W}=\langle W \rangle$ are solutions, up to order $\sigma$, of
the renormalized hydrodynamic equation
  \begin{subequations}%
  \begin{align}%
    \label{eq:hydro_Voronoi_ren_1}
    \partial_t \tilde{\rho} + v\partial_x \tilde{W} &= 0 \\
    \label{eq:hydro_Voronoi_ren_2}
    \partial_t \tilde{W} + \frac{\hat{\lambda}}{\tilde{\rho}} \tilde{W} \partial_x \tilde{W} &= -\frac{v}{2} \nabla \tilde{\rho} + D \partial_{xx}\tilde{W} - \hat{\alpha}\tilde{W} - \frac{\hat{\gamma}}{\tilde{\rho}^2}\tilde{W}^3 \; ,
\end{align}\label{eq:hydro_Voronoi_ren}%
\end{subequations}%
  where $\hat{\alpha}$, $\hat{\gamma}$ and $\hat{\lambda}$ are the
  renormalized coefficients due to fluctuations. In particular, we
  find that the renormalized mass of the polar field is given by
  \begin{align}
    \label{eq:alpha_ren_TT}
    \hat{\alpha} = \alpha +\sigma\frac{3\gamma}{\tilde{\rho}}\frac{2D+\alpha \epsilon}{4D\sqrt{D|\alpha|}} \, ,
  \end{align}
  and is now density dependent. Based on the stability analysis of
  Sec.~\ref{subsec:lin_stab_toner_tu}, we thus conclude that
  Eq.~\eqref{eq:hydro_Voronoi_ren} exhibits a discontinuous emergence
  of collective motion with travelling bands at onset.

  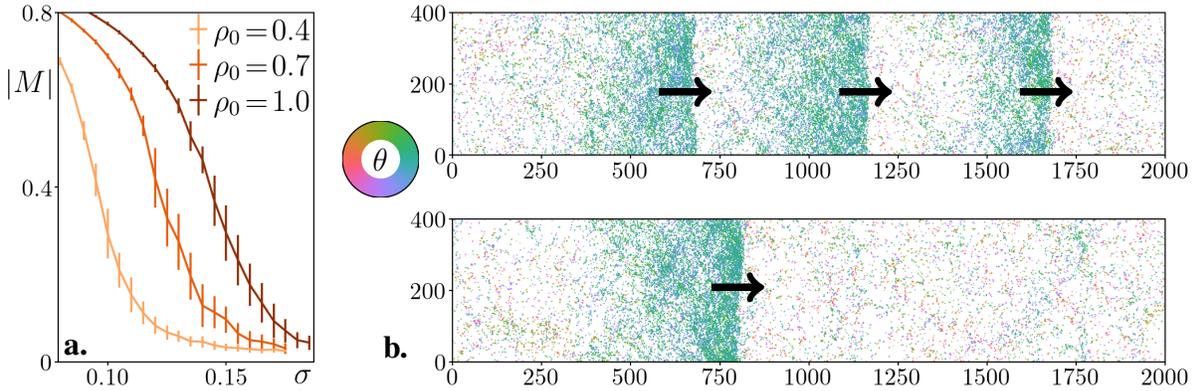
\begin{figure}
    \flushright
    \includegraphics{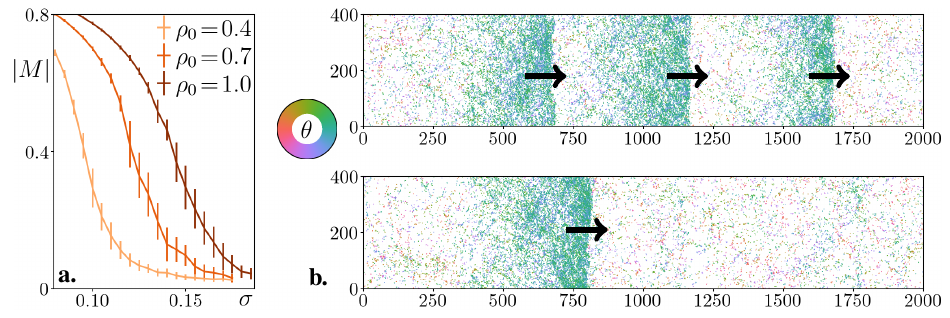}
    \if{
    \begin{tikzpicture}
    \node at (0,0) {\includegraphics[width=\textwidth]{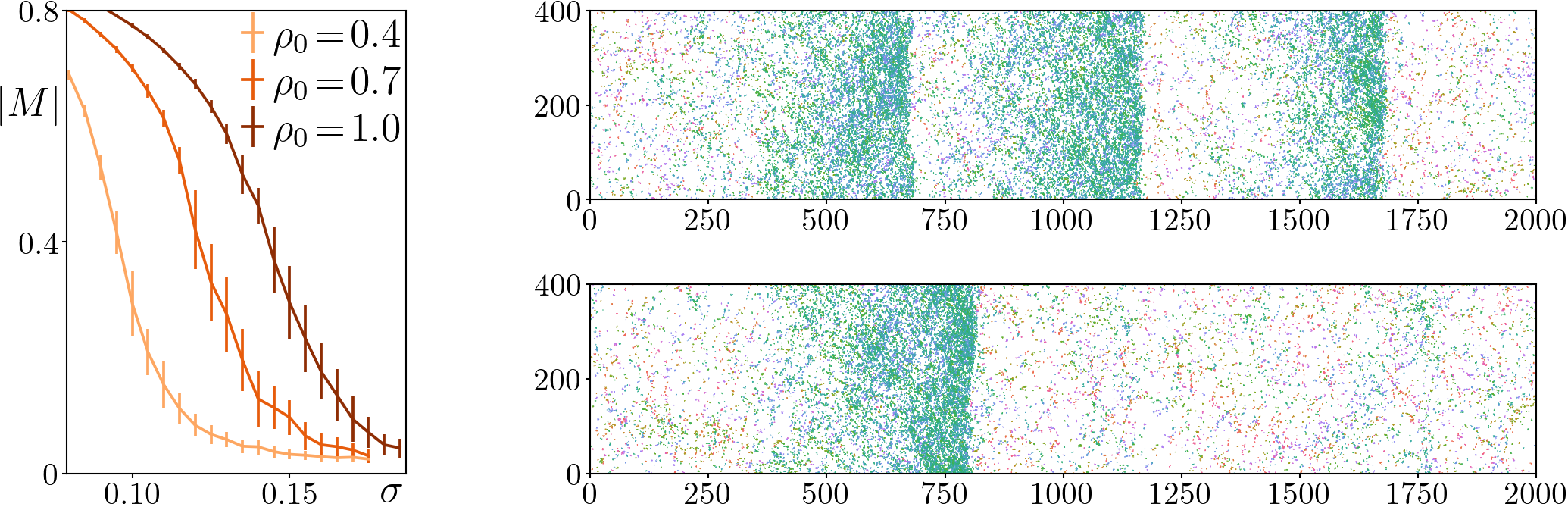}};
    \def\dx{-2.9}
    \def\dy{0.5}
    \node at (\dx,\dy) {\includegraphics[width=1cm]{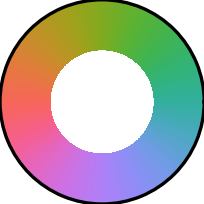}};
    \node at (\dx,\dy) {$\theta$};
    \node[font=\bf] at (-6.95,-2) {a.};
    \node[font=\bf] at (-2.7,-2) {b.};
    
    {\draw[line width = 1mm,black,->] (3.2,1.4) -- ++(0.7,0);}
    {\draw[line width = 1mm,black,->] (0.8,1.4) -- ++(0.7,0);}
    {\draw[line width = 1mm,black,->] (5.6,1.4) -- ++(0.7,0);}
    
    {\draw[line width = 1mm,black,->] (1.5,-1.2) -- ++(0.7,0);}
    
    \end{tikzpicture}
    }\fi
    \caption{Simulations of the Vicsek Model with $k$-nearest
      neighbors alignment in 2D \eref{eq:Vicsek_micro}. At low
      density, the system remains disordered. Increasing the density
      then leads to an onset of order accompanied by propagating
      bands.  {\bf a.} Magnitude $|M|$ of the average total
      magnetization as a function of the noise strength $\sigma$ for
      different values of the density. The curves show the dependency
      of the onset of order on the density.  Parameters: $L_x = 400$,
      $L_y=400$, $k=3$, $v=0.2$, $\Delta t=1$. Errorbars correspond to
      the standard deviation over 48 realizations.  {\bf b.}
      Snapshots of the particle-based system showing travelling
      bands emerging at the onset of order. The black arrows indicate the direction of motion of the bands. Parameters: $L_x =
      2000$, $L_y=400$, $\sigma=0.08$, $k=3$, $v=0.2$, $\Delta t=1$,
      $\rho_0=0.4$ (top) and $\rho_0=0.25$ (bottom). Figure partially
      adapted from~\cite{martin2021fluctuation}.  }
    \label{fig:vicsek_topological_neigh_microscopic}
  \end{figure}

\subsection{Numerical results for topological Vicsek models}
\label{subsec:numerical_topol_vicsek}
The results of the sections above suggest that topological Vicsek models should also undergo first-order phase transitions. 
To verify this prediction in microscopic models, we performed  particle-based simulations of the Vicsek Model with both $k$-nearest
  and Voronoi alignments according to Eq.~\eref{eq:Vicsek_micro}. 
  
  In the case of $k$-nearest alignment, we observed both a density-dependent onset of order and, in large-enough systems, the formation of travelling bands (see  \Fref{fig:vicsek_topological_neigh_microscopic}), thereby
  confirming our ﬁeld-theoretical predictions.

\begin{figure}
    \begin{center}
    \flushright
      \includegraphics{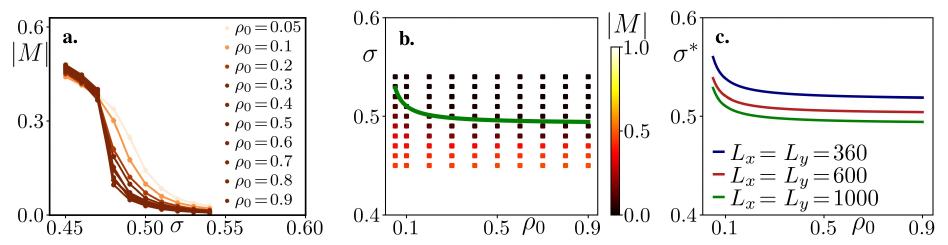}
      \if{
      \beginpgfgraphicnamed{Fig9}
    \begin{tikzpicture}
            
  \def\x{-9.2}

  \path (\x,0) node {\includegraphics[]{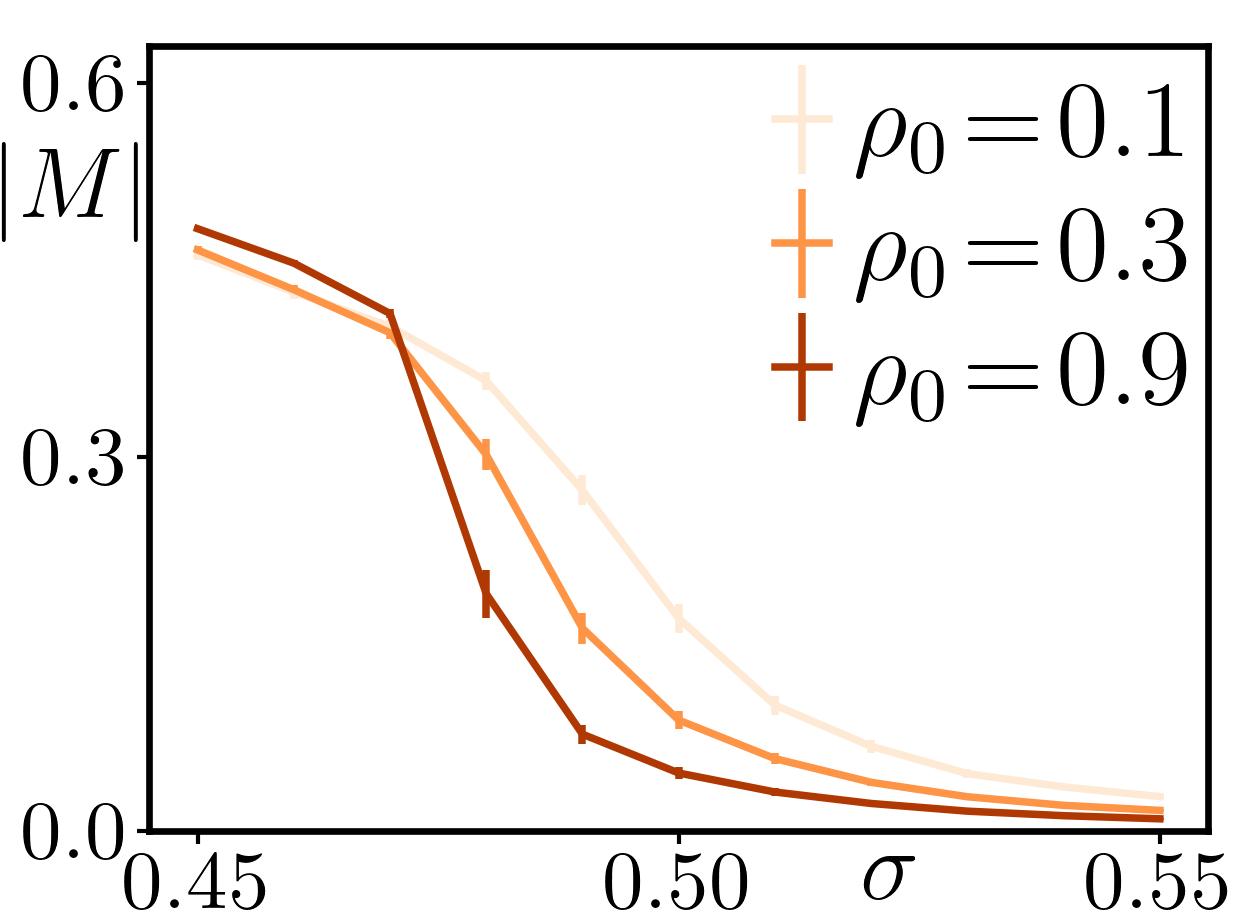}};

  \def\xlabel{-1.65}
  \def\ylabel{1.42}
  \path ( \x + \xlabel, \ylabel) node {\color{black}\fontsize{12}{2}\selectfont \textbf{a.}};

  \def\x{-3.5}

  \path (\x,0) node {\includegraphics[]{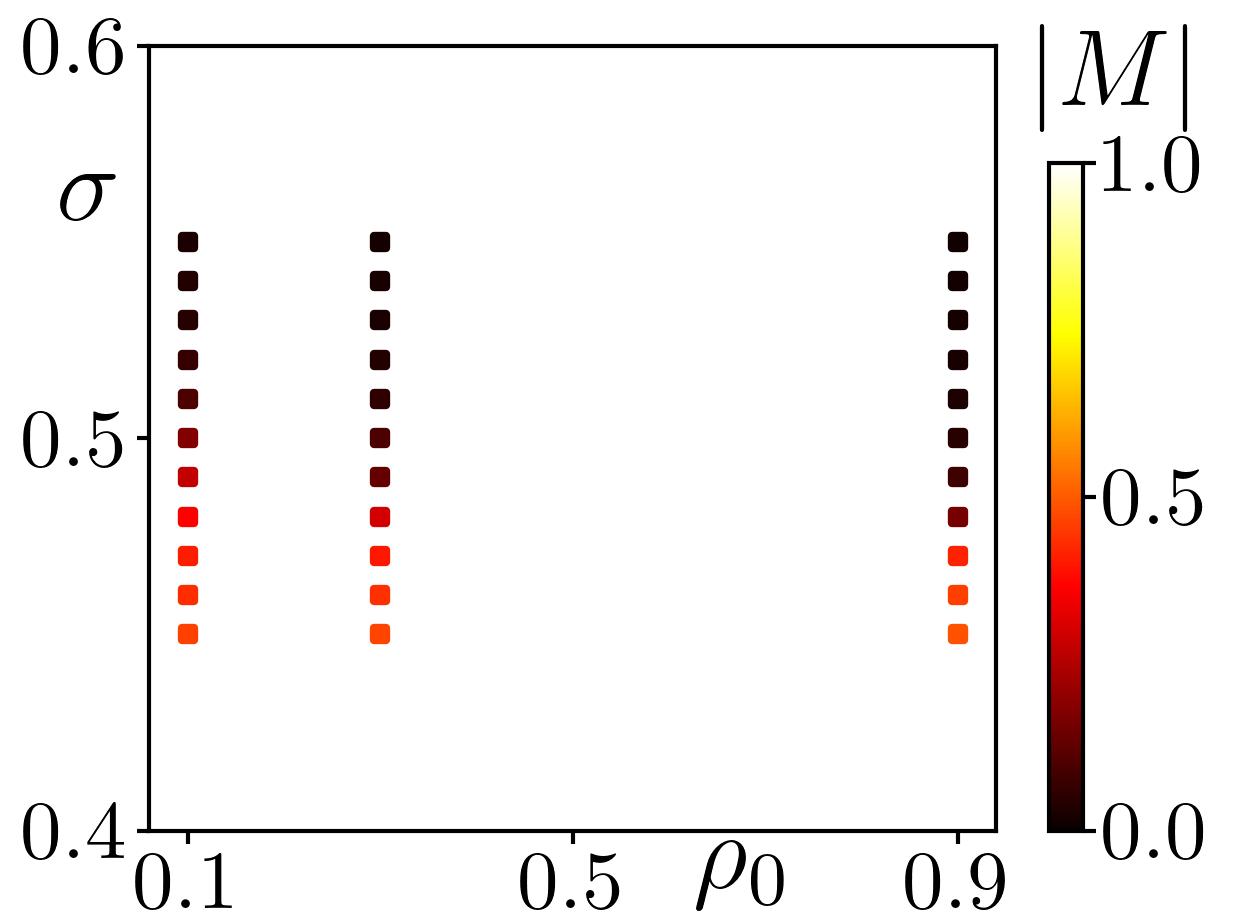}};

  \def\dwx{0.17}
  \def\dwy{0.2}
  \filldraw[white] (\x+\xlabel-\dwx,  + \ylabel-\dwy) rectangle (\x+\xlabel+\dwx, \ylabel+\dwy);
  \path (+\x+ \xlabel, \ylabel) node {\color{black}\fontsize{12}{2}\selectfont \textbf{b.}};

  \def\x{2}
  \def\xinset{3}
  \def\yinset{-1}

    \end{tikzpicture}
    \endpgfgraphicnamed{Fig9}
    }\fi
\end{center}
    \caption{Simulations of the Vicsek Model with Voronoi alignment. {\bf a.} Magnitude $|M|$ of the average total magnetization as a function of the noise strength $\sigma$. Different colors correspond to different densities $\rho_0$. {\bf b.}  The color code indicates the magnitude of the average magnetization in the $(\sigma, \rho_0)$ plane.  The solid green line is a fit of the boundary between the `ordered' region (defined as $|M|>M^{\rm c}$) and the `disordered' one ($|M|<M^{\rm c}$) to the function $\sigma^*(\rho_0) = \sigma_0 + \sigma_1/\rho_0$. We use $M^{\rm c}=0.05$ and find $\sigma_0\simeq 0.492$ and $\sigma_1\simeq 0.002$. 
      Parameters:
      $L_x = L_y=1000$, $v=0.5$, $\Delta t=1$. {\bf c.} Threshold $\sigma^*(\rho_0)$ between the 'ordered' and 'disordered' regions for three different system sizes. The onset of order is shifted towards lower noise values as the system size increases, but the density dependence of the threshold remains.}
    \label{fig:vicsek_topological_microscopic}
  \end{figure}
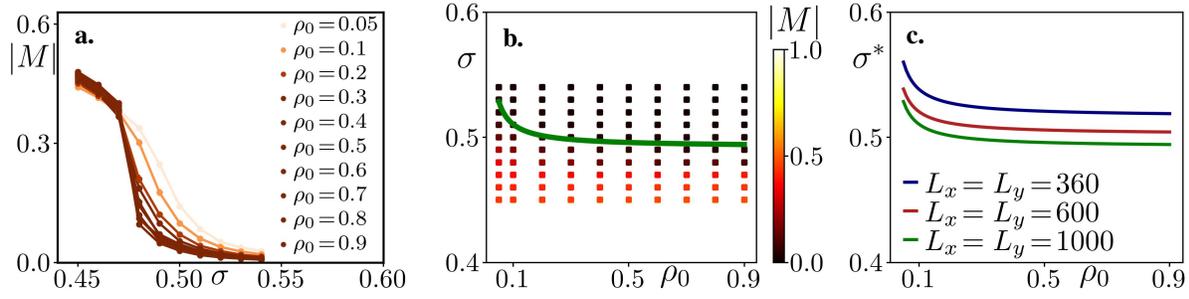

  In the case of Voronoi alignment, we measured the critical
  temperature at fixed system size. Our results are consistent with a density-dependent critical temperature, albeit with a much weaklier dependence than in the other models
  considered so far (See
  \Fref{fig:vicsek_topological_microscopic}). This weak dependence
  suggests that very large sizes would be needed to observe travelling bands and we failed at doing so. Furthermore, our simulations suggest that $\alpha'(\rho) >0$ in this case, contrary to the other models we studied. 
  We note, in particular, that the sign of $\alpha'(\rho)$ is opposite to what the renormalization done in Sec.~\ref{subsec:vicsek_voronoi_renorm} predicts. 
  This may signal that the mean-field hydrodynamic description proposed in~\cite{peshkov2012continuous} is not a good starting point to carry out this computation or that a simple Gaussian white noise field is not sufficient to capture the trend of $\alpha(\rho)$ quantitatively. 
  All in all, the linear instability leading to travelling bands should still be present, but this unexpected difference is worth noting and calls for a great care in assessing our numerical results. More extensive numerical simulations are definitely required to confirm these results. 
  
\section{Conclusion}
In this article, we have shown analytically and numerically that
complementing mean-field theories of flocking models with noise
generically leads to a density-dependent renormalization of the mass of
the polar field. In turn, this dependency on the density triggers a
feedback loop between the ordering dynamics and the advection of the
density fields that leads to the emergence of traveling patterns. This
fluctuation-induced first order transition (FIFOT) is generic and
applies both to metric and topological models of collective motion.

For topological models, we confirmed our field-theoretical predictions
using particle-level simulations with $k$-nearest-neighbour and Voronoi
alignments. Whether the ordering field is vectorial (Vicsek-like) or
discrete (Ising), we measured a density-dependent onset of order. In
all cases but the Voronoi-based Vicsek models, we could reach system
sizes large enough to observe traveling bands.

We also extended our approach to two-dimensional systems and showed
that the FIFOT mechanism still holds with an interesting twist: in 1D,
our method shows that different microscopic models sharing the same
mean-field description have the same `universal' density-dependent
correction of the polar-field mass. In 2D, we expect that the UV
divergence of our theory makes particle-level differences relevant.

Moreover, we characterized the sensitivity of the flocking transition to finite-size effects and detailed how the weakly first order transition could be mistaken for a continuous one. Since the instability of the homogeneous ordered phase is a finite-wavelength instability~\cite{bertin2009hydrodynamic}, one must indeed simulate systems larger than a critical size to observe the
discontinuous nature of the transition. As we show, this critical size may become very large when the polar-field mass depends weakly on the density. We derived 
criteria to predict the sizes beyond which one may expect to see the signature traveling bands, see Eqs.~\eref{eq:Lstar} and~\eqref{eq:LstartVM}. These 
criteria allowed us to find the system sizes above which bands
could be observed for the AIM with Voronoi alignment (see
Fig.~\ref{fig:aim_voronoi}). For the Vicsek model with Voronoi
alignment, the density-dependence of the polar-field mass is very weak
and we were not able to observe bands. 
Whether this is just a computational limitation of the simulations we and others performed~\cite{peshkov2012continuous,ginelli2010relevance} or the transition remains second order is still an open question. Indeed, while our theory shows that a density-dependent mass is created by fluctuations, this density dependence might vanish under the Renormalisation Group flow, hence re-establishing a second-order transition. 
An analogous scenario is known to happen in the Potts model in low spatial dimensions in which strong fluctuations transform a first-order transition at mean-field level into a second-order one~\cite{wu1982potts,newman1984q,sanchez2023q}.
Finding traveling bands in the Vicsek-Voronoi model thus
remains an open challenge that could benefit from using recently developed simulation methods~\cite{sussman2017cellgpu,pinto2022hierarchical}.

Note that, while we have focused on the framework of flocking models
in this article, our methods to capture the impact of fluctuations can
be extended to a broad range of systems, from bifurcations in
dynamical systems to phase transitions in active gels and Boltzmann
kinetic evolutions. We also emphasize that our formalism is lighter
than the usual path integral approach, as one need not compute Feynman
diagrams nor dynamical actions. This novel approach still leads to the
same result as the path integral method, and the obtained renormalized
hydrodynamics allows for quantitative predictions of noise-induced
phenomena. To illustrate the usefulness of our approach, we highlight
that it has been successfully applied to the case of active
particles experiencing nematic torques~\cite{spera2024nematic}.  In
this case, fluctuations renormalize the mass of the polar field,
leading to an increase of particles' persistence and hence
facilitating the onset of the motility-induced phase separation.

\ack

D.M. acknowledges support from the Kadanoff Center for Theoretical
Physics and from the France-Chicago Center through the grant FACCTS.
C.D. and G.S. acknowledges the support of the LabEx “Who~Am~I?”
(ANR-11-LABX-0071) and of the Université Paris Cité IdEx
(ANR-18-IDEX-0001) funded by the French Government through its
“Investments for the Future” program. CN acknowledges the
support of the ANR grant PSAM and the INP-IRP grant IFAM. FvW and JT acknowledge the
support of ANR grant THEMA. We thank Jose Armengol-Collado who was
involved in early numerical simulations and Daniel Sussman for several
insightful discussions.

\clearpage
\addcontentsline{toc}{section}{Appendices}
\appendix


\section{Fourier series}

Throughout the appendices, we use Fourier series for $L$-periodic functions $f$ with the convention
\begin{equation}
  \label{eq:fourier_serie}
  f^q=\frac{1}{L}\int_0^{L}f(x)e^{-iqx}\ , \qquad f(x)=\sum_{n\in\mathbb{Z}}f^q e^{iqx}\; ,
\end{equation}
where $q=2\pi n/L$.
We note that for an arbitrary stochastic function $f(x)$ verifying $\langle f^qf^{q^{\prime}}\rangle \propto  \langle f^q f^{-q} \rangle L^{-1}\delta_{q+q^{\prime},0}$, $\langle f^2(x) \rangle$ is given in the large-system-size limit $L\rightarrow\infty$ as
\begin{equation}
  \label{eq:limit_fourier_serie}
  \lim_{L\rightarrow\infty}  \langle f^2(x)\rangle =\lim_{L\rightarrow\infty} \sum_{q,q^{\prime}} \langle f^qf^{q^{\prime}}\rangle e^{iqx+iq^{\prime}x} = \lim_{L\rightarrow\infty} \sum_{q}\frac{1}{L} \langle f^qf^{-q} \rangle = \int_{-\infty}^{+\infty}\frac{dq}{2\pi} \langle f^qf^{-q}\rangle\ .
\end{equation}

\section{The Active Ising Model with metric alignment}

\subsection{Linear stability analysis}
\label{app:linear_stab}

In this appendix, we generically analyze the stability of the perturbations $\delta\rho$ and $\delta m$ for the metric AIM.
We need to handle two different linearized dynamics: the one obtained from the mean-field metric AIM \eref{eq:MF_AIM_final} 
and the one obtained from the renormalized metric AIM \eref{eq:MF_AIM_renormalized_generic}.
As the former evolution can be obtained from the later by setting $\alpha(\rho)=cst$, we will only detail the derivation for \eref{eq:MF_AIM_renormalized_generic}.
Extending the results of this appendix to the case of the mean-field metric AIM \eref{eq:MF_AIM_final}  
is easily performed by setting $\alpha^{\prime}=0$ in what follows.
Following the approach discussed in Sec.~\ref{sec:linearStab_meanField}, we perform the analysis in $1$D and adimensionalize \eref{eq:MF_AIM_renormalized_generic} into 
\begin{subequations}\label{eq:active_Ising_renorm_adim_app}%
\begin{align}%
  \partial_{\tilde{t}}\rho &=  \partial_{\tilde{x}\tilde{x}}\rho-\partial_{\tilde{x}} m \\
  \partial_{\tilde{t}} m &= \partial_{\tilde{x}\tilde{x}} m - \partial_{\tilde{x}} \rho - \tilde{\alpha}(\rho)m-\tilde{\gamma} \frac{m^{3}}{\rho^{2}}\;,
\end{align}%
\end{subequations}%
where $\tilde{x}=x v/D$, $\tilde{t}=v^{2}/D$, $\tilde{\alpha}(\rho)=D\alpha(\rho)/v^{2}$ and $\tilde{\gamma}=D\gamma/v^{2}$.
In the following, we drop the tilde to lighten the notation.
The linearized dynamics of the perturbations $\delta m$ and $\delta \rho$ around the homogeneous solutions $m_{0}$ and $\rho_{0}$ reads
\begin{equation}
  \label{eq:dynamic_alpha_rho_app}
  \partial_t
\begin{pmatrix}
\delta {\rho_q} \\
\delta {m_q}
\end{pmatrix} = \begin{pmatrix}
- q^{2} & -\rmi q \\
-\rmi q  - \sqrt{|\alpha_0|/\gamma}\big{(}\alpha^{\prime}_0\rho_0+2\alpha_0\big{)} & -q^{2} + 2\alpha_0
\end{pmatrix}
\begin{pmatrix}
\delta \rho_q \\
\delta m_q
\end{pmatrix}\;,
\end{equation}
where $q$ is the wave vector, $\alpha_0 = \alpha(\rho_0)$, $\alpha^{\prime}_0 = \alpha^{\prime}(\rho_0)$, and we have used the relation $m_0=\rho_0\sqrt{|\alpha_0|/\gamma}$.
The time evolution of $\delta\rho$ and $\delta m$ in Fourier space is governed by the eigenvalues of the stability matrix appearing in \eref{eq:dynamic_alpha_rho_app}. 
These eigenvalues read:
\begin{equation}
\lambda_{\pm}=\dfrac{-\left(2q^{2} - 2\alpha_0 \right)\pm \sqrt{\Delta}}{2}\ ,
\end{equation}
where the discriminant $\Delta $ is
\begin{equation}
\Delta = \left(2q^{2} - 2\alpha_0 \right)^{2}+4 {\rm i} q\left( {\rm i} q + \sqrt{|\alpha_0|/\gamma}\big{(}\alpha^{\prime}_0\rho_0+2\alpha_0 \big{)}\right)-4q^{2}\left(q^{2}-2\alpha_0 \right)\ .
\end{equation}
The stability of the homogeneous solution is determined by the sign of the real part of the eigenvalues $\lambda^\pm$.
An unstable mode exists as soon as $|2q^2-2\alpha_0| < |\Re (\sqrt\Delta)|$. This is equivalent to requiring the expression
\begin{subequations}
\begin{equation}\label{eq:stability_alpha_rho_sign}%
2\Re(\sqrt{\Delta})^2-2\Re\left(2q^{2} - 2\alpha_0 \right)^2 = -a(q) + \sqrt{a^2(q)+b(q)}
\end{equation}%
to be positive. In Eq.~\eref{eq:stability_alpha_rho_sign}, we have defined the real numbers $a(q)$ and $b(q)$ as
\begin{align}%
  a(q) =& 4\left(q^2-\alpha_0 \right)^2+4 \left(q^4-2 \alpha_0  q^2+q^2\right) \\
  b(q) =& 16 q^2 \left(\frac{|\alpha_0| (2 \alpha_0 +\alpha'_0  \rho_0)^2}{\gamma }+4 (2 \alpha_0 -1) \alpha_0 ^2\right)\notag\\
  &\label{eq:bq_app}-64 \alpha_0  (5 \alpha_0 -2) q^4+64 (4 \alpha_0 -1) q^6-64 q^8  \ .
\end{align}%
\end{subequations}%
Because $a(q)$ is always positive, the instability condition that Eq.~\eref{eq:stability_alpha_rho_sign} is positive amounts to $b(q)>0$, which is the condition given in main text.
Note that the density instability occurs in the ordered phase, where we have $\alpha_0<0$.
This implies that only the term of order $q^2$ in $b(q)$ can change sign and become positive to trigger the instability.
Finally, since we are interested in the behavior of the system at the onset of order where $|\alpha_0|\ll 1$, the necessary condition for an instability at lowest order in  $|\alpha_0|$ reads
\begin{equation}
\label{eq:stab_cond_app}
  \rho_0^2 (\alpha'_0)^2 >0 \, ,
\end{equation}
which is Eq.~\eref{eq:stab_cond} of the main text.

\subsection{Fluctuation-induced renormalization of the mass}
\label{app:AIM_FIFOT}

In this Appendix we detail the derivation of the renormalized mass in the metric AIM. Our starting point is Eq.~\eqref{eq:AIM_noise_1d}, that we rewrite here for convenience:
\begin{subequations}\label{eq:AIM_noise_app}%
\begin{align}%
  \partial_t \rho &= D \partial_{xx}\rho -v\partial_x m \, , \\
  \partial_t m &= D \partial_{xx} m - v \partial_x \rho -\mathcal{F}(\rho,m) + \sqrt{2 \sigma \rho} \, \eta
  \;.
\end{align}%
\end{subequations}%
The role of fluctuations is now discussed by constructing the dynamics of the average fields $\tilde{\rho}(x,t)=\langle \rho(x,t)\rangle$ and $\tilde{m}(x,t)=\langle m(x,t) \rangle$ to leading order in the noise strength $\sigma$.
Calling $\rho_0(x,t)$ and $m_0(x,t)$ the solution of Eq.~\eref{eq:AIM_noise_app} in the absence of noise (\textit{i.e.} when $\sigma=0$), we introduce their deviations $\Delta\rho$ and $\Delta m$ from this mean-field solution as series in $\sigma^{1/2}$
\begin{subequations}\label{eq:expans_Delta_app}%
\begin{align}%
  \Delta \rho &= \rho - \rho_0 = \sigma^{\frac{1}{2}}\delta\rho_1 + \sigma\delta\rho_2 + \cdots \; ,\\
  \Delta m &= m - m_0 = \sigma^{\frac{1}{2}}\delta m_1 + \sigma\delta m_2 + \cdots \, .
\end{align}%
\end{subequations}%
Note that the $\delta \rho_k$ and $\delta m_k$ are stochastic fields while $\rho_0$ and $m_0$ are deterministic.

Inserting the expansions \eref{eq:expans_Delta_app} in Eq.~\eref{eq:AIM_noise_app} and equating terms of order $\sigma^{k/2}$ yields the evolution equation for $\delta\rho_k$ and $\delta m_k$.
For $k=1$, it yields:
\begin{subequations}\label{eq:delta_fields_1}
\begin{align}%
  \label{eq:delta_rho_1}
  \partial_t \delta\rho_1 &= D \partial_{xx}\delta\rho_1 -v\partial_x \delta m_1 \\
  \label{eq:delta_m_1}
  \partial_t \delta m_1 &= D \partial_{xx} \delta m_1 - v \partial_x \delta\rho_1 -\frac{\partial\cF}{\partial\rho}\delta \rho_1 -\frac{\partial\cF}{\partial m}\delta m_1 + \sqrt{2 \rho_0} \, \eta
  \; ,
\end{align}%
\end{subequations}%
while for $k=2$ we obtain:
\begin{subequations}\label{eq:delta_fields_2}
\begin{align}%
  \label{eq:delta_rho_2}
  \partial_t \delta\rho_2 =& D \partial_{xx}\delta\rho_2 -v\partial_x \delta m_2 \\
  \nonumber
  \partial_t \delta m_2 =& D \partial_{xx} \delta m_2 - v \partial_x \delta\rho_2 -\frac{\partial\cF}{\partial\rho}\delta \rho_2 -\frac{\partial\cF}{\partial m}\delta m_2 - \frac{\partial^{2} \mathcal{F}}{\partial m^{2}} \frac{\delta m_1^{2}}{2}
  -  \frac{\partial^{2} \mathcal{F}}{\partial \rho^{2}} \frac{\delta \rho_1^{2}}{2} - \frac{\partial^{2} \mathcal{F}}{\partial m \partial \rho} \delta m_1\delta \rho_1 \\
  \label{eq:delta_m_2}
  &+ \frac{\delta\rho_1}{\sqrt{2\rho_0}} \, \eta
  \; .
\end{align}%
\end{subequations}%
Note that in Eqs.~\eqref{eq:delta_fields_1}, we will consider that both $\rho_0(x,t)$ and $m_0(x,t)$ are constant in time and space.
Indeed, we assume that $\delta \rho_1$ and $\delta m_1$ are fast modes varying on lengthscales and timescales much smaller than the ones relevant for $\rho_0$ and $m_0$.
This adiabatic approximation allows us to compute the correlators in terms of $\rho_0$ and $m_0$ as parameters and to re-establish their dependency on $x$ and $t$ a posteriori.

Averaging Eqs.~\eref{eq:delta_fields_1} over the noise with It\=o prescription then gives
\begin{subequations}%
\begin{align}%
  \label{eq:delta_rho_1_av}
  \partial_t \langle \delta\rho_1 \rangle &= D \partial_{xx}\delta\langle\rho_1\rangle -v\partial_x \langle\delta m_1 \rangle\\
  \label{eq:delta_m_1_av}
  \partial_t \delta \langle m_1 \rangle &= D \partial_{xx} \delta \langle m_1 \rangle - v \partial_x \langle \delta\rho_1 \rangle -\frac{\partial\cF}{\partial\rho}\langle\delta \rho_1\rangle -\frac{\partial\cF}{\partial m}\langle\delta  m_1\rangle
  \; ,
\end{align}%
\end{subequations}%
while averaging Eqs.~\eref{eq:delta_fields_2} yields
\begin{subequations}%
\begin{align}%
  \label{eq:delta_rho_2_av}
  \partial_t \langle\delta\rho_2\rangle =& D \partial_{xx}\langle\delta\rho_2\rangle -v\partial_x \langle\delta m_2\rangle \\
  \nonumber
  \partial_t \langle\delta m_2\rangle =& D \partial_{xx} \langle\delta m_2\rangle - v \partial_x \langle\delta\rho_2\rangle -\frac{\partial\cF}{\partial\rho}\langle\delta \rho_2\rangle -\frac{\partial\cF}{\partial m}\langle\delta m_2\rangle - \frac{\partial^{2} \mathcal{F}}{\partial m^{2}} \frac{\langle\delta m_1^{2}\rangle}{2}
  -  \frac{\partial^{2} \mathcal{F}}{\partial \rho^{2}} \frac{\langle\delta \rho_1^{2}\rangle}{2} \\
  \label{eq:delta_m_2_av}
  & - \frac{\partial^{2} \mathcal{F}}{\partial m \partial \rho} \langle\delta m_1\delta \rho_1\rangle \ .
\end{align}%
\end{subequations}%
Summing together Eq.~\eref{eq:delta_rho_1_av} multiplied by $\sigma^{\frac{1}{2}}$ and  Eq.~\eref{eq:delta_rho_2_av} multiplied by $\sigma$ gives the evolution of $\tilde{\rho}$ up to order $\sigma$ as
\begin{align}%
  \label{eq:bar_rho_app}
  \partial_t \tilde{\rho} =& D \partial_{xx}\tilde{\rho} -v\partial_x \tilde{m}\;,
\end{align}
while summing together  Eq.~\eref{eq:delta_m_1_av} multiplied by $\sigma^{\frac{1}{2}}$ and  Eq.~\eref{eq:delta_m_2_av} multiplied by $\sigma$ yields the evolution of $\tilde{m}$ up to order $\sigma$
\begin{align}
  \nonumber
  \partial_t \tilde{m} =& D \partial_{xx} \tilde{m} - v \partial_x \tilde{\rho} -\cF(\rho_0,m_0)-\sigma^{\frac{1}{2}}\frac{\partial\cF}{\partial\rho}\langle\delta\rho_1\rangle-
  \sigma^{\frac{1}{2}}\frac{\partial\cF}{\partial m}\langle \delta m_1\rangle-
  \sigma\frac{\partial\cF}{\partial\rho}\langle\delta\rho_2\rangle \\
  &-\sigma\frac{\partial\cF}{\partial m}\langle \delta m_2\rangle -\sigma \frac{\partial^{2} \mathcal{F}}{\partial m^{2}} \frac{\langle\delta m_1^{2}\rangle}{2}
  \label{eq:bar_m_interm}
  -  \sigma\frac{\partial^{2} \mathcal{F}}{\partial \rho^{2}} \frac{\langle\delta \rho_1^{2}\rangle}{2} - \sigma \frac{\partial^{2} \mathcal{F}}{\partial m \partial \rho} \langle\delta m_1\delta \rho_1\rangle \ ,
\end{align}%
where the derivatives of $\cF$ are evaluated at $\rho_0, m_0$. To have an evolution equation for $\tilde{m}$ that depends only on $\tilde{m}$ and $\tilde{\rho}$, we finally use the fact that, to order $\sigma$, the Landau term $\cF(\tilde{\rho},\tilde{m})$ taken at the averaged density and magnetization can be written as:
\begin{align}
\begin{split}
  \cF(\tilde{\rho},\tilde{m}) =& \cF(\rho_0,m_0) + \sigma^{\frac{1}{2}}\frac{\partial \cF}{\partial \rho}\langle\delta\rho_1\rangle + \sigma^{\frac{1}{2}}\frac{\partial \cF}{\partial m}
  \langle\delta m_1\rangle + \sigma \frac{\partial \cF}{\partial \rho}\langle\delta\rho_2\rangle + \sigma \frac{\partial \cF}{\partial m}\langle\delta m_2\rangle \\
  \label{eq:dev_cF_AIM}
  &+ \frac{\sigma}{2}\frac{\partial^2 \cF}{\partial^2 \rho}\langle\delta \rho_1\rangle^2
  +\frac{\sigma}{2}\frac{\partial^2 \cF}{\partial^2 m}\langle\delta m_1\rangle^2 + \sigma\frac{\partial^2 \cF}{\partial \rho\partial m}\langle\delta \rho_1\rangle\langle\delta m_1\rangle + \cO(\sigma^{\frac{3}{2}})\;.
\end{split}
\end{align}
Using expression \eref{eq:dev_cF_AIM} into \eref{eq:bar_m_interm} simplifies the time-evolution of $\tilde{m}$ as
\begin{align}
  \nonumber
  \partial_t \tilde{m} =& D \partial_{xx} \tilde{m} - v \partial_x \tilde{\rho} -\cF(\tilde{\rho},\tilde{m})-\sigma \frac{\partial^{2} \mathcal{F}}{\partial m^{2}} \left(\frac{\langle\delta m_1^{2}\rangle -\langle\delta m_1\rangle^{2}}{2}\right) \\
  \label{eq:bar_m_app}
  &-  \sigma\frac{\partial^{2} \mathcal{F}}{\partial \rho^{2}} \left(\frac{\langle\delta \rho_1^{2}\rangle-\langle\delta \rho_1\rangle^{2}}{2}\right) - \sigma \frac{\partial^{2} \mathcal{F}}{\partial m \partial \rho} \left(\langle\delta m_1\delta \rho_1\rangle-\langle\delta m_1\rangle \langle\delta \rho_1\rangle\right) \ .
\end{align}
The renormalized hydrodynamic equations~\eref{eq:bar_rho_app} and~\eref{eq:bar_m_app} correspond to Eqs.~\eref{eq:bar_rho} and~\eref{eq:bar_m} in the main text.
We detail the computation of the correlators $\langle \delta \rho_1^2\rangle$, $\langle \delta m_1^2\rangle$ and $\langle \delta m_1\delta\rho_1\rangle$ in the next Appendix.

\subsection{Derivation of the correlators in the high temperature phase}
\label{app:correlators_AIM}

This Appendix is devoted to the computation of the correlators $\langle \delta \rho_1^2\rangle$, $\langle \delta m_1^2\rangle$ and $\langle \delta m_1\delta\rho_1\rangle$ which are needed to close the renormalized evolution of the metric AIM, \textit{i.e.} Eqs.~\eref{eq:bar_rho} and~\eref{eq:bar_m} of the main text. This computation relies on the assumption that the profiles $\rho_0$ and $m_0$ vary over large enough scales that they can be approximated as homogeneous in space. As in the main text, we denote the corresponding averages by $\langle \dots \rangle_0$.

First, we cast the It\=o-stochastic equations~\eref{eq:delta_fields_1} for $\delta\rho_1$ and $\delta m_1$ into Fourier space with the Fourier convention defined in \eref{eq:fourier_serie}.
We obtain:
\begin{align}
\label{eq:linear_perturbation_AIM_app}
\partial_{t}\begin{pmatrix}
\delta \rho_1^{q} \\
\delta m_1^{q}
\end{pmatrix} = \begin{pmatrix}
M_{11}^q & M_{12}^q\\
M_{21}^q & M_{22}^q
\end{pmatrix}
\begin{pmatrix}
\delta \rho_1^{q} \\
\delta m_1^{q}
\end{pmatrix} + \begin{pmatrix}
0 \\
\sqrt{2\rho_{0}}\, \eta^{q}
\end{pmatrix}\;,
\end{align}
where $\eta^q$ is the $q$-th Fourier mode of the Gaussian white noise with correlations $\langle\eta^q(t)\eta^{q^{\prime}}(t^{\prime})\rangle=L^{-1}\delta_{q+q^{\prime},0}\delta(t-t^{\prime})$ and the matrix coefficients $M_{11}^{q}$, $M_{12}^{q}$, $M_{21}^{q}$, $M_{22}^{q}$ are given by
\begin{align}
  \label{eq:matrix_coeff_AIM}
  M_{11}^{q}=&-Dq^2\;, & M_{12}^{q}=& -ivq\;, & M_{21}^{q}=&-iqv +2\gamma\frac{m_{0}^{3}}{\rho_{0}^{3}}\;, & M_{22}^q =& -D q^{2} - \alpha - 3\gamma\frac{m_{0}^{2}}{\rho_{0}^{2}}\ .
\end{align}
To compute the equal-time two-point correlation functions at steady state, we first use It\=o calculus on the stochastic system \eref{eq:linear_perturbation_AIM_app} to get the following closed system of equations
\begin{subequations}\label{eq:Ito_correlator}%
\begin{align}%
  \label{eq:Ito_correlator_AIM_1_no_SS}
\frac{\dd }{\dd t}\langle \delta\rho_1^q \delta \rho_1^{q'} \rangle_0 =&(M_{11}^{q}+M^{q^{\prime}}_{11})\langle\delta\rho_1^{q}\delta\rho_1^{q^{\prime}}\rangle_0 + M_{12}^{q}\langle\delta m_1^{q}\delta\rho_1^{q^{\prime}}\rangle_0 + M_{12}^{q^{\prime}}\langle\delta \rho_1^{q}\delta m_1^{q^{\prime}}\rangle_0 \, , \\
\label{eq:Ito_correlator_AIM_2_no_SS}
\frac{\dd }{\dd t}\langle \delta m_1^q \delta \rho_1^{q'} \rangle_0 =&(M_{22}^{q}+M^{q^{\prime}}_{11})\langle\delta m_1^{q}\delta\rho_1^{q^{\prime}}\rangle_0 + M_{21}^{q}\langle\delta \rho_1^{q}\delta\rho_1^{q^{\prime}}\rangle_0 + M_{12}^{q^{\prime}}\langle\delta m_1^{q}\delta m_1^{q^{\prime}}\rangle_0 \, , \\
\label{eq:Ito_correlator_AIM_3_no_SS}
\frac{\dd }{\dd t}\langle \delta m_1^q \delta m_1^{q^{\prime}} \rangle_0 =&(M_{22}^{q}+M^{q^{\prime}}_{22})\langle\delta m_1^{q}\delta m_1^{q^{\prime}}\rangle_0 + M_{21}^{q}\langle\delta \rho_1^{q}\delta m_1^{q^{\prime}}\rangle_0 + M_{21}^{q^{\prime}}\langle\delta m_1^{q}\delta \rho_1^{q^{\prime}}\rangle_0 + \frac{2\rho_{0}}{L}\delta_{q+q^{\prime},0} \, .
\end{align}%
\end{subequations}%
Using the assumptions that the correlation functions are fast variables, we evaluate Eqs.~\eqref{eq:Ito_correlator} at steady state to obtain:
\begin{subequations}\label{eq:fourier_correlators_metric_AIM}
\begin{align}%
\begin{split}
\langle\delta m_1^{q}\delta m_1^{q^{\prime}}\rangle_0 =& \rho_0\bigg{[}-\frac{ 3\gamma \left(2 D^2 q^2+\alpha  D+v D q+v^2\right) \left(2 D^2 q^2+D (\alpha -q v)+v^2\right)}{ \left(\alpha +2 D q^2\right)^2 \left(D^2 q^2+\alpha  D+v^2\right)^2}\frac{m_{0}^{2}}{\rho_0^2}  \\
\label{eq:fourier_m_m_correlation_AIM}
&+\frac{2 D^2 q^2+\alpha  D+v^2}{\left( 2 D q^2+\alpha \right) \left(D^2 q^2+\alpha  D+v^2\right)}\bigg{]}\frac{\delta_{q+q^{\prime},0}}{L}+\cO\left(m_0^3\right) \, ,
\end{split} \\
\label{eq:fourier_rho_rho_correlation_AIM}
\langle\delta \rho_1^{q}\delta \rho_1^{q^{\prime}}\rangle_0 =& \frac{\rho_0 v^2}{\left(2 D q^2+\alpha \right) \left(D^2 q^2+\alpha  D+v^2\right)}\frac{\delta_{q+q^{\prime},0}}{L}+\cO\left(m_0^2\right) \, ,\\
\label{eq:fourier_m_rho_correlation_AIM}
\langle\delta m_1^{q}\delta \rho_1^{q^{\prime}}\rangle_0 =& \frac{i \rho_0 v  D q}{\left(2 D q^2+\alpha \right) \left(D^2 q^2+\alpha  D+v^2\right)}\frac{\delta_{q+q^{\prime},0}}{L}+\cO\left(m_0^2\right) \, .
\end{align}%
\end{subequations}%
Note that in \eref{eq:fourier_correlators_metric_AIM} we discarded higher order terms in $m_0$ since we are interested at the onset of collective motion where $m_0 \ll 1$.
The real-space correlators are then obtained by inverse Fourier transform using Eq.~\eref{eq:limit_fourier_serie} and performing the integral over $q$ analytically. We obtain:
\begin{subequations}%
\label{eq:corr_delta2}
\begin{align}%
\langle\delta m_1^2 \rangle_0 =& \rho_0 \frac{ v^2 \sqrt{\frac{2\alpha }{D}}+\alpha  \sqrt{v^2+\alpha  D}}{4 \alpha  v^2+2 \alpha ^2 D} - \rho_0\frac{3 \gamma  D \left(\frac{\alpha  D}{\sqrt{v^2+\alpha  D}}
+\frac{\sqrt{2} v^2 \left(2 v^2+3 \alpha  D\right)}{(\alpha  D)^{3/2}}\right)}{4 \left(\alpha  D+2 v^2\right)^2}\frac{m_0^2}{\rho_0^2} + \cO(m_0^3)\label{eqdeltam2} \\
\langle\delta \rho_1^2 \rangle_0 =&\rho_0 \frac{v^2 \left( \sqrt{\frac{2\alpha }{D}}-\frac{\alpha }{\sqrt{v^2+\alpha  D}}\right)}{2 \alpha  \left(\alpha  D+2 v^2\right)} + \cO(m_0^2)\,,\qquad \langle\delta \rho_1\delta m_1 \rangle_0 = 0 + \cO(m_0^2) \, ,
\label{eqdeltam3}
\end{align}%
\end{subequations}%
which correspond to Eqs.~\eref{eq:corr_deltam2}-\eref{eq:corr_deltarho2} of the main text.

  \subsection{Ordered phase}
  \label{app:AIM_FIFOT_ordered}

In this Appendix, we discuss the fluctuation-induced renormalization in the ordered (low temperature) phase for the metric AIM \eref{eq:AIM_noise_1d}.
In the low temperature phase ($\alpha < 0$), the fluctuations $\delta m_1$ and $\delta \rho_1$ have no mass and the computation cannot be straightforwardly extended from the high temperature case detailed in appendices \ref{app:AIM_FIFOT} and \ref{app:correlators_AIM}.
To circumvent this issue, we change variable for the magnetization and define the field $w$:
\begin{equation}
  \label{eq:def_ml}
  w = m - \sqrt{\frac{|\alpha|}{\gamma}}\rho
\end{equation}
In term of this new field, the time-evolution~\eref{eq:AIM_noise_1d} is rewritten as:
\begin{subequations}%
\begin{align}%
  \partial_t\rho &= D \partial_{xx} \rho - v \partial_{x}  w -v\sqrt{\frac{|\alpha|}{\gamma}}\partial_x\rho \label{eq:MFrho_LT}\\
  \label{eq:MFml_LT}
  \partial_t  w &= D \partial_{xx} w + v\sqrt{\frac{|\alpha|}{\gamma}}\partial_x w -
  v\left(1-\frac{|\alpha|}{\gamma}\right)\partial_{x} \rho - 2|\alpha| w
  - 3\sqrt{\gamma|\alpha|}\frac{w^2}{\rho} - \gamma w^3/\rho^2 + \sqrt{2 \sigma \rho} \, \eta \, ,
\end{align}%
\end{subequations}%
where a mass term $-2|\alpha|w$ is indeed present in Eq.~\eref{eq:MFml_LT}.
Introducing $v_1=v\sqrt{|\alpha|/\gamma}$, $v_2 = v(1-|\alpha|/\gamma)$, $a_1=2|\alpha|$, $a_2=3\sqrt{\gamma |\alpha|}$ and $\cF_{\ell}(w,\rho)=a_1 w + a_2 w^2/\rho + \gamma w^3/\rho^2$, we cast the above evolution into
\begin{subequations}%
\begin{align}%
  \partial_t\rho &= D \partial_{xx} \rho - v \partial_{x}  w -v_1\partial_x\rho \label{eq:FD_rho_LT}\\
  \label{eq:FD_ml_LT}
  \partial_t  w &= D \partial_{xx} w + v_1\partial_x w -
  v_2\partial_{x} \rho - \cF_\ell(w,\rho) + \sqrt{2\sigma\rho}\;\eta \ ,
\end{align}%
\end{subequations}%
Using the expansions $\rho = \rho_0 + \sqrt{\sigma}\delta\rho_1 + .. $ and $w = w_0 + \sqrt{\sigma}\delta w_1 + .. $, we can now repeat the computations detailed in appendix \ref{app:AIM_FIFOT} for the high temperature phase.
By doing so, we obtain the renormalized hydrodynamic for $\tilde{\rho}=\langle\rho\rangle$ and $\tilde{w}=\langle w\rangle$ to order $\sigma$ as
\begin{subequations}%
\begin{align}%
  \label{eq:tilde_rho_LT}
  \partial_t \tilde{\rho} =& D \partial_{xx}\tilde{\rho} -v\partial_x \tilde{w} - v_1\partial_x\tilde{\rho} \\
  \nonumber
  \partial_t \tilde{w} =& D \partial_{xx} \tilde{w} + v_1\partial_x \tilde{w} - v_2 \partial_x \tilde{\rho} -\cF_l(\tilde{\rho},\tilde{w})-\sigma \frac{\partial^{2} \mathcal{F}_\ell}{\partial w^{2}} \left(\frac{\langle\delta w_1^{2}\rangle -\langle\delta w_1\rangle^{2}}{2}\right) \\
  \label{eq:tilde_w_LT}
  &-  \sigma\frac{\partial^{2} \mathcal{F}_\ell}{\partial \rho^{2}} \left(\frac{\langle\delta \rho_1^{2}\rangle-\langle\delta \rho_1\rangle^{2}}{2}\right) - \sigma \frac{\partial^{2} \mathcal{F}_\ell}{\partial w \partial \rho} \left(\langle\delta w_1\delta \rho_1\rangle-\langle\delta w_1\rangle \langle\delta \rho_1\rangle\right) \ ,
\end{align}%
\end{subequations}%
where the derivatives of $\cF_{\ell}$ are evaluated at $\tilde{\rho},\tilde{w}$.
As detailed in \ref{app:correlators_AIM} for the high temperature case, we now need to compute the correlators involving $\delta w_1$ and $\delta \rho_1$ by using their linearized stochastic evolution.
We obtain (see App.~\ref{app:ren_LT} for details):
\begin{subequations}\label{eq:corr_final_LT}
\begin{align}%
  \label{eq:corr_w_w_final_LT}
\langle\delta w_1^2 \rangle_0 =& \frac{\rho_0 }{\sqrt{D a_1}}f_w\left(\frac{v v_2}{Da_1},\frac{v_1^2}{D a_1}\right) + \cO(w_0)\;, \\
  \label{eq:corr_rho_rho_final_LT}
\langle\delta \rho_1^2 \rangle_0 =& \frac{\rho_0}{\sqrt{Da_1}} f_{\rho}\left(\frac{v^2}{Da_1},\frac{v v_2}{D a_1},\frac{v_1^2}{D a_1}\right) + \cO(w_0)\;,\\
\langle\delta \rho_1\delta w_1 \rangle_0 =& 0 + \cO(w_0)\; ,
\end{align}%
\end{subequations}%
where $f_w(s,u)$ and $f_{\rho}(s,u,z)$ are defined as:
\begin{subequations}%
\begin{align}%
  f_w(s,u)=& \int_{-\infty}^{+\infty} \frac{d\tilde{q}}{2\pi}\frac{ s\left(1 + 2\tilde{q}^2\right) + (1+2\tilde{q}^2)^2+4u\tilde{q}^2}{s\left(1+2  \tilde{q}^2\right)^2
  +\left(1+\tilde{q}^2 \right) \left(\left(1+2 \tilde{q}^2 \right)^2+4 u \tilde{q}^2 \right)} \\
  f_\rho(s,u,z)=& \int_{-\infty}^{+\infty} \frac{d\tilde{q}}{2\pi}\frac{s\left(1 + 2\tilde{q}^2\right)}{u \left(1+2  \tilde{q}^2\right)^2+\left(1+\tilde{q}^2 \right) \left(\left(1+2 \tilde{q}^2 \right)^2+4 \tilde{q}^2 z\right)}\; .
\end{align}%
\end{subequations}%
Inserting \eref{eq:corr_w_w_final_LT}-\eref{eq:corr_rho_rho_final_LT} into \eref{eq:tilde_w_LT}, we obtain:
\begin{align}
  \partial_t \tilde{w} =& D \partial_{xx} \tilde{w} + v_1\partial_x \tilde{w} - v_2 \partial_x \tilde{\rho} -a_1\tilde{w} - \frac{a_2}{ \sqrt{D a_1}}\sigma f_w
  + \cO(\sigma^{\frac{3}{2}}) + \cO(\sigma\tilde{w})  \, .\label{eq:final_w_ren_LT}
\end{align}
Note that $\tilde{w}=-\sigma a_2 vf_w/(v_2 a_1^2)$ and $\tilde{\rho}=\rho_0$ are the homogeneous solutions of the above renormalized evolution Eqs~\eref{eq:final_w_ren_LT} and~\eref{eq:tilde_rho_LT}.
Such a solution, which is only valid up to order $\sigma$, is consistent with the terms discarded in \eref{eq:final_w_ren_LT} which are all of order $\sigma^2$ at least. Note also that, due to mass conservation, fluctuations cannot generate any correction to a homogeneous solution for $\tilde{\rho}$.

Finally, considering the definition of $w$ given in Eq.~\eref{eq:def_ml} and inserting the expression of $v_1$, $v_2$, $a_2$ and $a_1$ in terms of $|\alpha|$ and $\gamma$, we find that fluctuations lower the homogeneous magnetization in the ordered phase according to
\begin{align}
  \label{eq:prediction_m_low_temp_app}
  \tilde{m} = \sqrt{\frac{|\alpha|}{\gamma}}\rho_0 - \sigma \frac{3\sqrt{\gamma}}{2\sqrt{2D}|\alpha|}f_w\left(\frac{v^2(\gamma-|\alpha|)}{2D|\alpha|\gamma},\frac{v^2}{2D\gamma}\right) + \cO(\sigma^{\frac{3}{2}}) \; ,
\end{align}
which corresponds to Eq.~\eqref{eq:prediction_m_low_temp} in the main text. This lowering of the magnetization in the ordered phase can be absorbed in a renormalization of $|\alpha|/\gamma$ according to $\tilde{m}=\sqrt{|\hat{\alpha}|/\hat{\gamma}}\;\rho_0$, with
\begin{align}
  \frac{|\hat{\alpha}|}{\hat{\gamma}}=\frac{|\alpha|}{\gamma}-\frac{\sigma}{\rho_0}\frac{3}{ \sqrt{2D|\alpha|}}f_w\left(\frac{v^2(\gamma-|\alpha|)}{2D|\alpha|\gamma},\frac{v^2}{2D\gamma}\right)+ \cO(\sigma^{\frac{3}{2}})\; .
\end{align}
Note that, consistently with our result \eref{eq:RenAlpha} obtained in the high temperature phase, the fluctuation-induced correction of $\alpha/\gamma$ also scales as $\rho_0^{-1}$ in the low temperature phase.

\subsection{Renormalization in the low temperature phase}
\label{app:ren_LT}

In this Appendix, we detail the computation of the correlators $\langle \delta\rho_1^2\rangle$, $\langle \delta w_1^2\rangle$ and $\langle \delta \rho_1 \delta w_1 \rangle$ involved in \eref{eq:tilde_rho_LT}-\eref{eq:tilde_w_LT}.
Inserting the expansion $\rho = \rho_0 + \sqrt{\sigma} \delta\rho_1 + .. $, $w = w_0 + \sqrt{\sigma} \delta w_1 + .. $ into \eref{eq:FD_rho_LT}-\eref{eq:FD_ml_LT} and identifying terms of order $\sigma^{1/2}$ yields the time-evolution of the stochastic fields $\delta\rho_1$ and $\delta w_1$ as
\begin{subequations}\label{eq:delta_fields_LT}
\begin{align}
  \label{eq:delta_rho_1_LT}
  \partial_t \delta\rho_1 &= D \partial_{xx}\delta\rho_1 -v\partial_x \delta w_1 - v_1 \partial_x \delta \rho \, , \\
  \label{eq:delta_w_1_LT}
  \partial_t \delta w_1 &= D \partial_{xx} \delta w_1 + v_1 \partial_x \delta w_1- v_2 \partial_x \delta\rho_1 -\frac{\partial\cF_\ell}{\partial\rho}\delta \rho_1 -\frac{\partial\cF_\ell}{\partial w}\delta w_1 + \sqrt{2 \rho_0} \, \eta
  \; .
\end{align}%
\end{subequations}%
where $\eta$ is a Gaussian white noise such that $\langle \eta(x,t)\eta(x',t')=\delta(x-x')\delta(t-t')\rangle$.
In Fourier space, using the Fourier convention~\eref{eq:fourier_serie}, the above system \eref{eq:delta_fields_LT} reads
\begin{subequations}%
\begin{align}%
\label{eq:linear_perturbation_AIM_LT}
\partial_{t}\begin{pmatrix}
\delta \rho_1^{q} \\
\delta w_1^{q}
\end{pmatrix} = \begin{pmatrix}
B_{11}^q & B_{12}^q\\
B_{21}^q & B_{22}^q
\end{pmatrix}
\begin{pmatrix}
\delta \rho_1^{q} \\
\delta w_1^{q}
\end{pmatrix} + \begin{pmatrix}
0 \\
\sqrt{2\rho_{0}}\, \eta^{q}
\end{pmatrix}\;,
\end{align}
where $\eta^q$ is the $q$-th Fourier mode of the Gaussian white noise with correlations $\langle\eta^q(t)\eta^{q^{\prime}}(t^{\prime})\rangle=L^{-1}\delta_{q+q^{\prime},0}\delta(t-t^{\prime})$ and the matrix coefficients $B_{11}^{q}$, $B_{12}^{q}$, $B_{21}^{q}$, $B_{22}^{q}$ are given by
\begin{align}
  B_{11}^{q}=&-Dq^2-iv_1 q\;, & B_{12}^{q}=& -ivq\;, \\ B_{21}^{q}=&-iqv_2 + a_2 \frac{w_0^2}{\rho_0^2} +2\gamma\frac{w_{0}^{3}}{\rho_{0}^{3}}\;, &
  B_{22}^q =& -D q^{2} + v_1 iq - a_1 - 2 a_2\frac{w_0}{\rho_0}- 3\gamma\frac{w_{0}^{2}}{\rho_{0}^{2}}\ .
\end{align}%
\end{subequations}%
Following the same steps as for the high-temperature case, we use It\=o calculus on Eq.~\eref{eq:linear_perturbation_AIM_LT} to obtain the dynamics of the correlators. Evaluating these dynamics at steady state, we obtain a closed system of equations for the correlators, whose solution reads:
\begin{subequations}%
\begin{align}%
\begin{split}
  \langle\delta w_1^{q}\delta w_1^{q^{\prime}}\rangle_0 =& \langle\delta w_1^{q}\delta w_1^{-q}\rangle_0 \frac{\delta_{q+q^{\prime},0}}{L} + \cO\left(w_0\right)\;, \quad \langle\delta \rho_1^{q}\delta \rho_1^{q^{\prime}}\rangle_0 = \langle\delta \rho_1^{q}\delta \rho_1^{-q}\rangle_0 \frac{\delta_{q+q^{\prime},0}}{L} + \cO\left(w_0\right)\\
  \langle\delta w_1^{q}\delta \rho_1^{q^{\prime}}\rangle_0 =& \langle\delta w_1^{q}\delta \rho_1^{-q}\rangle_0 \frac{\delta_{q+q^{\prime},0}}{L} + \cO\left(w_0\right)\; ,
\end{split}
\end{align}
where $\langle\delta w_1^{q}\delta w_1^{-q}\rangle_0$, $\langle\delta \rho_1^{q}\delta \rho_1^{-q}\rangle_0$ and $\langle\delta w_1^{q}\delta \rho_1^{-q}\rangle_0$ are given by
\begin{align}
\label{eq:fourier_w_w_correlation_AIM_LT}
\langle\delta \rho_1^{q}\delta \rho_1^{-q}\rangle_0 =& \frac{\rho_0 v^2 \left(a_1+2 D q^2\right)}{v v_2 \left(a_1+2 D q^2\right)^2+D \left(a_1+D q^2\right) \left(\left(a_1+2 D q^2\right)^2+4 q^2 v_1^2\right)} \\
\label{eq:fourier_rho_rho_correlation_AIM_LT}
\langle\delta w_1^{q}\delta w_1^{-q}\rangle_0 =& \frac{\rho_0 \left(v v_2 \left(a_1+2 D q^2\right)+D \left(\left(a_1+2 D q^2\right)^2+4 q^2 v_1^2\right)\right)}{v v_2 \left(a_1+2 D q^2\right)^2+D \left(a_1+D q^2\right) \left(\left(a_1+2 D q^2\right)^2+4 q^2 v_1^2\right)}\\
\label{eq:fourier_w_rho_correlation_AIM_LT}
\langle\delta w_1^{q}\delta \rho_1^{-q}\rangle_0 =& \frac{i D q \rho_0 v \left(a_1+2 q (D q+i v_1)\right)}{v v_2 \left(a_1+2 D q^2\right)^2+D \left(a_1+D q^2\right) \left(\left(a_1+2 D q^2\right)^2+4 q^2 v_1^2\right)}\;.
\end{align}%
\end{subequations}%
Because $v$, $v_1$, $v_2$, $a_1$, $a_2$ and $\gamma$ are positive, the denominators in the above expressions remains positive as well.
This allows us to integrate over $q$, and, using Eq.~\eref{eq:limit_fourier_serie}, we obtain the real-space correlators given in Eqs.~\eref{eq:corr_final_LT}.


\subsection{Renormalization with conserved noise on the density field}
\label{app:renorm_conserved_noise}

This appendix is devoted to the study of a generalization of Eq.~\eref{eq:fluctuating_mf_AIM} where we consider an additional conservative noise acting on the evolution of the density
\begin{subequations}\label{eq:hydro_AIM_conserved_noise}
\begin{align}
  \label{eq:hydro_AIM_conserved_noise_1}
  \partial_t \rho &= D \partial_{xx}\rho -v\partial_x m +\partial_x(\sqrt{\sigma_\rho\rho}\; \eta_\rho) \, ,\\
  \label{eq:hydro_AIM_cocnserved_noise_2}
  \partial_t m &= D \partial_{xx} m - v \partial_x \rho -\mathcal{F}(\rho,m) + \sqrt{2 \sigma_m \rho} \, \eta_m
  \;.
\end{align}
\end{subequations}
In particular, we show here that the later noise does not affect the FIFOT scenario described in \ref{subsec:renorm_AIM}: the renormalized linear mass of \eref{eq:hydro_AIM_conserved_noise} is also density-dependent.
To generalize the method detailed in Sec.~\ref{app:AIM_FIFOT}, we expand the deviations $\Delta \rho$ and $\Delta m$ from the mean-field solutions as 
\begin{subequations}
    \begin{align}
        \Delta \rho & = \rho - \rho_0 = \sigma^{\frac{1}{2}}_\rho \delta \rho_\rho + \sigma^{\frac{1}{2}}_m \delta \rho_m + \cdots  \\ 
        \Delta m & = m - m_0 = \sigma^{\frac{1}{2}}_\rho \delta m_\rho  + \sigma^{\frac{1}{2}}_m \delta m_m  + \dots  
    \end{align}
\end{subequations}
where we neglected higher powers of $\sigma_{\rho},\sigma_{m}$ since they do not contribute to the renormalization of the mass at first order. 
The dynamics of the average fields $\tilde \rho$ and $\tilde m$ then read:
\begin{eqnarray}
  \label{eq:bar_rho_cons}
  \partial_t \tilde{\rho} &=& D \partial_{xx}\tilde{\rho} -v\partial_x \tilde{m}\;, \\
    \partial_t \tilde{m} &=& D \partial_{xx} \tilde{m} - v \partial_x \tilde{\rho} -\cF(\tilde{\rho},\tilde{m})\notag\\&& -\frac{1}{2} \frac{\partial^{2} \mathcal{F}}{\partial m^{2}} \left( \sigma_\rho \langle\delta m_\rho^{2}\rangle_c + \sigma_m \langle\delta m_\rho^{2}\rangle_c + \sqrt{\sigma_\rho \sigma_m} \langle \delta m_\rho\notag \delta m_m \rangle\right) \\
    &&-  \frac{1}{2} \frac{\partial^{2} \mathcal{F}}{\partial \rho^{2}} \left( \sigma_\rho \langle \delta \rho_\rho \delta \rho_\rho \rangle_c + \sigma_m \langle \delta \rho_m \delta \rho_m \rangle_c + \sqrt{\sigma_\rho \sigma_m } \langle \delta \rho_\rho \delta \rho_m \rangle_c \right) \notag\\
    && 
    -\frac{\partial^{2} \mathcal{F}}{\partial m \partial \rho} (  \sigma_\rho \langle \delta \rho_\rho \delta m_\rho  \rangle_c + \sigma_m \langle \delta \rho_m \delta m_m  \rangle_c + 
    \sqrt{\sigma_\rho \sigma_m } \langle \delta \rho_\rho \delta m_m\rangle_c\notag \\
    && + 
    \sqrt{\sigma_\rho \sigma_m } \langle \delta \rho_m \delta m_\rho  \label{eq:bar_m_cons}
\rangle_c ) \, ,
\end{eqnarray}
where we introduced the connected correlation functions $ \langle a b  \rangle_c = \langle ab\rangle - \langle a\rangle\langle b\rangle$ to lighten the notation and the derivatives of $\mathcal{F}$ are evaluated at $\tilde \rho, \tilde m$. \\
The time evolution of the stochastic fields $\delta \rho_\rho, \delta \rho_m, \delta m_\rho, \delta m_m$ at order $\sigma^{\frac{1}{2}}_{\rho}, \sigma^{\frac{1}{2}}_{m}$ in Fourier space read 
\begin{align}
\partial_{t}\begin{pmatrix}
\delta \rho_\rho^{q} \\
\delta m_\rho^{q} \\ 
\delta \rho_m^{q} \\ 
\delta m_m^{q} \\ 
\end{pmatrix} = \begin{pmatrix}
M_{11}^q & M_{12}^q & 0 & 0 \\
M_{21}^q & M_{22}^q &  0 & 0 \\
 0 & 0 & M_{21}^q & M_{22}^q \\
0 & 0 & M_{21}^q & M_{22}^q  \\
\end{pmatrix}
\begin{pmatrix}
\delta \rho_\rho^{q} \\
\delta m_\rho^{q} \\ 
\delta \rho_m^{q} \\ 
\delta m_m^{q} \\ 
\end{pmatrix} + \begin{pmatrix}
iq\sqrt{2\rho_0}\; \eta_\rho^{q} \\
0\\ 
0\\
\sqrt{2\rho_{0}}\; \eta_m^{q}
\end{pmatrix}\;,
\end{align}
where $\eta_\rho^q$ and $\eta_m^q$ are the $q$-th Fourier modes of the Gaussian white noises with correlations $\langle\eta_j^q(t)\eta_k^{q^{\prime}}(t^{\prime})\rangle=L^{-1}\delta_{q+q^{\prime},0}\delta(t-t^{\prime})\delta_{j,k}$ and the matrix coefficients $M_{11}^{q}$, $M_{12}^{q}$, $M_{21}^{q}$, $M_{22}^{q}$
are still given by \eref{eq:matrix_coeff_AIM}.

We note that fluctuations due to $\sigma_\rho$ and $\sigma_m$ are decoupled. Therefore, the cross correlators proportional to $\sqrt{\sigma_\rho \sigma_m }$
in Eq.~\eqref{eq:bar_m_cons} vanish, and the renormalization contributions proportional to $\sigma_m$ are identical to those computed in section~\ref{app:AIM_FIFOT}, namely 
\begin{align}
\langle\delta\rho_m^2 \rangle_c=&\langle\delta\rho_1^2 \rangle_0, & \langle\delta m_m^2\rangle_c=&\langle\delta m_1^2\rangle_0\;,
\end{align}
where $\langle\delta\rho_1^2\rangle_0$ and $\langle\delta m_1^2 \rangle_0$ are given by \eqref{eq:corr_delta2}.
We thus need to compute only the contributions of the correlators involving fluctuations induced by the conserved noise $\sigma_\rho$. 
In Fourier space, their dynamics read 
\begin{subequations} 
\begin{align}
  \label{eq:Ito_correlator_AIM_cons_noise_no_SS}
\frac{d\langle \delta\rho_\rho^q \delta \rho_\rho^{q'}\rangle_0}{dt} =&(M_{11}^{q}+M^{q^{\prime}}_{11})\langle\delta\rho_\rho^{q}\delta\rho_\rho^{q^{\prime}}\rangle_0 + M_{12}^{q}\langle\delta m_\rho^{q}\delta\rho_\rho^{q^{\prime}}\rangle_0 + M_{12}^{q^{\prime}}\langle\delta \rho_\rho^{q}\delta m_\rho^{q^{\prime}}\rangle_0 - \frac{2\rho_0 q^2}{L}\delta_{q+q^{\prime},0}\, ,\\
\label{eq:Ito_correlator_AIM_2_cons_noise_no_SS}
\frac{d \langle \delta m_\rho^q \delta \rho_\rho^{q'} \rangle_0}{dt} = &(M_{22}^{q}+M^{q^{\prime}}_{11})\langle\delta m_\rho^{q}\delta\rho_\rho^{q^{\prime}}\rangle_0 + M_{21}^{q}\langle\delta \rho_\rho^{q}\delta\rho_\rho^{q^{\prime}}\rangle_0 + M_{12}^{q^{\prime}}\langle\delta m_\rho^{q}\delta m_\rho^{q^{\prime}}\rangle_0, , \\
\label{eq:Ito_correlator_AIM_3_cons_noise_no_SS}
\frac{d\langle \delta m_\rho^q \delta m_\rho^{q^{\prime}} \rangle_0}{dt} =&(M_{22}^{q}+M^{q^{\prime}}_{22})\langle\delta m_\rho^{q}\delta m_\rho^{q^{\prime}}\rangle_0 + M_{21}^{q}\langle\delta \rho_\rho^{q}\delta m_\rho^{q^{\prime}}\rangle_0 + M_{21}^{q^{\prime}}\langle\delta m_\rho^{q}\delta \rho_\rho^{q^{\prime}}\rangle_0\;.
\end{align}
\end{subequations}
In the steady state, one has that
\begin{equation}
    \frac{d\langle \delta\rho_\rho^q \delta \rho_\rho^{q'}\rangle_0}{dt} =\frac{d \langle \delta m_\rho^q \delta \rho_\rho^{q'} \rangle_0}{dt}=\frac{d\langle \delta m_\rho^q \delta m_\rho^{q^{\prime}} \rangle_0}{dt} =0\;.
\end{equation}
Inverting the corresponding system of equations then yields
\begin{align}
\label{eq:fourier_corr_scaling_AIM_cons_noise}
\langle\delta m_\rho^{q}\delta m_\rho^{q^{\prime}}\rangle_0 =& g_{mm}(q)\frac{\delta_{q,-q^{\prime}}}{L},  &
\langle\delta \rho_\rho^{q}\delta \rho_\rho^{q^{\prime}}\rangle_0 =& g_{\rho\rho}(q)\frac{\delta_{q,-q^{\prime}}}{L}, &
\langle\delta m_\rho^{q}\delta \rho_\rho^{q^{\prime}}\rangle_0 =& g_{m\rho}(q)\frac{\delta_{q,-q^{\prime}}}{L},
\end{align}
where, to first order in $m_0$, the functions $g_{mm}(q)$, $g_{\rho\rho}(q)$ and $g_{m\rho}(q)$ are given by
\begin{align}
\label{eq:fourier_m_m_corr_AIM_cons_noise}
g_{mm}(q) =& - \rho_0 \frac{ q^2 v^2}{\left(\alpha +2 D q^2\right) \left(D \left(\alpha +D q^2\right)+v^2\right)}+\cO\left(m_0^2\right)\; ,  \\
\label{eq:fourier_rho_rho_corr_AIM_cons_noise}
g_{\rho\rho}(q) =& -\rho_0\frac{ \left(\alpha ^2+2 D^2 q^4+3 \alpha  D q^2+q^2 v^2\right)}{\left(\alpha +2 D q^2\right) \left(D \left(\alpha +D q^2\right)+v^2\right)}+\cO\left(m_0^2\right)\;, \\
\label{eq:fourier_m_rho_corr_AIM_cons_noise}
g_{m\rho}(q) =& \rho_0\frac{i q v  \left(\alpha +D q^2 \right)}{\left(\alpha +2 D q^2\right) \left(D \left(\alpha +D q^2\right)+v^2\right)}+\cO\left(m_0^2\right)\ .
\end{align}
At this point we note that $g_{mm}(q)$, $g_{\rho\rho}(q)$ and $g_{m\rho}(q)$ contain a polynomial part in $q$ and a rational part. For instance, $g_{\rho\rho}$ can be rewritten as
\begin{equation}
  g_{\rho\rho}=-\frac{\rho_0}{D} +\frac{\rho_0}{D}\frac{\alpha v^2 + D q^2 v^2}{\left(\alpha +2 D q^2\right) \left(D \left(\alpha +D q^2\right)+v^2\right)}+\cO\left(m_0^2\right)\;. \label{eq:fourier_m_rho_corr_AIM_cons_noise2} \\
\end{equation}
Upon Fourier transforming back into real space, these polynomial parts  yield  divergent contributions that takes the form of a sum of delta peaks since $\int dq \sum_{j=0}^{N} a_j q^j e^{iqx} \simeq \sum_{j=0}^N b_j \nabla^j \delta(x)$. 
The emergence of these divergences can be traced back to the presence of the conserved noise in the continuum field theory: the latter lacks a microscopic cutoff, which is usually determined by the lattice spacing of the underlying microscopic model.  
Therefore, this diverging part is an artifact that results from the continuum nature of the description \eref{eq:hydro_AIM_conserved_noise}. The contribution of this polynomial part to the dressed coefficients thus depends on the microscopic regularization that one uses and its computation is beyond the scope of our theory.
On the contrary, the rational parts of $g_{mm}(q)$, $g_{\rho\rho}(q)$ and $g_{m\rho}(q)$ are converging upon integrating over $q$: their contributions are universal and do not depend on the chosen regularization. As we show below, they lead to a density-dependent correction to the mass of the polar field and thus suffice to predict the emergence of a fluctuation-induced first-order transition.
To this aim, we discard the polynomial parts in~\eqref{eq:fourier_m_rho_corr_AIM_cons_noise2} to consider:
\begin{align}
\label{eq:fourier_rho_rho_corr_AIM_cpns_noise_reg}
g_{\rho\rho}(q) =& \frac{\rho_0}{D}\frac{\alpha v^2+Dq^2  v^2}{\left(\alpha +2 D q^2\right) \left(D \left(\alpha +D q^2\right)+v^2\right)}+\cO\left(m_0^2\right)\;.
\end{align}
We further remark that we can integrate $g_{mm}(q)$, $g_{\rho\rho}(q)$ and $g_{m\rho}(q)$ over $q$ only if $\alpha > 0$, which means that we have to restrict our study to the high temperature phase where such a condition is respected.
Performing the integration in such a regime, we finally obtain
\begin{subequations}\label{eq:corr_cons_noise}
\begin{align}
  \label{eq:corr_cons_noise_rho_rho}
  \langle \delta\rho_\rho^2 \rangle_0 &=\int \frac{dq}{2\pi} g_{\rho\rho}(q) = \frac{\rho_0\alpha^2}{v^3}f_3\left(\frac{\alpha D}{v^2}\right) + \cO(m_0^2)\;, \\
  \label{eq:corr_cons_noise_m_m}
  \langle \delta m_\rho^2 \rangle_0 &=\int \frac{dq}{2\pi} g_{mm}(q) = - \frac{\rho_0\alpha^2}{v^3}f_2\left(\frac{\alpha D}{v^2}\right) + \cO(m_0^2) \;, \\
  \label{eq:corr_cons_noise_rho_m}
  \langle \delta m_\rho\delta\rho_\rho \rangle_0 &=\int \frac{dq}{2\pi} g_{m\rho}(q) = 0 + \mathcal{O}(m_0^2)\; ,
\end{align}
\end{subequations}
where the functions $f_2$, and $f_3$ are given by
\begin{align}
  f_2(u)=&\frac{2u-\sqrt{2u(u+1)}+2}{4u^2\sqrt{u+1}(u+2)}\,, 
  \quad
  f_3(u)=\frac{\sqrt{2u(u+1)}+2}{4u^2\sqrt{u+1}(u+2)}\, .
\end{align}
Finally, inserting \eref{eq:corr_cons_noise} and \eref{eq:corr_delta2} into \eref{eq:bar_m} then yields the following renormalized dynamics for $\tilde{m}$
\begin{align}
  \label{eq:dyn_m_cons_noise_renorm}
  \partial_t \tilde{m} =& D \partial_{xx} \tilde{m} - v \partial_x \tilde{\rho} -\cF(\tilde{\rho},\tilde{m})- \frac{3\gamma \tilde{m}}{\tilde{\rho} v}\left[\sigma_m f_1\left(\frac{\alpha D}{v^2}\right)- \sigma_\rho\frac{\alpha^2}{v^2}f_2\left(\frac{\alpha D}{v^2}\right)\right] + \mathcal{O}(\sigma \tilde{m}^2)\, ,
\end{align}
where we introduced the function $f_1$ defined as 
\begin{align}
    f_1(u) = \frac{\sqrt{2/u} + \sqrt{1+u} }{2(2+u)} = \frac{\sqrt{2u(u+1)} + u (u+1) }{2u \sqrt{u+1} (u+2)}\;.
\end{align}
From \eref{eq:dyn_m_cons_noise_renorm}, we deduce the expression of the renormalized linear Landau term $\hat{\alpha}$ at first order in $\sigma_{m}$ and $\sigma_{\rho}$ as
\begin{align}\label{eq:RenAlphaCons}
  \hat{\alpha} = \alpha + \frac{3\gamma}{\tilde{\rho} v}\left[\sigma_m f_1\left(\frac{\alpha D}{v^2}\right) -\sigma_\rho \frac{\alpha^2}{v^2}f_2\left(\frac{\alpha D}{v^2}\right)\right]\, .
\end{align}
Note that, in the absence of conserved noise ($\sigma_\rho=0$), Eq.\eref{eq:RenAlphaCons} reduces to the renormalized mass given by Eq.\eref{eq:RenAlpha} of the main text. 
Note also that Eq.~\eqref{eq:RenAlphaCons} should be complemented by the corrections due to the polynomial contribution to $g_{\rho\rho}$. 
The latter is not universal and will not cancel the density dependence reported in~\eqref{eq:RenAlphaCons}.\\
All in all, the renormalized linear mass term $\hat{\alpha}$ has become density-dependent due to the effect of fluctuations, even in the presence of a conserved noise acting on the density field.
Therefore, according to the stability analysis performed in \ref{sec:LinStab}, $\hat{\alpha}(\tilde{\rho})$ will turn the continuous transition predicted by the mean-field evolution \eref{eq:hydro_AIM_conserved_noise} into the standard discontinuous phase transition with travelling flocks.

\subsection{Renormalization in 2D}
\label{sec:AIM_2D}

This appendix is devoted to the computation of the correlators $\langle \delta \rho_1^2\rangle$, $\langle \delta m_1^2\rangle$ and $\langle \delta m_1\delta\rho_1\rangle$ in dimension 2.
As in Appendix \ref{app:correlators_AIM}, we cast the stochastic evolution of $\delta\rho_1$ and $\delta m_1$ into Fourier space with the Fourier convention defined in \eref{eq:fourier_serie}.
We obtain
\begin{align}
\label{eq:linear_perturbation_AIM_2D_app}
\partial_{t}\begin{pmatrix}
\delta \rho_1^{\bq} \\
\delta m_1^{\bq}
\end{pmatrix} = \begin{pmatrix}
M_{11}^{\bq} & M_{12}^{\bq}\\
M_{21}^{\bq} & M_{22}^{\bq}
\end{pmatrix}
\begin{pmatrix}
\delta \rho_1^{\bq} \\
\delta m_1^{\bq}
\end{pmatrix} + \begin{pmatrix}
0 \\
\sqrt{2\rho_{0}}\; \eta^{\bq}
\end{pmatrix}\;,
\end{align}
where $\eta^{\bq}$ is the $\bq$-th Fourier mode of the Gaussian white noise with correlations $\langle\eta^{\bq}(t)\eta^{{\bq}^{\prime}}(t^{\prime})\rangle=L^{-1}\delta_{\bq+\bq^{\prime},0}\delta(t-t^{\prime})$ and the matrix coefficients $M_{11}^{\bq}$, $M_{12}^{\bq}$, $M_{21}^{\bq}$, $M_{22}^{\bq}$
are now given by
\begin{align}
  \label{eq:matrix_coeff_AIM_2D}
  M_{11}^{\bq}=&-D\bq^2\;, & M_{12}^{\bq}=& -ivq_x\;, & M_{21}^{\bq}=&-iq_x v +2\gamma\frac{m_{0}^{3}}{\rho_{0}^{3}}\;, & M_{22}^{\bq} =& -D \bq^{2} - \alpha - 3\gamma\frac{m_{0}^{2}}{\rho_{0}^{2}}\ .
\end{align}
To compute the equal-time two-point correlation functions, we use It\=o calculus on the stochastic system \eref{eq:linear_perturbation_AIM_2D_app} and obtain the following closed system of equations
\begin{subequations}
\begin{align}
  \label{eq:Ito_correlator_AIM_2D_no_SS}
\frac{d}{dt}\langle \delta\rho_1^{\bq} \delta \rho_1^{\bq^{\prime}} \rangle_0 =&(M_{11}^{\bq}+M^{\bq^{\prime}}_{11})\langle\delta\rho_1^{\bq}\delta\rho_1^{\bq^{\prime}}\rangle_0 + M_{12}^{\bq}\langle\delta m_1^{\bq}\delta\rho_1^{\bq^{\prime}}\rangle_0 + M_{12}^{\bq^{\prime}}\langle\delta \rho_1^{\bq}\delta m_1^{\bq^{\prime}}\rangle_0 =0\, ,\\
\label{eq:Ito_correlator_AIM_2_2D_SS}
\frac{d}{dt}\langle \delta m_1^{\bq} \delta \rho_1^{\bq^{\prime}} \rangle_0 =&(M_{22}^{\bq}+M^{\bq^{\prime}}_{11})\langle\delta m_1^{\bq}\delta\rho_1^{\bq^{\prime}}\rangle_0 + M_{21}^{\bq}\langle\delta \rho_1^{\bq}\delta\rho_1^{\bq^{\prime}}\rangle_0 + M_{12}^{\bq^{\prime}}\langle\delta m_1^{\bq}\delta m_1^{\bq^{\prime}}\rangle_0=0\, , \\
\label{eq:Ito_correlator_AIM_3_2D_SS}
\frac{d}{dt}\langle \delta m_1^{\bq} \delta m_1^{\bq^{\prime}} \rangle_0 =&(M_{22}^{\bq}+M^{\bq^{\prime}}_{22})\langle\delta m_1^{\bq}\delta m_1^{\bq^{\prime}}\rangle_0 + M_{21}^{\bq}\langle\delta \rho_1^{\bq}\delta m_1^{\bq^{\prime}}\rangle_0 + M_{21}^{\bq^{\prime}}\langle\delta m_1^{\bq}\delta \rho_1^{\bq^{\prime}}\rangle_0 + \frac{2\rho_{0}}{L}\delta_{\bq+\bq^{\prime},0}=0\, .
\end{align}
\end{subequations}
where the last equality stems from working in the steady state.
Inverting the above system yields
\begin{align}
\label{eq:fourier_corr_scaling_AIM_2D}
\langle\delta m_1^{\bq}\delta m_1^{\bq^{\prime}}\rangle_0 =& g_{mm}(\bq)\frac{\delta_{\bq+\bq^{\prime},0}}{L}\;,  &
\langle\delta \rho_1^{\bq}\delta \rho_1^{\bq^{\prime}}\rangle_0 =& g_{\rho\rho}(\bq)\frac{\delta_{\bq+\bq^{\prime},0}}{L}\;, &
\langle\delta m_1^{\bq}\delta \rho_1^{\bq^{\prime}}\rangle_0 =& g_{m\rho}(\bq)\frac{\delta_{\bq+\bq^{\prime},0}}{L}\ ,
\end{align}
where, to first order in $m_0$, the functions $g_{mm}(q)$, $g_{\rho\rho}(q)$ and $g_{m\rho}(q)$ are given by
\begin{subequations}
\begin{align}
\label{eq:fourier_m_m_corr_AIM_2D}
g_{mm}(\bq) =&\rho_0\frac{2 D^2 \bq^4+\alpha  D \bq^2+q_x^2 v^2}{ \left(\alpha +2 D \bq^2\right) \left(D^2 \bq^4+\alpha  D\bq^2+q_x^2 v^2\right)} +\cO\left(m_0^2\right)\;, \\
\label{eq:fourier_rho_rho_corr_AIM_2D}
g_{\rho\rho}(\bq) =& \rho_0\frac{q_x^2 v^2}{\left(\alpha +2 D \bq^2\right) \left(D^2 \bq^4+\alpha  D\bq^2+q_x^2 v^2\right)}+\cO\left(m_0^2\right)\;, \\
\label{eq:fourier_m_rho_corr_AIM_2D}
g_{m\rho}(\bq) =& \rho_0\frac{i D q_x v \bq^2}{\left(\alpha +2 D \bq^2\right) \left(D^2 \bq^4+\alpha  D\bq^2+q_x^2 v^2\right)}+\cO\left(m_0^2\right)\,.
\end{align}
\end{subequations}
The real-space correlators are then obtained by inverse Fourier transform using Eq.~\eref{eq:limit_fourier_serie} and performing the integral
over $q$
\begin{subequations}
\begin{align}%
  \label{eq:corr_aim_2D_app}
  \langle\delta m_1^2 \rangle_0
  =& \frac{\rho_0 \alpha}{v^2}h_1\left(\frac{\alpha D}{v^2}\right) + \mathcal{O}(m_0^2)\; , \quad \langle\delta \rho_1^2 \rangle_0
  =  \frac{\rho_0 \alpha}{v^2}h_2\left(\frac{\alpha D}{v^2}\right) + \mathcal{O}(m_0^2)\;, \\
  \langle\delta \rho_1\delta m_1 \rangle_0 =& 0 + \mathcal{O}(m_0^2)\; ,
\end{align}
\end{subequations}
where $h_1(u)$ and $h_2(u)$ are given by \eref{eq:func_corr_AIM_2d} and were obtained using the change of variable $q=\tilde{q} \alpha/v$.

\section{The Active Ising Model with topological alignment}
\label{app:topological_AIM}

We recall below the definition of the mean-field evolution of the AIM for a generic aligning field $\bar{m}$ in $1$D
\begin{subequations}\label{eq:field_topological_generic_app}
\begin{align}
\label{eq:rho_topological_pde}
\partial_{t}\rho &= D\partial_{xx} \rho -v\partial_x m \;, \\
\label{eq:m_topological_pde}
\partial_{t}m &= D\partial_{xx} m -v\partial_x \rho +2\Gamma\rho\beta\bar{m}\left(1+\frac{\beta^{2}\bar{m}^{2}}{6}\right)-2\Gamma m\left(1+\frac{\beta^{2}}{2}\bar{m}^{2}\right) \;.
\end{align}
\end{subequations}
In the case of $k$-nearest neighbors alignment, we recall the expression of the aligning field $\bar{m}$
\begin{align}
  \label{eq:k_nearest_field_app}
\bar{m}=\frac{1}{k}\int^{x+y(x)}_{x-y(x)}m(x)dx\;,
\end{align}
where $y(x)$ is defined implicitly through $k=\int^{x+y(x)}_{x-y(x)}\rho(x)dx$.

\subsection{Linear stability analysis for $k$-nearest alignment}
\label{app:lin_stab_topo}
In this Appendix, we study the linear stability of homogeneous solutions $\rho=\rho_{0}$, $m=m_{0}$, $y=y_0$ and $\bar{m}=\bar{m}_0$ of Eqs.~\eref{eq:field_topological_generic_app} with $k$-nearest neighbors alignment \eref{eq:k_nearest_field_app}.
Because $y(x)$ and $\bar{m}(x)$ are enslaved to $\rho(x)$ and $m(x)$ through \eref{eq:k_nearest_field_app}, $\delta \rho$ and $\delta m$ are the only two independent perturbations in the system and we further have that $y_0=k/(2\rho_0)$ and that $\bar{m}_0=m_0/\rho_0$.
We first relate $\delta \bar{m}$ to $\delta \rho$ and $\delta m$.
To first order,
\begin{align}
  \delta \bar{m} =& \frac{\int^{x+y_{0}+\delta y}_{x-y_{0}-\delta y}(m_{0}+\delta m(z))dz}{k} -\frac{m_0}{\rho_0}
  \label{eq:lin_m_bar}
  = \int^{x+y_{0}}_{x-y_{0}}\frac{\delta m(z)dz}{k} + 2\delta y \frac{m_{0}}{k} + \cO(\delta m^2) + \cO(\delta y^2)\ .
\end{align}
We then express $\delta y$ as a function of $\delta \rho$ using the implicit equation $k=\int^{x+y(x)}_{x-y(x)}\rho(x)dx$ :
\begin{align}
\label{eq:lin_delta_y}
\delta y=&-\frac{1}{2\rho_0}\int^{x+y_{0}}_{x-y_{0}}\delta \rho(z)dz + \cO(\delta\rho^2)\ .
\end{align}
Interestingly, the fluctuations of $\rho$ impact $\bar m$ through $\delta y$, despite the topological nature of the alignment.
To study the linear stability of \eref{eq:field_topological_generic_app} in Fourier space, we express $\delta y^q$ and $\delta \bar{m}^q$ in terms of $\delta \rho^q$ and $\delta m^q$.
To this aim, we compute the Fourier transform of \eref{eq:lin_m_bar} and \eref{eq:lin_delta_y}.
We start by determining $\delta y^q$ as
\begin{align}
  \label{eq:delta_y_q}
  \delta y^q =& -\frac{y_0}{\rho_0}\sinc(q y_0)\delta\rho^q \ .
\end{align}
We then get $\delta m_q$ as
\begin{align}
  \label{eq:delta_m_bar_q_1}
  \delta \bar{m}^q =& \frac{y_0}{k}2\sinc(q y_0)\delta m^q + 2\frac{m_0}{k}\delta y^q \ .
\end{align}
Reinserting \eref{eq:delta_y_q} in \eref{eq:delta_m_bar_q_1} and further using $k=2y_0\rho_0$ we obtain
\begin{align}
  \label{eq:delta_m_bar_q_2}
  \delta \bar{m}^q =& \frac{1}{\rho_0}\sinc(q y_0)\delta m^q - \frac{m_0}{\rho_{0}^2}\sinc(q y_0)\delta\rho^q \ .
\end{align}
Linearizing \eref{eq:field_topological_generic_app} to first order in $\delta m$, $\delta \rho$, $\delta \bar{m}$, multiplying both sides by $e^{-iqx}$, and integrating over $q$ yields the dynamics for the $q$-th Fourier modes $\delta m^q$, $\delta \rho^q$ and $\delta \bar{m}^q$.
Further using \eref{eq:delta_m_bar_q_2}, we obtain the following dynamics for $\delta m^q$ and $\delta \rho^q$
\begin{equation}
\label{eq:dynamic_topo}
\partial_t \begin{pmatrix}
\delta {\rho^q} \\
\delta {m^q}
\end{pmatrix} = \begin{pmatrix}
M_{11}^q & M_{12}^q \\
M_{21}^q  & M_{22}^q
\end{pmatrix}
\begin{pmatrix}
\delta \rho^q \\
\delta m^q
\end{pmatrix}\;,
\end{equation}
where the matrix coefficients $M_{11}^q$, $M_{12}^q$, $M_{21}^q$ and $M_{22}^q$ are given by:
\begin{subequations}\label{eq:mat_coeff_knearest}
\begin{align}
  \label{eq:mat_coeff_topo_11}
  M_{11}^q =& -Dq^2\;,\qquad
  M_{12}^q = -ivq\\
  \label{eq:mat_coeff_topo_21}
  M_{21}^q =& \frac{\Gamma}{3}\left(\frac{\beta m_{0}}{\rho_{0}}\right)^{3}+2\Gamma\beta\frac{m_{0}}{\rho_{0}}-iqv+\Gamma\sinc(qy_0)\frac{m_{0}}{\rho_{0}}\left(\left(\frac{\beta m_{0}}{\rho_{0}}\right)^{2}\left(-\beta+2\right)-2\beta\right) \\
  \label{eq:mat_coeff_topo_22}
  M_{22}^q =& -2\Gamma-\Gamma\left(\frac{\beta m_{0}}{\rho_{0}}\right)^{2}+\Gamma\sinc(qy_0)\left(\left(\frac{\beta m_{0}}{\rho_{0}}\right)^{2}\left(\beta-2\right)+2\beta\right)-Dq^{2}\ .
\end{align}
\end{subequations}
We are interested in the onset of the transition where $m_0$ is small and $\beta=1+\cO(m_0)$.
To leading order in $m_0$, the matrix coefficients simplify into
\begin{align}
  M_{11}^q =& -Dq^2 + \cO(m_0) &
  M_{12}^q =& -iqv + \cO(m_0) \\
  M_{21}^q =& -iqv + \cO(m_0) &
  M_{22}^q =& 2\Gamma\left(\sinc(qy_0)-1\right) - Dq^2 + \cO(m_0)\ .
\end{align}
Within this regime, the eigenvalues of the linearized dynamic \eref{eq:dynamic_topo} are given by
\begin{equation}
\lambda_{\pm}=\dfrac{-\left(2Dq^{2} + 2\Gamma\left(1-\sinc(qy_0)\right) \right)\pm \sqrt{\Delta}}{2}\;,
\end{equation}
with the discriminant $\Delta $ reading
\begin{equation}
\Delta = \left(2Dq^{2} + 2\Gamma\left(1-\sinc(qy_0)\right) \right)^{2}-4v^2q^2-4Dq^{2}\left(Dq^{2}+2\Gamma\left(1-\sinc(qy_0)\right) \right)\;.
\end{equation}
An instability emerges when at least one of the eigenvalues $\lambda_{\pm}$ has a positive real part, which translates into
\begin{equation}
\label{eq:stability_topo_cond}
2\Re(\sqrt{\Delta})^2-2\Re\left(2Dq^{2} + 2\Gamma\left(1-\sinc(qy_0)\right) \right)^2 >0 \ .
\end{equation}
This condition can be simplified by noticing that
\begin{align}
  \label{eq:stability_topo_sign}
  2\Re(\sqrt{\Delta})^2-2\Re\left(2Dq^{2} + 2\Gamma\left(1-\sinc(qy_0)\right) \right)^2 = -a + \sqrt{a^2+b(q)}>0 \ ,
\end{align}
where $a$ and $b$ reads:
\begin{align}
  a(q) =& \left[2Dq^2+2\Gamma(1-\sinc(q y_0))\right]^2 + 4v^2 q^2 + 4Dq^2\left[Dq^2 + 2\Gamma(1-\sinc(q y_0))\right] \\
  \nonumber
  \frac{b(q)}{64q^2} =& -\Gamma^2\left(\sinc(qy_0)-1\right)^2 \left[v^2+2 D\Gamma \left(1-\sinc(qy_0)\right)\right] - D^2 q^4 \left[4 D\Gamma \left(1-\sinc(qy_0)\right)+v^2\right]\\
  \label{eq:b_topo_stab}
  & - q^2 D \Gamma\left(1-\sinc(qy_0)\right) \left[5 D\Gamma \left(1-\sinc(qy_0)\right)+2 v^2\right] - D^4 q^6
\end{align}
We remark that $a(q)$ is a positive term and thus that the stability is determined by the sign of $b(q)$.
All the terms in its expression \eqref{eq:b_topo_stab} are negative; the homogeneous ordered solution is thus always stable at onset.
The mean-field theory \eqref{eq:field_topological_generic_app} with $k$-nearest neighbors alignment thus predicts a continuous transition to collective motion without phase separation.

\subsection{Renormalization for $k$-nearest alignment}
\label{sec:renormalization_generic}
In this appendix, we compute the renormalization of the linear mass term due to the addition of noise term in \eref{eq:AIM_generic_noisy}.
To remain as general as possible, we first compute this renormalization for a generic unconstrained $\bar{m}$, which we take to be a functional of $\rho$ and $m$: $\bar{m}=\cG(x,\{\rho,m\})$.
At the end of this appendix, we apply our results for the specific case of $k$-nearest neighbors alignment where $\bar{m}$ is defined by \eref{eq:k_nearest_field_app} and obtain expression \eref{eq:delta_Ft_renorm_topo} of main text.
We first recall the stochastic evolution of the fields
\begin{align}
  \label{eq:rho_topological_pde_noise}
  \partial_{t}\rho &= D\nabla^{2} \rho -v\nabla m\;, \qquad
  \partial_{t}m = D\nabla^{2} m -v\nabla \rho - \cF(\bar{m},m,\rho) + \sqrt{2\sigma\rho}\ \eta \ ,
\end{align}
where $\cF(\bar{m},m,\rho)$ is given by \eref{eq:F_onset}. Following section \ref{subsec:renorm_AIM}, we perform a perturbative expansion of the fields $m$ and $\rho$ as
\begin{align}
  \label{eq:expans_rho_topol}
  \rho =& \rho_0 + \sqrt{\sigma}\delta\rho_1 + \sigma \delta \rho_2 + .. +\;, \qquad
  m = m_0 + \sqrt{\sigma}\delta m_1 + \sigma \delta m_2 + .. +
\end{align}
where $\rho_0$ and $m_0$ are solutions of \eref{eq:rho_topological_pde_noise} with $\sigma=0$
\begin{align}
  \label{eq:delta_rho_0_topo}
  \partial_{t}\rho_0 &= D\nabla^{2} \rho_0 -v\nabla m_0\;, \\
  \label{eq:delta_m_0_topo}
  \partial_{t}m_0 =& D\nabla^{2} m_0 -v\nabla \rho_0 - \Ft(\bar{m}_0,m_0,\rho_0)\; .
\end{align}
Noting that $\bar{m}$ is enslaved to $\rho$ and $m$ through $\bar{m}=\cG(x,\{\rho,m\})$, it can be expanded in powers of $\sigma^{1/2}$ as
\begin{align}
  \label{eq:expans_bar_m_topol}
  \bar{m}=&\bar{m}_0+\sqrt{\sigma}\left[\int \frac{\delta \bar{m}(x)}{\delta m(z)}\delta m_1(z) dz + \int \frac{\delta \bar{m}(x)}{\delta \rho(z)}\delta \rho_1(z) dz \right] \\
  \nonumber
  &+ \sigma\bigg{[}\int \bigg{(}\frac{1}{2}\frac{\delta^2 \bar{m}(x)}{\delta m(s)\delta m(z)}\delta m_1(s) \delta m_1(z) + \frac{1}{2}\frac{\delta^2 \bar{m}(x)}{\delta \rho(s)\delta \rho(z)}\delta \rho_1(s)\delta \rho_1(z)\bigg{)}dsdz  \\
  \nonumber
  & + \int \frac{\delta^2 \bar{m}(x)}{\delta \rho(s)\delta  m(z)}\delta \rho_1(s)\delta m_1(z) dsdz + \int \frac{\delta \bar{m}(x)}{\delta m(z)}\delta m_2(z) dz
  + \int \frac{\delta \bar{m}(x)}{\delta \rho(z)}\delta \rho_2(z) dz \bigg{]}\\
  \nonumber
  & + .. +.
\end{align}
Inserting \eref{eq:expans_rho_topol} and \eref{eq:expans_bar_m_topol} into \eref{eq:rho_topological_pde_noise} and projecting on $\sigma^{n/2}$ gives the evolution equation for $\delta\rho_n$ and $\delta m_n$.
For $n=1$, it yields
\begin{subequations}\label{eq:delta_fields_evolution_generic}
\begin{align}
  \label{eq:delta_rho_1_topo}
  \partial_{t}\delta \rho_1 =& D\nabla^{2} \delta \rho_1 -v\nabla \delta m_1 \\
  \label{eq:delta_m_1_topo}
  \partial_{t} \delta m_1 =& D\nabla^{2} \delta m_1 -v\nabla \delta \rho_1 - \frac{\partial \cF}{\partial m} \delta m_1 - \frac{\partial \cF}{\partial \rho} \delta \rho_1 \\
  \nonumber
  &- \frac{\partial \cF}{\partial \bar{m}} \left[\int \frac{\delta \bar{m}(x)}{\delta m(z)}\delta m_1(z) dz + \int \frac{\delta \bar{m}(x)}{\delta \rho(z)}\delta \rho_1(z) dz \right] + \sqrt{2\rho_0}\ \eta \ .
\end{align}
\end{subequations}
Noting that $\partial_{\rho\rho}\cF=\partial_{mm}\cF=\partial_{m\rho}\cF=0$, we obtain for $n=2$
\begin{align}
  \label{eq:delta_rho_2_topo}
  \partial_{t}\delta \rho_2 &= D\nabla^{2} \delta \rho_2 -v\nabla \delta m_2 \\
  \nonumber
  \partial_{t} \delta m_2 &= D\nabla^{2} \delta m_2 -v\nabla \delta \rho_2 - \frac{\partial \cF}{\partial m} \delta m_2 - \frac{\partial \cF}{\partial \rho} \delta \rho_2 - \frac{\partial \cF}{\partial \bar{m}} \delta \bar{m}_2
  \\
  \nonumber
  &- \frac{1}{2}\frac{\partial^2 \Ft}{\partial^2 \bar{m}} \left[\int \frac{\delta \bar{m}(x)}{\delta m(z)}\delta m_1(z) dz + \int \frac{\delta \bar{m}(x)}{\delta \rho(z)}\delta \rho_1(z) dz \right]^2 \\
  \nonumber
  & - \frac{\partial^2 \Ft}{\partial\bar{m} \partial m } \delta m_1\left[\int \frac{\delta \bar{m}(x)}{\delta m(z)}\delta m_1(z) dz + \int \frac{\delta \bar{m}(x)}{\delta \rho(z)}\delta \rho_1(z) dz \right] \\
  \nonumber
  & - \frac{\partial^2 \Ft}{\partial \bar{m} \partial\rho} \delta \rho_1\left[\int \frac{\delta \bar{m}(x)}{\delta m(z)}\delta m_1(z) dz + \int \frac{\delta \bar{m}(x)}{\delta \rho(z)}\delta \rho_1(z) dz \right] \\
  \nonumber
  & - \frac{\partial \Ft}{\partial \bar{m}} \left[\int \frac{\delta \bar{m}(x)}{\delta m(z)}\delta m_2(z) dz + \int \frac{\delta \bar{m}(x)}{\delta \rho(z)}\delta \rho_2(z) dz \right] \\
  \nonumber
  & - \frac{\partial \Ft}{\partial \bar{m}} \int \bigg{(}\frac{1}{2}\frac{\delta^2 \bar{m}(x)}{\delta m(s)\delta m(z)}\delta m_1(s) \delta m_1(z) + \frac{1}{2}\frac{\delta^2 \bar{m}(x)}{\delta \rho(s)\delta \rho(z)}\delta \rho_1(s)\delta \rho_1(z)\bigg{)}dsdz \\
  \label{eq:delta_m_2_topo}
  & - \frac{\partial \Ft}{\partial \bar{m}} \int \frac{\delta^2 \bar{m}(x)}{\delta \rho(s)\delta  m(z)}\delta \rho_1(s)\delta m_1(z) dsdz + \frac{\delta\rho_1}{\sqrt{2\rho_0}}\ \eta \ .
\end{align}
Summing together \eref{eq:delta_rho_0_topo}, \eref{eq:delta_rho_1_topo} multiplied by $\sqrt{\sigma}$, and \eref{eq:delta_rho_2_topo} multiplied by $\sigma$ gives the stochastic evolution of $\rho$ up to order $\sigma$.
Averaging the result over the noise using It\=o prescription, we obtain the time evolution of $\tilde{\rho}=\langle\rho\rangle$ to order $\sigma$ as
\begin{align}
  \label{eq:tilde_rho_topo_1}
  \partial_{t}\tilde{\rho} &= D\nabla^{2} \tilde{\rho} -v\nabla \tilde {m}\;.
\end{align}
Likewise, summing \eref{eq:delta_m_0_topo}, \eref{eq:delta_m_1_topo} multiplied by $\sqrt{\sigma}$, and \eref{eq:delta_m_2_topo} multiplied by $\sigma$ gives the stochastic evolution of $m$ to order $\sigma$.
Averaging the result over the noise using It\=o prescription, we obtain the time evolution of $\tilde{m}=\langle m\rangle$ to order $\sigma$ as
\begin{align}
  \label{eq:tilde_rho_topo_2}
  \partial_{t} \tilde{m} = D\nabla^{2} \tilde {m} -v\nabla \delta \tilde{\rho} - \cF(\tilde{m},\tilde{\rho},\tilde{\bar{m}}) - \sigma \Delta\cF \; .
\end{align}
In \eref{eq:tilde_rho_topo_2}, $\tilde{\bar{m}}$ is the topological field constructed with $\tilde{\rho}$ and $\tilde{m}$, \textit{ie} $\tilde{\bar{m}}=\cG(x,\{\tilde{m},\tilde{\rho}\})$,
and $\Delta\cF$ is given by
\begin{align}
  \label{eq:delta_Ft}
  \Delta\cF = \frac{1}{2}\frac{\partial^2 \cF}{\partial^2 \bar{m}^2} C_1 + \frac{\partial^2 \cF}{\partial \bar{m}\partial m} C_2 + \frac{\partial^2 \cF}{\partial \bar{m}\partial\rho} C_3 + \frac{\partial \cF}{\partial \bar{m}} C_4 \ ,
\end{align}
where the $C_i$'s are correlators given by
\begin{subequations}\label{eq:generic_correlators_app}
\begin{align}
  \nonumber
  C_1 =& \int \frac{\delta \bar{m}(x)}{\delta m(s)}\frac{\delta \bar{m}(x)}{\delta m(z)}\big{[}\langle\delta m_1(s)\delta m_1(z)\rangle- \langle\delta m_1(s)\rangle\langle\delta m_1(z)\rangle\big{]}dsdz \\
  \nonumber
  &+ \int \frac{\delta \bar{m}(x)}{\delta \rho(s)}\frac{\delta \bar{m}(x)}{\delta \rho(z)}\big{[}\langle\delta \rho_1(s)\delta \rho_1(z)\rangle- \langle\delta \rho_1(s)\rangle\langle\delta \rho_1(z)\rangle\big{]}dsdz \\
  \label{eq:C_1}
  &+ 2\int \frac{\delta \bar{m}(x)}{\delta \rho(s)}\frac{\delta \bar{m}(x)}{\delta m(z)}\big{[}\langle\delta \rho_1(s)\delta m_1(z)\rangle- \langle\delta \rho_1(s)\rangle\langle\delta m_1(z)\rangle\big{]}dsdz \\
  \nonumber
  C_2 =& \int \frac{\delta \bar{m}(x)}{\delta m(z)}\big{[}\langle\delta m_1(z)\delta m_1(x)\rangle- \langle\delta m_1(z)\rangle\langle\delta m_1(x)\rangle\big{]}dz  \\
  \label{eq:C_2}
  & + \int \frac{\delta \bar{m}(x)}{\delta \rho(z)}\big{[}\langle\delta \rho_1(z)\delta m_1(x)\rangle- \langle\delta \rho_1(z)\rangle\langle\delta m_1(x)\rangle\big{]}dz \\
  \nonumber
  C_3 = & \int \frac{\delta \bar{m}(x)}{\delta m(z)}\big{[}\langle\delta m_1(z)\delta \rho_1(x)\rangle- \langle\delta m_1(z)\rangle\langle\delta m_1(x)\rangle\big{]}dz \\
  \label{eq:C_3}
  & + \int \frac{\delta \bar{m}(x)}{\delta \rho(z)}\big{[}\langle\delta \rho_1(z)\delta \rho_1(x)\rangle- \langle\delta \rho_1(z)\rangle\langle\delta \rho_1(x)\rangle\big{]}dz \\
  \nonumber
  C_4 =& \int\frac{1}{2}\frac{\delta^2\bar{m}(x)}{\delta m(s)\delta m(z)}\big{[}\langle\delta m_1(s)\delta m_1(z)\rangle-\langle\delta m_1(s)\rangle\langle\delta m_1(z)\rangle\big{]}dsdz \\
  \nonumber
  & + \int\frac{1}{2}\frac{\delta^2\bar{m}(x)}{\delta \rho(s)\delta \rho(z)}\big{[}\langle\delta \rho_1(s)\delta \rho_1(z)\rangle-\langle\delta \rho_1(s)\rangle\langle\delta \rho_1(z)\rangle\big{]}dsdz \\
  \label{eq:C_4}
  & + \int \frac{\delta^2\bar{m}(x)}{\delta \rho(s)\delta m(z)}\big{[}\langle\delta \rho_1(s)\delta m_1(z)\rangle-\langle\delta \rho_1(s)\rangle\langle\delta m_1(z)\rangle\big{]}dsdz
\end{align}
\end{subequations}
To close \eref{eq:generic_correlators_app}, we need to derive the correlators involving $\delta\rho_1$ and $\delta m_1$ by using their stochastic evolution \eref{eq:delta_fields_evolution_generic}.
We readily find that $\langle\delta m_1\rangle=\langle\delta\rho_1\rangle=0$.
We now use the assumption that $\rho_0$, $m_0$ and $\bar{m}_0=\cG(x,\{\rho_0,m_0\})$ can be considered homogeneous in the dynamics of $\delta\rho_1$ and $\delta m_1$.
Under such an assumption, \eref{eq:delta_fields_evolution_generic} becomes a linear system and its dynamic in Fourier space reads
\begin{equation}
\label{eq:dynamic_delta_topo_generic}
\partial_t \begin{pmatrix}
\delta {\rho^q_1} \\
\delta {m^q_1}
\end{pmatrix} = \begin{pmatrix}
A_{11}^q & A_{12}^q \\
A_{21}^q  & A_{22}^q
\end{pmatrix}
\begin{pmatrix}
\delta \rho^q_1 \\
\delta m^q_1
\end{pmatrix} + \begin{pmatrix}
0 \\
\sqrt{2\rho_0}\;\eta^q
\end{pmatrix}\ ,
\end{equation}
where $\eta^q$ is the $q$-th Fourier mode of the Gaussian white noise, with correlations $\langle\eta^q(t)\eta^{q^{\prime}}(t^{\prime})\rangle=L^{-1}\delta_{q+q^{\prime},0}\delta(t-t^{\prime})$.
To pursue our derivation, we won't need the exact expression of the $A_{ij}^q$'s, so we just remark that they generically depend on the homogeneous fields $\rho_0$, $m_0$, $\bar{m}_0$ as well as on the Fourier transform of the functional derivatives $\delta \bar{m}(x)/\delta \rho(y)$ and $\delta \bar{m}(x)/\delta m(y)$.
The steady-state correlators of \eref{eq:dynamic_delta_topo_generic} are obtained by using It\=o calculus.
They satisfy
\begin{subequations}\label{eq:ss_corr_generic}
\begin{align}
  \label{eq:ss_corr_generic_1}
  0=&(A_{11}^{q}+A^{q^{\prime}}_{11})\langle\delta\rho^{q}_1\delta\rho^{q^{\prime}}_1\rangle_0 + A_{12}^{q}\langle\delta m^{q}_1\delta\rho^{q^{\prime}}_1\rangle_0 + A_{12}^{q^{\prime}}\langle\delta \rho^{q}_1\delta m^{q^{\prime}}_1\rangle_0 \\
  \label{eq:ss_corr_generic_2}
  0=&(A_{22}^{q}+A^{q^{\prime}}_{11})\langle\delta m^{q}_1\delta\rho^{q^{\prime}}_1\rangle_0 + A_{21}^{q}\langle\delta \rho^{q}_1\delta\rho^{q^{\prime}}_1\rangle_0 + A_{12}^{q^{\prime}}\langle\delta m^{q}_1\delta m^{q^{\prime}}_1\rangle_0 \\
  \label{eq:ss_corr_generic_3}
  0=&(A_{22}^{q}+A^{q^{\prime}}_{22})\langle\delta m^{q}_1\delta m^{q^{\prime}}_1\rangle_0 + A_{21}^{q}\langle\delta \rho^{q}_1\delta m^{q^{\prime}}_1\rangle_0 + A_{21}^{q^{\prime}}\langle\delta m^{q}_1\delta \rho^{q^{\prime}}_1\rangle_0 + \frac{2\rho_{0}}{L}\delta_{q+q^{\prime},0}\ ,
\end{align}
\end{subequations}
which can be solved as
\begin{subequations}\label{eq:fourrier_correlators_generic_app}
\begin{align}
\label{eq:fourier_m_m_correlation}
\langle\delta m^{q}_1\delta m^{q^{\prime}}_1\rangle_0 =&\langle\delta m^{q}_1\delta m^{-q}_1\rangle_0\frac{\delta_{q+q^{\prime},0}}{L}\\
\label{eq:fourier_rho_rho_correlation}
\langle\delta \rho^{q}_1\delta \rho^{q^{\prime}}_1\rangle_0 =& \langle\delta \rho^{q}_1\delta \rho^{-q}_1\rangle_0 \frac{\delta_{q+q^{\prime},0}}{L}\\
\label{eq:fourier_m_rho_correlation}
\langle\delta m^{q}_1\delta \rho^{q^{\prime}}_1\rangle_0 =& \langle\delta m^{q}_1\delta \rho^{-q}_1\rangle_0 \frac{\delta_{q+q^{\prime},0}}{L}\ ,
\end{align}
\end{subequations}
where $\langle \delta m_1^q \delta m_1^{-q}\rangle_0$, $\langle \delta \rho_1^q \delta \rho_1^{-q}\rangle_0$ and $\langle \delta m_1^q \delta \rho_1^{-q}\rangle_0$ are functions depending on the $A_{ij}^q$ whose explicit expression will not be needed.
Using our convention \eref{eq:fourier_serie}, the Fourier development of $\delta m_1$ and $\delta \rho_1$ reads
\begin{align}
  \label{eq:Fourier_dev_rho_m}
  \delta m_1 = \sum_{q} \delta m_1^q e^{iqx} \qquad \delta \rho_1 = \sum_{q} \delta \rho_1^q e^{iqx}\ .
\end{align}
Injecting this development \eref{eq:Fourier_dev_rho_m} into \eref{eq:generic_correlators_app}, and using the scalings \eref{eq:fourrier_correlators_generic_app} allow us to compute the expression of the $C_i$'s.
For $C_1$, we obtain
\begin{align}
  \label{eq:C1_q}
  C_1 =& \frac{1}{L}\sum_q \bar{m}^q_{m}\bar{m}^{-q}_{m}\langle\delta m_1^q\delta m_1^{-q}\rangle_0 +\bar{m}^q_{\rho}\bar{m}^{-q}_{\rho}\langle\delta \rho_1^q\delta \rho_1^{-q}\rangle_0 + \bar{m}^q_{m}\bar{m}^{-q}_{\rho}\langle\delta m_1^q\delta \rho_1^{-q}\rangle_0\ ,
\end{align}
where $\bar{m}_{\rho}^q$ and $\bar{m}_{m}^q$ are the Fourier transforms of the functional derivatives of $\bar{m}$ with respect to $\rho$ and $m$ respectively,
\begin{align}
  \bar{m}_{\rho}^q=\int_0^L e^{-iqz}\frac{\delta \bar{m}}{\delta \rho}(z)\ dz , \qquad \bar{m}_{m}^q=\int_0^L e^{-iqz}\frac{\delta \bar{m}}{\delta m}(z)\ dz\ .
\end{align}
For $C_2$ and $C_3$, we get
\begin{align}
  \label{eq:C2_q}
  C_2 &= \frac{1}{L}\sum_q \bar{m}^q_{m}\langle\delta m_1^q\delta m_1^{-q}\rangle_0 +\bar{m}^{-q}_{\rho}\langle\delta m_1^{q}\delta \rho_1^{-q}\rangle_0 \\
  \label{eq:C3_q}
  C_3 &= \frac{1}{L}\sum_q \bar{m}^q_{\rho}\langle\delta \rho_1^q\delta \rho_1^{-q}\rangle_0 +\bar{m}^{q}_{m}\langle\delta m_1^{q}\delta \rho_1^{-q}\rangle_0\ .
\end{align}
Finally, $C_4$ reads
\begin{align}
  \label{eq:C4_q}
  C_4 =&  \frac{1}{2L}\sum_q \bar{m}^{\; q,-q}_{m,m}\langle\delta m_1^q\delta m_1^{-q}\rangle_0 +\bar{m}^{\; q,-q}_{\rho,\rho}\langle\delta \rho_1^q\delta \rho_1^{-q}\rangle_0 + 2\bar{m}^{\; q,-q}_{m,\rho}\langle\delta m_1^q\delta \rho_1^{-q}\rangle_0\ ,
\end{align}
where $\bar{m}^{\; q,q^{\prime}}_{m,m}$, $\bar{m}^{\; q,q^{\prime}}_{\rho,\rho}$ and $\bar{m}^{\; q,q^{\prime}}_{m,\rho}$ are the Fourier transform of the second functional derivatives of $\bar{m}$ with respect to $\rho$ and $m$
\begin{align}
  \nonumber
  \bar{m}^{\; q,q^{\prime}}_{m,m} =& \int_0^L\int_0^L  e^{-iqs-iq^{\prime}z}\frac{\delta^2\bar{m}}{\delta m \delta m}(s,z)\ ds dz\;, & \bar{m}^{\; q,q^{\prime}}_{\rho,\rho} =& \int_0^L\int_0^L  e^{-iqs-iq^{\prime}z}\frac{\delta^2\bar{m}}{\delta \rho \delta \rho}(s,z)\ ds dz  \\
  \bar{m}^{\; q,q^{\prime}}_{m,\rho} =& \int_0^L\int_0^L  e^{-iqs-iq^{\prime}z}\frac{\delta^2\bar{m}}{\delta m \delta \rho}(s,z)\ ds dz\;. & &
\end{align}
The last step is to take the large system size limit $L\rightarrow\infty$ in \eref{eq:C1_q}-\eref{eq:C2_q}-\eref{eq:C3_q}-\eref{eq:C4_q} to obtain the $C_i$'s as integrals over Fourier modes $q$
\begin{subequations}\label{eq:Ci_final}
\begin{align}
  \label{eq:C1_final}
  C_1 =& \int \frac{dq}{2\pi} |\bar{m}^q_{m}|^2\langle\delta m_1^q\delta m_1^{-q}\rangle_0 +|\bar{m}^q_{\rho}|^2\langle\delta \rho_1^q\delta \rho_1^{-q}\rangle_0 +\bar{m}^q_{m}\bar{m}^{-q}_{\rho}\langle\delta m_1^q\delta \rho_1^{-q}\rangle_0 \\
  \label{eq:C2_final}
  C_2 =& \int \frac{dq}{2\pi} \bar{m}^q_{m}\langle\delta m_1^q\delta m_1^{-q}\rangle_0 +\bar{m}^{-q}_{\rho}\langle\delta m_1^{q}\delta \rho_1^{-q}\rangle_0 \\
  \label{eq:C3_final}
  C_3 =& \int \frac{dq}{2\pi} \bar{m}^q_{\rho}\langle\delta \rho_1^q\delta \rho_1^{-q}\rangle_0 +\bar{m}^{q}_{m}\langle\delta m_1^{q}\delta \rho_1^{-q}\rangle_0 \\
  \label{eq:C4_final}
  C_4 =&  \int \frac{dq}{4\pi}\bar{m}^{\; q,-q}_{m,m}\langle\delta m_1^q\delta m_1^{-q}\rangle_0 +\bar{m}^{\; q,-q}_{\rho,\rho}\langle\delta \rho_1^q\delta \rho_1^{-q}\rangle_0 + 2\bar{m}^{\; q,-q}_{m,\rho}\langle\delta m_1^q\delta \rho_1^{-q}\rangle_0\;.
\end{align}
\end{subequations}
So far, the correction $\Delta\cF$ in \eref{eq:delta_Ft} together with the expression of the $C_i$'s in \eref{eq:Ci_final} is valid for any aligning field $\bar{m}=\cG(x,\{\rho,m\})$ where $\cG$ is a functional.
We now apply this result for the specific case of $k$-nearest neighbors alignment for which $\bar{m}$ given by \eref{eq:k_nearest_field_app}.
For homogeneous $\rho_0$ and $m_0$, we can compute the functional derivatives of $\bar{m}=\int^{x+y}_{x-y}m(z)dz/k$ as
\begin{subequations}\label{eq:func_deriv_knearest}
\begin{align}
  \frac{\delta \bar{m}(x)}{\delta m(z)} =& \frac{\Theta(x+y_0-z)\Theta(z-x+y_0)}{2y_0\rho_0}\;, \qquad \quad\ \ \frac{\delta^2 \bar{m}(x)}{\delta m(z) \delta m(s)} = 0\;, \\
  \frac{\delta \bar{m}(x)}{\delta \rho(z)} =& -\frac{m_0\Theta(x+y_0-z)\Theta(z-x+y_0)}{2y_0\rho_0^2}\;, \qquad
  \frac{\delta^2 \bar{m}(x)}{\delta \rho(z) \delta \rho(s)} = 0\;, \\
  \frac{\delta^2 \bar{m}(x)}{\delta \rho(z) \delta m(s)} =& -\frac{\Theta(x+y_0-z)\Theta(z-x+y_0)}{4y_0\rho_0^2}\big{[}\delta(s-x-y_0)+\delta(s-x+y_0)\big{]}\;,
\end{align}
\end{subequations}
where $\Theta(u)$ is the Heaviside function which is equal to $0$ if $u<0$ or equal to $1$ if $u>0$.
Going into Fourier space, we obtain
\begin{subequations}\label{eq:func_der_knearest}
\begin{align}
  \label{eq:func_der_m}
  \bar{m}^{q}_{m} &= -\frac{\sinc(qy_0)}{\rho_0}\;, &
  \bar{m}^{q}_{\rho} &= \frac{m_0}{\rho_0}\sinc(qy_0)\;, \\
  \label{eq:func_der_m_rho}
  \bar{m}^{\;q,q^{\prime}}_{\rho,\rho} &= \bar{m}^{\;q,q^{\prime}}_{m,m} = 0\;, & \bar{m}^{\;q,q^{\prime}}_{\rho,m} &= \frac{\sinc(qy_0)}{\rho_0}\cos(q^{\prime}y_0)\; .
\end{align}
\end{subequations}
To compute the $C_i$'s of \eref{eq:generic_correlators_app}, we still need the expression of the correlators of the fields $\delta\rho_1$ and $\delta m_1$.
For the specific case of $k$-nearest neighbors alignment \eref{eq:k_nearest_field_app} at hand, the matrix coefficients $A_{ij}^q$ of the linearized dynamics \eref{eq:dynamic_delta_topo_generic} are equal to the $M_{ij}^q$ defined in \eref{eq:mat_coeff_knearest} and we get
\begin{equation}
\label{eq:dynamic_delta_topo}
\partial_t \begin{pmatrix}
\delta {\rho^q_1} \\
\delta {m^q_1}
\end{pmatrix} = \begin{pmatrix}
M_{11}^q & M_{12}^q \\
M_{21}^q  & M_{22}^q
\end{pmatrix}
\begin{pmatrix}
\delta \rho^q_1 \\
\delta m^q_1
\end{pmatrix} + \begin{pmatrix}
0 \\
\sqrt{2\rho_0}\;\eta^q
\end{pmatrix}\; .
\end{equation}
Using the system \eref{eq:ss_corr_generic} with $A_{ij}^q=M_{ij}^q$, we obtain the expressions of the steady-state correlators in Fourier space to first order in $m_0$
\begin{subequations}\label{eq:corr_knearest}
\begin{align}
\label{eq:m_m_correlation_topo}
\langle\delta m^{q}_1\delta m^{-q}_1\rangle_0 =&\rho_{0}\frac{2D\Gamma-2D\Gamma\beta \sinc(qy_{0})+2D^{2}q^{2}+v^{2}}{2(\Gamma-\Gamma\beta \sinc(qy_{0})+Dq^{2})\left(2D\Gamma-2D\Gamma\beta \sinc(qy_{0})+D^{2}q^{2}+v^{2}\right)}+\cO\left(m_0^2\right)\\
\label{eq:rho_rho_correlation_topo}
\langle\delta \rho^{q}_1\delta \rho^{-q}_1\rangle_0 =&\frac{\rho_{0}v^{2}}{2(\Gamma-\Gamma\beta \sinc(qy_{0})+Dq^{2})\left(2D\Gamma-2D\Gamma\beta \sinc(qy_{0})+D^{2}q^{2}+v^{2}\right)} +\cO\left(m_0^2\right) \\
\nonumber
\langle\delta m^{q}_1\delta \rho^{-q}_1\rangle_0 =&\frac{iqDv\rho_0}{2(\Gamma-\Gamma\beta \sinc(qy_{0})+Dq^{2})\left(2D\Gamma-2D\Gamma\beta \sinc(qy_{0})+D^{2}q^{2}+v^{2}\right)} \\
\label{eq:m_rho_correlation_topo}
& + \frac{\beta v^2 \Gamma\left(1-\sinc\left(q y_0\right)\right)m_0}{2 \left(\Gamma -\Gamma\beta  \sinc \left(q y_0\right) + D q^2 \right){}^2 \left( 2D\Gamma-2D\Gamma\beta \sinc \left(q y_0\right)+D^2 q^2+v^2\right)} + \cO\left(m_0^2\right)\ .
\end{align}
\end{subequations}
We now have all the ingredients needed to compute $\Delta\cF$ in \eref{eq:delta_Ft}.
First, we have to inject the correlators \eref{eq:corr_knearest} and the functional derivatives \eref{eq:func_der_knearest} into the expressions of the $C_i$'s in \eref{eq:generic_correlators_app}.
Once we have the $C_i$'s, we just plug them back into the expression \eref{eq:delta_Ft} for $\Delta\cF$.
A lengthy but straightforward computation then gives
\begin{equation}
\label{eq:final_landau_topol}
\Delta \cF = \frac{m_0}{y_0\rho_0}\left[2\beta(c_1-c_4)+\beta^2(1-\beta)c_2 +2\beta^2 c_3\right] + \cO(m_0^2)\ .
\end{equation}
Where $c_1$, $c_2$, $c_3$, and $c_4$ are given by
\begin{subequations}
\begin{align}
  \label{eq:c1_topo_app}
c_1 =& \int\frac{d\tilde{q}}{2\pi}\frac{v^{2}\sinc(\tilde{q})}{2\left(1-\beta \sinc(\tilde{q})+\frac{D}{\Gamma y_0^2}\tilde{q}^{2}\right)\left(2D\Gamma-2D\Gamma\beta \sinc(\tilde{q})+D^{2}\frac{\tilde{q}^{2}}{y_0^2}+v^{2}\right)} \\
c_2 =& \int\frac{d\tilde{q}}{2\pi}\frac{\sinc^2(\tilde{q})\left[2D\Gamma-2D\Gamma\beta \sinc(\tilde{q})+2D^{2}\frac{\tilde{q}^{2}}{y_0^2}+v^{2}\right]}{2\left(1-\beta \sinc(\tilde{q})+\frac{D}{\Gamma y_0^2}\tilde{q}^{2}\right)\left(2D\Gamma-2D\Gamma\beta \sinc(\tilde{q})+D^{2}\frac{\tilde{q}^{2}}{y_0^2}+v^{2}\right)} \\
c_3 =& \int\frac{d\tilde{q}}{2\pi}\frac{\sinc(\tilde{q})\left[2D\Gamma-2D\Gamma\beta \sinc(\tilde{q})+2D^{2}\frac{\tilde{q}^{2}}{y_0^2}+v^{2}\right]}{2\left(1-\beta \sinc(\tilde{q})+\frac{D}{\Gamma y_0^2}\tilde{q}^{2}\right)\left(2D\Gamma-2D\Gamma\beta \sinc(\tilde{q})+D^{2}\frac{\tilde{q}^{2}}{y_0^2}+v^{2}\right)} \\
c_4 =& \int\frac{d\tilde{q}}{2\pi}\frac{\beta v^2 \left(\Gamma-\Gamma\sinc\left(\tilde{q}\right)\right)\sinc(\tilde{q})\left(1-\cos(\tilde{q})\right)}{2 \left(1 -\beta  \sinc \left(\tilde{q}\right)+ \frac{D}{\Gamma y_0^2} \tilde{q}^{2} \right)^2 \left( 2D\Gamma- 2D\Gamma\beta \sinc \left(\tilde{q}\right) + D^2 \frac{\tilde{q}^{2}}{y_0^2}+v^2\right)} \ .
\end{align}
\end{subequations}
Note that all the $c_{i}$'s are dimensionless and only depends on three independent parameters: $\beta$, $\Gamma D/v^2$, and $\Gamma k / (v \rho_0)$.
Consequently, they all assume the scaling form
\begin{align}
  \label{eq:scaling_forms_ci}
  c_i = \bar{c}_i\left(\beta,\tfrac{\Gamma k}{ v\rho_0},\tfrac{\Gamma D}{v^2}\right)\;.
\end{align}
For example, we deduce from \eref{eq:c1_topo_app} the expression for $\bar{c}_1$
\begin{align}
  \bar{c}_1(\mu,\nu,\omega)=\int \frac{du}{2\pi}\frac{\sinc(u)}{\left(1-\mu\sinc(u)+4u^2\tfrac{\omega^2}{\nu^2}\right)\left(1+2\omega-2\omega\mu\sinc(u)+4u^2\tfrac{\omega^2}{\nu^2}\right)}\ .
\end{align}
Using the scaling forms of \eref{eq:scaling_forms_ci} in \eqref{eq:final_landau_topol}, $\Delta\cF$ can be cast into
\begin{equation}
\Delta\cF = \frac{2 m_0}{k}g\left(\beta,\frac{\Gamma k}{v\rho_0 },\frac{\Gamma D}{v^2}\right) + \cO(m_0^2)\ .
\end{equation}
where $g$ is given by
\begin{align}
  \label{eq:g_app}
g\left(\beta,\frac{\Gamma k}{v\rho_0},\frac{\Gamma D}{v^2}\right)=2\beta (\bar{c}_1-\bar{c}_4)+\beta^2(1-\beta )\bar{c}_2 + 2\beta^2 \bar{c}_3\ .
\end{align}
Note that the convergence of the integrals $c_1$, $c_2$, $c_3$ and $c_4$ is ensured
in the high temperature phase when $\beta < 1$.
\par
Importantly, the dependence of $g$ on $\rho_0$ cannot be eliminated, so that the renormalization of the linear mass indeed leads to a density-dependent onset of order.
Using the fact that $\beta<1$, $c_1>c_4$, and that $c_1$, $c_2$, $c_3$ are positive, we deduce that $g$ is a density-dependent positive function in the high temperature phase.
Consequently, the incorporation of microscopic noise should lower the critical temperature.

\subsection{Extension to generic alignment}
\label{app:generic_alignment}

This appendix is devoted to the computation of $\Delta\cF$ in
\eref{eq:rho_topological_general_ren} for a generic functional
alignment $\bar{m}$ as in \eref{eq:functional_topol_m}. To limit
repetitions, this appendix borrows many aspects of the methodology
presented in appendix~\ref{sec:renormalization_generic}, whose prior
reading is thus recommended.  \par As shown in \eref{eq:delta_Ft} of
appendix \ref{sec:renormalization_generic}, $\Delta\cF$ crucially
depends on the functional derivatives of $\bar{m}$ with respect to
$\rho$ and $m$ through the $C_i$'s whose expressions, given in
\eref{eq:generic_correlators_app}, hold for any field $\bar m$.  Given
$\delta \bar{m}/\delta \rho$ and $\delta \bar{m}/\delta m$, we can
derive the matrix coefficients $A_{ij}^q$ of
\eref{eq:dynamic_delta_topo_generic}, which in turn yields the
correlators $\langle \delta \rho_1^q \delta \rho_1^{-q} \rangle$,
$\langle \delta \rho_1^q \delta m_1^{-q} \rangle$ and $\langle \delta
m_1^q \delta m_1^{-q} \rangle$ in
\eref{eq:fourrier_correlators_generic_app}.  These correlators,
together with the second functional derivatives of $\bar{m}$, allow us
to compute the $C_i$'s through the use of \eref{eq:Ci_final}.  Thus,
if the functional derivatives of $\bar{m}$ are constrained to assume
some scaling law, such a scaling will also be reflected in
$\Delta\cF$.  \par
The first and second functional derivatives of $\bar{m}$ at $\rho_0$, $m_0$ are defined as
\begin{align}
  \nonumber
  \bar{m}(x) =& \cG(x,\{\rho_0,m_0\}) + \int_0^L \left[ \frac{\delta \bar{m}(x)}{\delta \rho(z)}\delta\rho(z)+ \frac{\delta \bar{m}(x)}{\delta m(z)} \delta m(z)\right] dz \\
  \nonumber
  & + \int_0^L\int_0^L \frac{1}{2}\left[\frac{\delta^2 \bar{m}(x)}{\delta \rho(s) \delta \rho(z)}  \delta \rho(s) \delta \rho(z)
  + \frac{\delta^2 \bar{m}(x)}{\delta m(s) \delta m(z)}  \delta m(s) \delta m(z)\right]dsdz \\
  & + \int_0^L\int_0^L \frac{\delta^2 \bar{m}(x)}{\delta \rho(s) \delta m(z)}  \delta \rho(s) \delta m(z)dsdz
  \label{eq:func_der_def} + o(\delta \rho^2,\delta m^2,\delta \rho \delta m)\ .
\end{align}
Because $\bar{m}$ is dimensionless, we read on the above equation that all its functional derivatives are dimensionless as well.
It entails that, in Fourier space, the first order functional derivatives of $\bar{m}$ scale as lengths $[L]$ while the second order ones scale as lengths squared $[L^2]$:
\begin{subequations}\label{eq:scaling_func_der_generic}
\begin{align}
  \label{eq:m_bar_q_def_1}
  \bar{m}_{\rho}^q=&\int_0^L e^{-iqz}\frac{\delta \bar{m}}{\delta \rho}(z)\ dz \propto [L] , \qquad \bar{m}_{m}^q=\int_0^L e^{-iqz}\frac{\delta \bar{m}}{\delta m}(z)\ dz \propto [L] \\
  \label{eq:m_bar_q_def_2}
  \bar{m}^{\; q,q^{\prime}}_{m,m} =& \int_0^L\int_0^L  e^{-iqs-iq^{\prime}z}\frac{\delta^2\bar{m}}{\delta m \delta m}(s,z)\ ds dz \propto [L]^2\\
  \label{eq:m_bar_q_def_3}
  \bar{m}^{\; q,q^{\prime}}_{\rho,\rho} =& \int_0^L\int_0^L  e^{-iqs-iq^{\prime}z}\frac{\delta^2\bar{m}}{\delta \rho \delta \rho}(s,z)\ ds dz \propto [L]^2 \\
  \label{eq:m_bar_q_def_4}
  \bar{m}^{\; q,q^{\prime}}_{m,\rho} =& \int_0^L\int_0^L  e^{-iqs-iq^{\prime}z}\frac{\delta^2\bar{m}}{\delta m \delta \rho}(s,z)\ ds dz \propto [L]^2 \ .
\end{align}
\end{subequations}
We expect $\bar{m}_\rho^q$, $\bar{m}_m^q$, $\bar{m}^{\; q,-q}_{\rho,\rho}$, $\bar{m}^{\; q,-q}_{m,m}$, $\bar{m}^{\; q,-q}_{m,\rho}$ to depend on $\rho_0$, $m_0$ and on the wavevector $q$.
As in the previous appendices, we suppose that $\rho_0$, $m_0$, and $\bar{m}_0=\cG(x,\{m_0,\rho_0\})$ remain homogeneous in space throughout the derivation.
As $\bar{m}$ is dimensionless, $\bar{m}_0$ will depend on the ratio $m_0/\rho_0$ by dimensional analysis.
Taking into account all these dependences, $\bar{m}_\rho^q$, $\bar{m}_m^q$, $\bar{m}^{\; q,-q}_{\rho,\rho}$, $\bar{m}^{\; q,-q}_{m,m}$, $\bar{m}^{\; q,-q}_{m,\rho}$, and $\bar{m}_0$ will have the following scaling forms
\begin{align}
  \nonumber
  \bar{m}^q_{\rho} =& \frac{1}{\rho_0}\bar{f}_{\rho}\left(\frac{m_0}{\rho_0},\frac{q}{\rho_0},\dots\right) & \bar{m}^q_{m} =& \frac{1}{\rho_0}\bar{f}_{m}\left(\frac{m_0}{\rho_0},\frac{q}{\rho_0},\dots\right) \\
  \nonumber
  \bar{m}^{\; q,-q}_{\rho,\rho} =& \frac{1}{\rho_0^2}\bar{f}_{\rho\rho}\left(\frac{m_0}{\rho_0},\frac{q}{\rho_0},\dots\right) &  \bar{m}^{\; q,-q}_{m,m} =& \frac{1}{\rho_0^2}\bar{f}_{mm}\left(\frac{m_0}{\rho_0},\frac{q}{\rho_0},\dots\right) \\
  \label{eq:adim_f}
  \bar{m}^{\; q,-q}_{m,\rho} =& \frac{1}{\rho_0^2}\bar{f}_{m\rho}\left(\frac{m_0}{\rho_0},\frac{q}{\rho_0},\dots\right) & \bar{m}_0 =& \bar{f}\left(\frac{m_0}{\rho_0},\dots\right)\ ,
\end{align}
where the dots refer to other dimensionless parameters entering in the definition of $\bar{m}$ (e.g., the number $k$ of nearest neighbours) and are omitted for clarity from now on.
Using \eref{eq:adim_f}, the $A_{ij}^q$ involved in the linearized hydrodynamics \eref{eq:dynamic_delta_topo_generic} for $\delta \rho_1$ and $\delta m_1$ are generically given by
\begin{subequations}\label{eq:A_generic_scaling}
\begin{align}
\label{eq:A11}
A_{11}^q =& -Dq^2\,, \qquad A_{12}^q = -ivq \\
\nonumber
A_{21}^q =& -iqv 
+2\Gamma\left[\beta\left(1+\frac{\beta^2}{2}\bar{f}\left(\tfrac{m_0}{\rho_0}\right)\right)- \tfrac{m_0}{\rho_0} \beta^2\bar{f}\left(\tfrac{m_0}{\rho_0}\right)\right]\bar f_{\rho}\left(\tfrac{m_0}{\rho_0},\tfrac{q}{\rho_0}\right)\\
\label{eq:A21}
&+2\Gamma\beta\bar{f}\left(\tfrac{m_0}{\rho_0}\right)\left[1+\frac{\beta^2}{6}\bar{f}\left(\tfrac{m_0}{\rho_0}\right)^2\right]\\
\nonumber
A_{22}^q =& - Dq^2
+ 2\Gamma\left[\beta \left(1+\frac{\beta^2}{2}\bar{f}\left(\tfrac{m_0}{\rho_0}\right)\right)- \tfrac{m_0}{\rho_0} \beta^2\bar{f}\left(\tfrac{m_0}{\rho_0}\right)\right]\bar f_{m}\left(\tfrac{m_0}{\rho_0},\tfrac{q}{\rho_0}\right) \\ 
\label{eq:A22}
& -2\Gamma\left(1+\frac{\beta^2}{2}\bar{f}\left(\tfrac{m_0}{\rho_0}\right)^2\right)
\end{align}
\end{subequations}
From \eref{eq:A_generic_scaling}, we observe that the $A_{ij}^q$'s all follow the generic scaling form
\begin{align}
  \label{eq:Aij_scalings}
  A_{ij}^q = \Gamma\; \bar{g}_{ij}\left(\frac{\Gamma D}{v^2}, \frac{\Gamma}{v\rho_0},\frac{q}{\rho_0},\frac{m_0}{\rho_0},\beta\right)
\end{align}
Because the system \eref{eq:ss_corr_generic} only involves the $A_{ij}^q$'s and $\rho_0$ as coefficients, its solution will assume a scaling form similar to \eref{eq:Aij_scalings}: a prefactor depending on $\rho_0$ and $\Gamma$ multiplied by a function depending on the dimensionless variables entering in $\bar{g}_{ij}$.
Noting that $\langle \delta m_1^q \delta m_1^{-q} \rangle$, $\langle \delta \rho_1^q \delta \rho_1^{-q} \rangle$ and $\langle \delta m_1^q \delta \rho_1^{-q} \rangle$ in \eref{eq:fourrier_correlators_generic_app} scale as $[T]/[L]$, this prefactor must be $\rho_0/\Gamma$ and we finally obtain
\begin{subequations}\label{eq:corr_scale_generic}
\begin{align}
  \label{eq:corr_scal_1}
  \langle \delta \rho_1^{q} \delta \rho_1^{-q}\rangle =& \frac{\rho_0}{\Gamma}\bar{g}_1\left(\frac{\Gamma D}{v^2}, \frac{\Gamma}{v\rho_0},\frac{q}{\rho_0},\frac{m_0}{\rho_0},\beta\right)\,, &
  \langle \delta m_1^{q} \delta m_1^{-q}\rangle =& \frac{\rho_0}{\Gamma} \bar{g}_2\left(\frac{\Gamma D}{v^2}, \frac{\Gamma}{v\rho_0},\frac{q}{\rho_0},\frac{m_0}{\rho_0},\beta\right) \\
  \label{eq:corr_scal_3}
  \langle \delta m_1^{q} \delta \rho_1^{-q}\rangle =& \frac{\rho_0}{\Gamma} \bar{g}_3\left(\frac{\Gamma D}{v^2}, \frac{\Gamma}{v\rho_0},\frac{q}{\rho_0},\frac{m_0}{\rho_0},\beta\right) \ . & &
\end{align}
\end{subequations}
Injecting \eref{eq:corr_scale_generic} together with \eref{eq:adim_f} into \eref{eq:Ci_final} yields the $C_i$'s as
\begin{subequations}\label{eq:Ci_scalings_generic}
\begin{align}
  \label{eq:C1_scalings}
  C_1 =& \int \frac{dq}{2\pi} \frac{1}{\rho_0\Gamma}\bar{h}_1\left(\frac{\Gamma D}{v^2}, \frac{\Gamma}{v\rho_0},\frac{q}{\rho_0},\frac{m_0}{\rho_0},\beta\right) &
  C_2 =& \int \frac{dq}{2\pi} \frac{1}{\Gamma}\bar{h}_2\left(\frac{\Gamma D}{v^2}, \frac{\Gamma}{v\rho_0},\frac{q}{\rho_0},\frac{m_0}{\rho_0},\beta\right) \\
  \label{eq:C3_scalings}
  C_3 =& \int \frac{dq}{2\pi} \frac{1}{\Gamma}\bar{h}_3\left(\frac{\Gamma D}{v^2}, \frac{\Gamma}{v\rho_0},\frac{q}{\rho_0},\frac{m_0}{\rho_0},\beta\right) &
  C_4 =&  \int \frac{dq}{2\pi}\frac{1}{\rho_0\Gamma}\bar{h}_4\left(\frac{\Gamma D}{v^2}, \frac{\Gamma}{v\rho_0},\frac{q}{\rho_0},\frac{m_0}{\rho_0},\beta\right)\ ,
\end{align}
\end{subequations}
where the $\bar{h}_i$'s are given by
\begin{align}
  \bar{h}_1 =& |\bar{f}_m|^2\;\bar{g}_2 + |\bar{f}_\rho|^2\;\bar{g}_1 + \bar{f}_m \bar{f}^{\star}_\rho \;\bar{g}_3\;, &
  \bar{h}_2 =& \bar{f}_m\;\bar{g}_2 + \bar{f}_\rho^{\star}\;\bar{g}_3\;, \\
  \bar{h}_3 =& \bar{f}_\rho\;\bar{g}_1 + \bar{f}_m\;\bar{g}_3\;, &
  \bar{h}_4 =& \bar{f}_{mm}\;\bar{g}_2 + \bar{f}_{\rho\rho}\;\bar{g}_1 + \bar{f}_{m\rho}\;\bar{g}_3\ .
\end{align}
Making the integrals dimensionless in \eref{eq:Ci_scalings_generic} by changing variable to $\tilde{q} = q/\rho_0$, we obtain the final scaling form of the $C_i$'s as
\begin{subequations}\label{eq:Ci_final_scalings_generic}
\begin{align}
  \label{eq:C1_final_scalings}
  C_1 =& \frac{1}{\Gamma} \bar{H}_1\left(\frac{\Gamma D}{v^2}, \frac{\Gamma}{v\rho_0},\frac{m_0}{\rho_0},\beta\right)\;, &
  C_2 =& \frac{\rho_0}{\Gamma}\bar{H}_2\left(\frac{\Gamma D}{v^2}, \frac{\Gamma}{v\rho_0},\frac{m_0}{\rho_0},\beta\right)\;, \\
  \label{eq:C3_final_scalings}
  C_3 =& \frac{\rho_0}{\Gamma} \bar{H}_3\left(\frac{\Gamma D}{v^2}, \frac{\Gamma}{v\rho_0},\frac{m_0}{\rho_0},\beta\right)\;, &
  C_4 =& \frac{1}{\Gamma} \bar{H}_4\left(\frac{\Gamma D}{v^2}, \frac{\Gamma}{v\rho_0},\frac{m_0}{\rho_0},\beta\right)\ .
\end{align}
\end{subequations}
Finally plugging the above scalings into the expression of $\Delta\cF$ \eref{eq:delta_Ft}, we obtain
\begin{align}
  \nonumber
  \Delta\cF =& \left(\beta^2 m_0- \beta^3\rho_0\bar{m}_0\right)\bar{H}_1 + 2\beta^2\bar{m}_0\rho_0\bar{H}_2 - \left(2\beta+\beta^3\bar{m}_0^2\right)\rho_0\bar{H}_3 + \left(2\beta^2\bar{m}_0m_0-\beta^3\bar{m}_0^2\rho_0-2\beta\rho_0\right)\bar{H}_4\ ,
\end{align}
which can be cast into the following scaling form
\begin{align}
  \Delta \cF = \rho_0 \bar{\cF}_1\left(\frac{\Gamma D}{v^2}, \frac{\Gamma}{v\rho_0},\frac{m_0}{\rho_0},\beta\right) + m_0 \bar{\cF}_2\left(\frac{\Gamma D}{v^2}, \frac{\Gamma}{v\rho_0},\frac{m_0}{\rho_0},\beta\right)\ .
\end{align}
As we are interested in the renormalization of the linear Landau term, we can taylor expand $\Delta\cF$ up to order $m_0$
\begin{align}
  \label{eq:delta_Ft_scaling_2}
  \Delta \cF =& \rho_0 \bar{\cF}_1\left(\frac{\Gamma D}{v^2}, \frac{\Gamma}{v\rho_0},0,\beta\right) + m_0\; \bar{\cF}_1^{\;(0,0,1,0)}\left(\frac{\Gamma D}{v^2}, \frac{\Gamma}{v\rho_0},0,\beta\right)
  + m_0\;\bar{\cF}_2\left(\frac{\Gamma D}{v^2}, \frac{\Gamma}{v\rho_0},0,\beta\right) + \cO(m_0^2)\;,
\end{align}
The first term on the right hand side of \eref{eq:delta_Ft_scaling_2} is unphysical as it breaks the parity $m_0 \rightarrow -m_0$ in the Landau potential.
Its presence inconsistently implies that a nonzero magnetization would subsist even in the very high temperature regime: we thus set this term to zero and obtain the final scaling form of $\Delta\cF$ as
\begin{align}
  \label{eq:delta_Ft_scaling_final}
  \Delta \cF = &\; m_0\; \bar{\cF}\left(\frac{\Gamma D}{v^2}, \frac{\Gamma}{v\rho_0},\beta\right) + \cO(m_0^2)\ ,
\end{align}
where $\bar{\cF}(u,v,w)=\bar{\cF}_1^{\;(0,0,1,0)}(u,v,0,w)+\bar{\cF}_2(u,v,0,w)$.
Without changing the zero-th order in the noise strength $\sigma$, we can replace $m_0$ and $\rho_0$ by $\tilde{m}=\langle m\rangle$ and $\tilde{\rho}=\langle\rho\rangle$ in \eref{eq:delta_Ft_scaling_final}.
This substitution gives back \eref{eq:delta_Ft_scaling_final_main} of main text:
\begin{align}
  \label{eq:delta_Ft_scaling_final_2}
  \Delta \cF = &\; \tilde{m}\; \bar{\cF}\left(\frac{\Gamma D}{v^2}, \frac{\Gamma}{v\tilde{\rho}},\beta\right) + \cO(\tilde{m}^2)\ ,
\end{align}

\section{Renormalization for fully connected alignment}
\label{app:renormalization_fully_connected}

This appendix is devoted to the computation of $\Delta\cF$ in \eref{eq:rho_topological_general_ren} for a fully connected AIM.
It is not self-contained and as such we advise a previous reading of appendix \ref{sec:renormalization_generic} and \ref{app:generic_alignment}.
We first recall the definition \eref{eq:bar_m_FC_main} of the alignment $\bar{m}$ for the fully connected AIM in $1$D
\begin{equation}
  \label{eq:bar_m_FC}
  \bar m = \frac{\int_0^L  m(z) dz}{N};\qquad\text{where}\qquad N=\int_0^L \rho(z) dz\;.
\end{equation}
For this specific choice of $\bar{m}$, the following simplifications occur in the Fourier transforms of the functional derivatives defined in \eref{eq:scaling_func_der_generic}
\begin{align}
  \label{eq:func_der_FC}
  \bar{m}_0=&\frac{m_0}{\rho_0}\ ,\quad
  \bar{m}^q_m=\frac{1}{\rho_0} \delta_{q,0}\ , \qquad
  \bar{m}^{q}_{\rho}= \bar{m}^{\;q,q^{\prime}}_{\rho,\rho}=\bar{m}^{\;q,q^{\prime}}_{m,m}=\bar{m}^{\;q,q^{\prime}}_{m,m}=0 \ .
\end{align}
Using the discrete expressions for the $C_i$'s in \eref{eq:C1_q}, \eref{eq:C2_q}, \eref{eq:C3_q} and \eref{eq:C4_q} together with \eref{eq:func_der_FC}, we obtain for the fully-connected version
\begin{align}
  \label{eq:ci_fully_connected}
  C_1= \frac{1}{N\rho_0}\langle\delta m_1^{0}\delta m_1^{-0}\rangle ,\
  C_2= \frac{1}{N}\langle\delta m_1^{0}\delta m_1^{-0}\rangle ,\
  C_3= \frac{1}{N}\langle\delta m_1^{0}\delta \rho_1^{-0}\rangle,\  C_4=0\ .
\end{align}
Note that to obtain \eref{eq:ci_fully_connected}, we have used that $N=L\rho_0$.
As mass is conserved in the system, we have that $\delta \rho_1^0\propto\int_0^L\delta\rho_1(z)dz=0$. Using \eref{eq:fourier_m_rho_correlation}, it first entails that $\langle\delta m_1^0\delta\rho_1^{-0}\rangle=0$ and we thus further simplify the $C_i$'s as
\begin{align}
  \label{eq:C_i_FC_final}
  C_1= \frac{1}{N\rho_0}\langle\delta m_1^{0}\delta m_1^{-0}\rangle ,\
  C_2= \frac{1}{N}\langle\delta m_1^{0}\delta m_1^{-0}\rangle ,\
  C_3=C_4=0\ .
\end{align}
Mass conservation also entails, using \eref{eq:dynamic_delta_topo_generic}, that $\delta m_1^0$ evolves according to the following Langevin equation
\begin{equation}
  \label{eq:delta_m_1_0}
  \dot{\delta m_1^0} = A_{22}^{0}\delta m_1^0 + \sqrt{2\rho_0}\;\eta^0\ ,
\end{equation}
where $\eta^0$ is a Gaussian white noise such that $\langle \eta^0(t) \eta^0(t^{\prime})\rangle=L^{-1}\delta(t-t^{\prime})$ and $A_{22}^0$ is given by \eref{eq:A22} as
\begin{align}
  A_{22}^0 =& -2\Gamma\left(1+\frac{\beta^2}{2}\frac{m_0^2}{\rho_0^2}\right)+ 2\Gamma\left[\beta \left(1+\frac{\beta^2}{2}\frac{m_0}{\rho_0}\right)-  \beta^2\frac{m_0^2}{\rho_0^2}\right]\ .
\end{align}
Using It\=o calculus on \eref{eq:delta_m_1_0}, and taking care of the factor $L^{-1}$ in \eref{eq:fourier_m_m_correlation}, we obtain
\begin{align}
  \langle\delta m_1^{0}\delta m_1^{-0}\rangle = \frac{\rho_0}{2\Gamma\left[1+\frac{\beta^2}{2}\frac{m_0^2}{\rho_0^2}- \beta \left(1+\frac{\beta^2}{2}\frac{m_0}{\rho_0}\right)+  \beta^2\frac{m_0^3}{\rho_0^3}\right]}\ .
\end{align}
Expanding $\langle \delta m_1^{0}\delta m_1^{-0} \rangle$ to first order in $m_0$, we obtain
\begin{align}
  \langle \delta m_1^{0}\delta m_1^{-0} \rangle = \frac{\rho_0}{2\Gamma(1-\beta)}\left(1+\frac{\beta^2}{2(1-\beta)}\frac{m_0}{\rho_0}+\cO(m_0^2)\right)
\end{align}
Inserting the above expansion into \eref{eq:C_i_FC_final}, and then injecting the resulting $C_i$'s into \eref{eq:delta_Ft} yields $\Delta\cF$ for the fully connected model
\begin{align}
  \label{eq:delta_Ft_FC_app}
  \Delta\cF =\frac{m_0}{N}\left(\frac{\beta^2}{2}+\frac{\beta^2}{1-\beta}\right) + \cO(m_0^2) \ .
\end{align}
Without changing the leading order in the noise strength $\sigma$ of $\Delta\cF$, we can replace $m_0$ and $\rho_0$ by $\tilde{m}=\langle m \rangle$ and $\tilde{\rho}=\langle \rho\rangle$ in \eref{eq:delta_Ft_FC_app}.
This substitution gives back \eref{eq:delta_ft_FC_main} of main text
\begin{align}
  \Delta\cF =\frac{\tilde{m}}{N}\left(\frac{\beta^2}{2}+\frac{\beta^2}{1-\beta}\right) + \cO(\tilde{m}^2) \ .
\end{align}

\section{The Toner-Tu model}
\subsection{Linear stability analysis}
\label{app:Caussin_stab}
In this appendix, we analyze the stability of the perturbations $\delta\rho$ and $\delta W$ for the Toner-Tu model \eref{eq:Caussin_W_adim}.
The growth rates of the perturbation are the eigenvalues of the stability matrix appearing in Eq.~\eqref{eq:Caussin_stabMatrix}. They read:
\begin{equation}
  \lambda_{\pm}=\dfrac{-\left(\xi_5 q^{2} + i\xi_4 q - \xi_3 \right)\pm \sqrt{\Delta}}{2} \, ,
\end{equation}
with
\begin{equation}
  \Delta = \left(\xi_5 q^{2} + i\xi_4 q - \xi_3 \right)^2-4v\left(iq\xi_1+\xi_2 q^2\right) \, .
\end{equation}
The stability of the homogeneous solution is determined by the sign of the real part of the eigenvalues $\lambda^\pm$.
An unstable mode exists as soon as $|\Re(\xi_5q^2+i\xi_4 q-\xi_3)| < |\Re (\sqrt\Delta)|$.
We note that
\begin{equation}
\label{eq:stability_alpha_rho_TT}
2\Re(\sqrt{\Delta})^2-2\Re\left(\xi_5q^{2}+i\xi_4 q - \xi_3 \right)^2 = -a(q) + \sqrt{a^2(q)+b(q)}\;,
\end{equation}
where $a(q)$ and $b(q)$ are real numbers given by
\begin{align}
  a(q) =& \xi_4 q^2 +(\xi_5 q^2-\xi_3)^2+4v \xi_2 q^2 \\
  b(q) =& 16 q^2 v \left(\xi_1^2 v+\xi_3 \xi_4\xi_1-\xi_2 \xi_3^2\right)+16 q^4 v \xi_5\; (2 \xi_2\; \xi_3-\xi_1\; \xi_4)-16 q^6 \left(\xi_2\; \xi_5^2\; v\right) \ .
\end{align}
Since $a(q)$ is always positive, we deduce that an unstable mode must have $b(q)>0$, which is the condition discussed in the main text.

\subsection{Renormalization for Voronoi-based alignment}
\label{app:ren_TT}

This appendix is devoted to the derivation of the linear mass in the renormalized hydrodynamics \eref{eq:hydro_Voronoi_ren} of the main text.
Following the scheme developed in section \ref{subsec:renorm_AIM}, we call $\rho_0$ and $W_0$ the mean-field solutions of \eref{eq:hydro_Voronoi_fluct} without noise (\textit{ie} with $\sigma=0$).
They are implicitly defined through
\begin{subequations}
\begin{align}%
  \label{eq:hydro_Voronoi_1_app}
  \partial_t \rho_0 + v\partial_x W_0 &= 0 \\
  \label{eq:hydro_Voronoi_2_app}
  \partial_t W_0 + \frac{\lambda}{\rho} W_0 \partial_x W_0 &= -\frac{v}{2} \nabla \rho_0 + D \partial_{xx}W_0 - \alpha W_0 -\frac{\gamma}{\rho_0^2} W_0^3 \; .
\end{align}%
\end{subequations}
We now introduce the perturbative series
\begin{align}
  \label{eq:perturbative_series_TT}
  \rho = \rho_0 + \sqrt{\sigma}\; \delta \rho_1 + \sigma \delta \rho_2 +..\;,\qquad W = W_0 + \sqrt{\sigma}\; \delta W_1 + \sigma \delta W_2 +..\; .
\end{align}
Injecting \eref{eq:perturbative_series_TT} into \eref{eq:hydro_Voronoi_fluct} and equating terms of order $\sigma^{k/2}$ yields the evolution equation for $\delta\rho_k$ and $\delta W_k$.
Further using the definition $\cF_{tt}=-\alpha W - \gamma W^3/\rho^2$, we obtain, for $k=1$
\begin{subequations}\label{eq:delta_fields_1_TT}
\begin{align}
  \label{eq:delta_rho_1_TT}
  \partial_t \delta\rho_1 =& -v\partial_x \delta W_1 + \partial_x\left(\sqrt{2\epsilon\rho_0}\eta_1\right) \\
  \nonumber
  \partial_t \delta W_1 =& D \partial_{xx} \delta W_1 - \frac{v}{2} \partial_x \delta\rho_1 -\frac{\lambda}{\rho_0} W_0 \partial_x W_1 -\frac{\lambda}{\rho_0} W_1 \partial_x W_0 + \frac{\lambda}{\rho_0^2}\delta\rho_1 W_0\partial_x W_0
  \\
  \label{eq:delta_W_1_TT}
  &+\frac{\partial\cF_{tt}}{\partial\rho}\delta \rho_1 +\frac{\partial\cF_{tt}}{\partial W}\delta W_1 + \sqrt{2 \rho_0} \, \eta_2
  \; .
\end{align}
\end{subequations}
while for $k=2$ we get
\begin{subequations}\label{eq:delta_fields_2_TT}
\begin{align}
  \label{eq:delta_rho_2_TT}
  \partial_t \delta\rho_2 =& -v\partial_x \delta W_2 + \partial_x\left(\frac{\sqrt{\epsilon}\delta\rho_1}{\sqrt{2\rho_0}}\eta_1\right) \\
  \nonumber
  \partial_t \delta W_2 =& D \partial_{xx} \delta W_2 - \frac{v}{2} \partial_x \delta\rho_2 +\frac{\partial\cF_{tt}}{\partial\rho}\delta \rho_2 +\frac{\partial\cF_{tt}}{\partial W}\delta W_2 + \frac{\partial^{2} \mathcal{F}_{tt}}{\partial W^{2}} \frac{\delta W_1^{2}}{2}+\frac{\partial^{2} \mathcal{F}_{tt}}{\partial \rho^{2}} \frac{\delta \rho_1^{2}}{2} \\ 
  \nonumber
  &+ \frac{\partial^{2} \mathcal{F}_{tt}}{\partial W \partial \rho} \delta W_1\delta \rho_1
  - \frac{\lambda}{\rho_0} \delta W_1\partial_x \delta W_1 -\frac{\lambda}{\rho_0^3} W_0\partial_x W_0 \delta\rho_1^2 - \frac{\lambda}{\rho_0} \delta W_2\partial_x W_0 - \frac{\lambda}{\rho_0} W_0\partial_x \delta W_2 \\ 
  \label{eq:delta_W_2_TT}
  & + \frac{\lambda}{\rho_0^2} \delta\rho_1 \delta W_1\partial_x W_0
  + \frac{\lambda}{\rho_0^2} W_0\delta \rho_1 \partial_x \delta W_1 + \frac{\lambda}{\rho_0^2}\delta \rho_2 W_0\partial_x W_0  + \frac{\delta\rho_1}{\sqrt{2\rho_0}} \, \eta_2
  \; .
\end{align}
\end{subequations}
Averaging \eref{eq:delta_fields_1_TT} over the noise with It\=o prescription gives
\begin{subequations}\label{eq:delta_fields_1_TT_av}
\begin{align}
  \label{eq:delta_rho_1_TT_av}
  \partial_t \langle \delta\rho_1\rangle =& -v \partial_x \langle \delta W_1 \rangle \\
  \nonumber
  \partial_t \langle \delta W_1 \rangle =& D \partial_{xx} \langle \delta W_1 \rangle- \frac{v}{2} \partial_x \langle \delta\rho_1 \rangle-\frac{\lambda}{\rho_0} W_0 \partial_x \langle W_1 \rangle-\frac{\lambda}{\rho_0} \langle W_1 \rangle \partial_x W_0 + \frac{\lambda}{\rho_0^2}\langle \delta\rho_1\rangle W_0\partial_x W_0
  \\
  \label{eq:delta_W_1_TT_av}
  &+\frac{\partial\cF_{tt}}{\partial\rho}\langle \delta \rho_1 \rangle+\frac{\partial\cF_{tt}}{\partial W}\langle \delta W_1 \rangle
  \; .
\end{align}
\end{subequations}
Doing likewise for \eref{eq:delta_fields_2_TT} yields
\begin{subequations}
\begin{align}
  \label{eq:delta_rho_2_TT_av}
  \partial_t \langle \delta\rho_2 \rangle=& -v\partial_x \langle\delta W_2 \rangle \\
  \nonumber
  \partial_t \langle\delta W_2\rangle =& D \partial_{xx} \langle \delta W_2\rangle - \frac{v}{2} \partial_x \langle \delta\rho_2\rangle +\frac{\partial\cF_{tt}}{\partial\rho}\langle\delta \rho_2 \rangle+\frac{\partial\cF_{tt}}{\partial W}\langle\delta W_2\rangle + \frac{\partial^{2} \mathcal{F}_{tt}}{\partial W^{2}} \frac{\langle\delta W_1^{2}\rangle}{2}+ \frac{\partial^{2} \mathcal{F}_{tt}}{\partial \rho^{2}} \frac{\langle\delta \rho_1^{2}\rangle}{2} \\
  & + \frac{\partial^{2} \mathcal{F}_{tt}}{\partial W \partial \rho} \langle \delta W_1\delta \rho_1\rangle
  \nonumber
  - \frac{\lambda}{\rho_0} \langle \delta W_1\partial_x \delta W_1 \rangle-\frac{\lambda}{\rho_0^3} W_0\partial_x W_0 \langle \delta\rho_1^2\rangle - \frac{\lambda}{\rho_0} \langle \delta W_2\rangle \partial_x W_0 \\
  &- \frac{\lambda}{\rho_0} W_0\partial_x \langle \delta W_2 \rangle + \frac{\lambda}{\rho_0^2} \langle\delta\rho_1 \delta W_1 \rangle\partial_x W_0
  \label{eq:delta_W_2_TT_av}
  + \frac{\lambda}{\rho_0^2} W_0 \langle \delta \rho_1 \partial_x \delta W_1\rangle  + \frac{\lambda}{\rho_0^2}\langle\delta \rho_2\rangle W_0\partial_x W_0
  \; ,
\end{align}
\end{subequations}
Summing together \eref{eq:hydro_Voronoi_1_app}, $\sqrt{\sigma}$ times \eref{eq:delta_rho_1_TT_av}, and $\sigma$ times \eref{eq:delta_rho_2_TT_av} gives the evolution of $\tilde{\rho}$ up to order $\sigma$
\begin{align}
  \label{eq:hydro_Voronoi_ren_corr}
  \partial_t \tilde{\rho} &= - v\partial_x \tilde{W}\;,
\end{align}
while adding \eref{eq:hydro_Voronoi_2_app}, $\sqrt{\sigma}$ times \eref{eq:delta_W_1_TT_av}, and $\sigma$ times \eref{eq:delta_W_2_TT_av} yields the evolution of $\tilde{m}$ up to order $\sigma$
\begin{align}
  \nonumber
  \partial_t \tilde{W}  =& D \partial_{xx}\tilde{W} - \frac{\lambda}{\tilde{\rho}} \tilde{W} \partial_x \tilde{W} -\frac{v}{2} \partial_x \tilde{\rho} + \alpha \tilde{W} -\frac{\gamma}{\tilde{\rho}^2} \tilde{W}^3 +\cF_{tt}(\tilde{\rho},\tilde{W})+\sigma \bigg{[}\frac{\partial^{2} \mathcal{F}_{tt}}{\partial W^{2}} \left(\frac{\langle\delta W_1^{2}\rangle -\langle\delta W_1\rangle^{2}}{2}\right) \\
  \nonumber
  & + \frac{\partial^{2} \cF_{tt}}{\partial \rho^{2}} \left(\frac{\langle\delta \rho_1^{2}\rangle-\langle\delta \rho_1\rangle^{2}}{2}\right) +  \frac{\partial^{2} \cF_{tt}}{\partial W \partial \rho} \left(\langle\delta m_1\delta \rho_1\rangle-\langle\delta W_1\rangle \langle\delta \rho_1\rangle\right)
  \\
  \label{eq:hydro_Voronoi_ren_corr_W}
  &-\frac{\lambda}{\rho_0}\left(\langle \delta W_1\partial_x \delta W_1\rangle-\langle \delta W_1\rangle\partial_x\langle \delta W_1\rangle\right) -\frac{\lambda}{\rho_0^3}W_0\partial_x W_0\left(\langle\delta\rho_1^2\rangle-\langle\delta\rho_1\rangle^2\right) \\
  \nonumber
  & + \frac{\lambda}{\rho_0^2}\partial_x W_0 \left(\langle\delta\rho_1\delta W_1\rangle - \langle \delta \rho_1\rangle\langle\delta W_1\rangle\right)+ \frac{\lambda}{\rho_0^2}W_0\left(\langle\delta\rho_1\partial_x\delta W_1\rangle - \langle \delta \rho_1\rangle \partial_x\langle W_1\rangle\right)\bigg{]}
  \; .
\end{align}
The terms proportional to $\partial_x W_0$ in \eref{eq:hydro_Voronoi_ren_corr_W} can only renormalize $\lambda$, not $\cF_{tt}$ so that we set them to zero and focus on computing
\begin{align}
  \nonumber
  \Delta \cF_{tt} =& \frac{\partial^{2} \mathcal{F}_{tt}}{\partial W^{2}} \left(\frac{\langle\delta W_1^{2}\rangle -\langle\delta W_1\rangle^{2}}{2}\right) + \frac{\partial^{2} \cF_{tt}}{\partial \rho^{2}} \left(\frac{\langle\delta \rho_1^{2}\rangle-\langle\delta \rho_1\rangle^{2}}{2}\right) +\frac{\lambda}{\rho_0^2}W_0\big(\langle\delta\rho_1\partial_x\delta W_1\rangle \\ 
  \nonumber
  & - \langle \delta \rho_1\rangle \partial_x\langle W_1\rangle\big)
  +  \frac{\partial^{2} \cF_{tt}}{\partial W \partial \rho} \left(\langle\delta m_1\delta \rho_1\rangle-\langle\delta W_1\rangle \langle\delta \rho_1\rangle\right) -\frac{\lambda}{\rho_0}\big(\langle \delta W_1\partial_x \delta W_1\rangle \\ 
  \label{eq:delta_F_tt}
  &-\langle \delta W_1\rangle\partial_x\langle \delta W_1\rangle\big) \;.
\end{align}
As we are solely interested in the renormalization of $\alpha$, we only have to derive the correlators involving $\delta \rho_1$ and $\delta W_1$ in $\Delta \cF_{tt}$ to order $W_0$.
To perform this derivation, we have to assume that $\delta \rho_1$ and $\delta W_1$ are fast mode varying on lengthscales much smaller than those of $\rho_0$ and $W_0$.
Under this assumption, $\rho_0(x,t)$, $W_0(x,t)$ and $\partial_x W_0(x,t)$ entering in the linearized evolution \eref{eq:delta_fields_1_TT} for $\delta \rho_1$ and $\delta W_1$ can be considered as constants.
This adiabatic approximation allows us to compute the correlators in terms of $\rho_0$, $W_0$ and $\partial_x W_0$ as constants and to re-establish their dependency on $x$ and $t$ a posteriori.
Note that any dependency of the correlators on $\partial_x W_0$ will renormalize $\lambda$, not $\alpha$.
Thus, to simplify the derivation of $\hat{\alpha}$ and get rid of any dependency on $\partial_x W_0$ in the correlators, we set $\partial_x W_0=0$ in the linearized evolution \eref{eq:delta_fields_1_TT}
\begin{subequations}
\begin{align}
  \label{eq:delta_rho_1_TT_simp}
  \partial_t \delta\rho_1 =& -v\partial_x \delta W_1 + \partial_x\left(\sqrt{2\epsilon\rho_0}\eta_1\right) \\
  \partial_t \delta W_1 =& D \partial_{xx} \delta W_1 - \frac{v}{2} \partial_x \delta\rho_1 -\frac{\lambda}{\rho_0} W_0 \partial_x W_1
  \label{eq:delta_W_1_TT_simp}
  +\frac{\partial\cF_{tt}}{\partial\rho}\delta \rho_1 +\frac{\partial\cF_{tt}}{\partial W}\delta W_1 + \sqrt{2 \rho_0} \, \eta_2
  \; .
\end{align}
\end{subequations}
Multiplying \eref{eq:delta_rho_1_TT_simp} and \eref{eq:delta_W_1_TT_simp} by $e^{iqx}/L$ and integrating over $x$ yields the time-evolution of the $q$-th Fourier modes
\begin{equation}
\label{eq:dynamic_delta_fourier_TT}
\partial_t \begin{pmatrix}
\delta {\rho^q_1} \\
\delta {W^q_1}
\end{pmatrix} = \begin{pmatrix}
T_{11}^q & T_{12}^q \\
T_{21}^q  & T_{22}^q
\end{pmatrix}
\begin{pmatrix}
\delta \rho^q_1 \\
\delta W^q_1
\end{pmatrix} + \begin{pmatrix}
\sqrt{2\epsilon\rho_0}\; iq\; \eta_1^q \\
\sqrt{2\rho_0}\;\eta_2^q
\end{pmatrix}\ ,
\end{equation}
where the convention for $\delta \rho_1^q$ and $\delta W_1^q$ is given in \eref{eq:fourier_serie}.
Note that $\eta_1^q$ and $\eta_2^q$ are the $q$-th Fourier modes of the two uncorrelated Gaussian white noises $\eta_1$ and $\eta_2$.
Their correlations reads $\langle \eta_k^q\eta_l^{q^{\prime}}\rangle=L^{-1}\delta_{k,l}\delta(t-t^{\prime})\delta_{q+q^{\prime}, 0}$.
Finally, the matrix coefficients $T_{kl}^q$ are given by
\begin{align}
  T_{11}^q &= 0\;, & T_{12}^q =& -ivq\;,  & T_{21}^q =& -iq\frac{v}{2} + 2 \gamma \frac{W_0^3}{\rho_0^3}\;, & T_{22}^q =&-\lambda iq \frac{W_0}{\rho_0} - Dq^2-\alpha -3\gamma \frac{W_0^2}{\rho_0^2}\;.
\end{align}
To compute the correlators involved in $\Delta \cF_{tt}$, we use It\=o calculus on the stochastic system \eref{eq:dynamic_delta_fourier_TT} to get the following closed system of equations
\begin{subequations}
\begin{align}
\frac{d}{dt}\langle \delta\rho_1^q \delta \rho_1^{q'} \rangle =&(T_{11}^{q}+T^{q^{\prime}}_{11})\langle\delta\rho_1^{q}\delta\rho_1^{q^{\prime}}\rangle + T_{12}^{q}\langle\delta W_1^{q}\delta\rho_1^{q^{\prime}}\rangle + T_{12}^{q^{\prime}}\langle\delta \rho_1^{q}\delta W_1^{q^{\prime}}\rangle - \frac{2\epsilon\rho_0 q^2}{L}\delta_{q+q^{\prime},0}=0\\
\frac{d}{dt}\langle \delta W_1^q \delta \rho_1^{q'} \rangle =&(T_{22}^{q}+T^{q^{\prime}}_{11})\langle\delta W_1^{q}\delta\rho_1^{q^{\prime}}\rangle + T_{21}^{q}\langle\delta \rho_1^{q}\delta\rho_1^{q^{\prime}}\rangle + T_{12}^{q^{\prime}}\langle\delta W_1^{q}\delta W_1^{q^{\prime}}\rangle=0 \\
\frac{d}{dt}\langle \delta W_1^q \delta W_1^{q^{\prime}} \rangle =&(T_{22}^{q}+T^{q^{\prime}}_{22})\langle\delta W_1^{q}\delta W_1^{q^{\prime}}\rangle + T_{21}^{q}\langle\delta \rho_1^{q}\delta W_1^{q^{\prime}}\rangle + T_{21}^{q^{\prime}}\langle\delta W_1^{q}\delta \rho_1^{q^{\prime}}\rangle + \frac{2\rho_{0}}{L}\delta_{q+q^{\prime},0}=0 \ ,
\end{align}
\end{subequations}
where the last equality once again stems from working in the steady state.
Inverting this system yields
\begin{align}
\label{eq:fourier_corr_scaling_TT}
\langle\delta W_1^{q}\delta W_1^{q^{\prime}}\rangle =& g_{ww}(q)\frac{\delta_{q+q^{\prime},0}}{L}\;,  &
\langle\delta \rho_1^{q}\delta \rho_1^{q^{\prime}}\rangle =& g_{\rho\rho}(q)\frac{\delta_{q+q^{\prime},0}}{L}\;, &
\langle\delta W_1^{q}\delta \rho_1^{q^{\prime}}\rangle =& g_{w\rho}(q)\frac{\delta_{q+q^{\prime},0}}{L}\ ,
\end{align}
where, to first order in $W_0$, the functions $g_{ww}(q)$, $g_{\rho\rho}(q)$ and $g_{w\rho}(q)$ are given by
\begin{subequations}\label{eq:fourier_corr_TT}
\begin{align}
\label{eq:fourier_W_W_corr_TT}
g_{ww}(q) =& \frac{\rho_0}{2D}\left[-\epsilon +\frac{2D+\epsilon\alpha}{D q^2+\alpha }\right]+\cO\left(W_0^2\right)  \\
\label{eq:fourier_rho_rho_corr_TT}
g_{\rho\rho}(q) =& \frac{\rho_0}{D}\left[-2 \frac{D^2q^2\epsilon}{v^2}-2\frac{D\epsilon\alpha}{v^2}-\epsilon+\frac{2D+\alpha\epsilon}{Dq^2+\alpha}\right]+\cO\left(W_0^2\right)\\
\label{eq:fourier_W_rho_corr_TT}
g_{w\rho}(q) =& \frac{\epsilon}{Dv}\left[iqD\rho_0+W_0\lambda-W_0\frac{\lambda\alpha}{Dq^2+\alpha} \right]+\cO\left(W_0^2\right)\ .
\end{align}
\end{subequations}
At this point, similarly to \eqref{eq:fourier_m_m_corr_AIM_cons_noise}-\eqref{eq:fourier_m_rho_corr_AIM_cons_noise}, we note that $g_{ww}(q)$, $g_{\rho\rho}(q)$ and $g_{w\rho}(q)$ also contain a polynomial part in $q$ and a rational part.
As discussed in, this polynomial part  yields a divergent contribution which depends on the microscopic regularization that one employs for the continuum description \eref{eq:hydro_Voronoi_1_app}-\eref{eq:hydro_Voronoi_2_app}.
We will therefore only conserve the rational parts of $g_{ww}(q)$, $g_{\rho\rho}(q)$ and $g_{w\rho}(q)$ which remain universal. As we show below, this leads to a density-dependent correction to the mass of the polar field and thus suffice to predict the emergence of a fluctuation-induced first-order transition.
Retaining only the rational parts in \eqref{eq:fourier_corr_TT} yields
\begin{align}
\label{eq:fourier_W_W_corr_TT_reg}
g_{ww}(q) =& \frac{\rho_0}{2D}\frac{2D+\epsilon\alpha}{D q^2+\alpha }\;,  &
g_{\rho\rho}(q) =& \frac{\rho_0}{D}\frac{2D+\alpha\epsilon}{Dq^2+\alpha}\;, &
g_{w\rho}(q) =& -\frac{\epsilon W_0}{Dv}\frac{\lambda\alpha}{Dq^2+\alpha}\ .
\end{align}
We further note that we can integrate $g_{ww}(q)$, $g_{\rho\rho}(q)$ and $g_{w\rho}(q)$ over $q$ only if $\alpha > 0$, which means that we have to restrict our study to the high temperature phase where such a condition is respected.
We finally obtain
\begin{align}
  \nonumber
  \langle \delta\rho_1^2 \rangle &=\int \frac{dq}{2\pi} g_{\rho\rho}(q) = \frac{\rho_0(2D+\alpha\epsilon)}{2 D\sqrt{D|\alpha|}}\;, \quad
  \langle \delta W_1^2 \rangle =\int \frac{dq}{2\pi} g_{ww}(q) = \frac{\rho_0(2D+\alpha\epsilon)}{4 D\sqrt{D|\alpha|}} \\
  \label{eq:final_corr_TT}
  \langle \delta W_1\delta\rho_1 \rangle &=\int \frac{dq}{2\pi} g_{w\rho}(q) = -W_0\frac{\epsilon\lambda\alpha}{2 Dv\sqrt{D|\alpha|}}\;, \quad \langle \delta W_1\partial_x \delta W_1 \rangle =\int \frac{dq}{2\pi} iq g_{ww}(q) = 0 \; .
\end{align}
Injecting the above expression \eref{eq:final_corr_TT} for the correlators in \eref{eq:delta_F_tt} yields the final expression for $\Delta \cF_{tt}$ as
\begin{align}
  \Delta \cF_{tt} = \frac{1}{2}\frac{\partial^{2} \mathcal{F}_{tt}}{\partial W^{2}}\langle\delta W_1^{2}\rangle + \cO(W_0^2) = -\frac{3\gamma}{\rho_0}\frac{2D+\alpha \epsilon}{4D\sqrt{D|\alpha|}} W_0 + \cO(W_0^2)\; .
\end{align}
We readily deduce from the above expression the formula \eref{eq:alpha_ren_TT} of main text
\begin{align}
\label{eq:RenAlphaConsTT}
  \hat{\alpha} = \alpha +\sigma\frac{3\gamma}{\tilde{\rho}}\frac{2 D+\alpha \epsilon}{4D\sqrt{D|\alpha|}}\;,
\end{align}
which shows that the renormalized mass acquired a dependence on the density.
Note that Eq.~\eqref{eq:RenAlphaConsTT} should be complemented by the corrections due to the polynomial contributions to $g_{\rho\rho}$, $g_{ww}$ and $g_{w\rho}$. 
The latters are not universal and will not cancel the density dependence reported in~\eqref{eq:RenAlphaConsTT}.

\section{Details of the numerical simulations}
\label{sec:numerics}
\subsection{Stochastic partial differential equations}
The numerical integrations of the stochastic PDEs were carried out using a semi-spectral method with a semi-implicit Euler scheme.
At every time step, the Gaussian noise fields are drawn in direct $q$ space. The latter being discretized, at every lattice site one draws a Gaussian random number of zero mean and variance $\sqrt{2Ddt/dx}$.
All the non-linearities are also computed in direct space.
All fields are then Fourier-transformed to $q$-space, where the time-stepping takes place.
Anti-aliasing with the standard $3/2$ rule was carried out after time-stepping.
\subsection{Microscopic simulations of the AIM}
\label{app:numerics_Voronoi}
The simulations of the microscopic AIM rely on a parallel update of all particles.
Each spin $\sigma_i$ has a probability $W(\sigma_i)dt$ to flip during a time-step $dt$, where $W(\sigma_i)$ is given by \eref{eq:flying_xy_alignment} and depends on the type of alignment considered: metric \eref{eq:OLAIM_dyn}, $k$-nearest \eref{eq:OLAIM_dyn_knearest} or Voronoi \eref{eq:OLAIM_dyn_voronoi}.
For the determination of the Voronoi neighbors, we used the CGAL package~\cite{cgaleb23b} for constructing the Voronoi tesselation. When computing the local magnetization in topological models, see Eq.~\eqref{eq:external_field}, we do not include the spin $i$ itself in the set of neighbours. Results do not depend qualitatively on this choice.
Note that topological simulations require an important numerical effort. Therefore, to simulate the topological AIM, we use a large time-step $dt = 0.5/ [ \Gamma {\rm exp}(\beta)]$ to reduce the simulation time while ensuring $W(\sigma_i) dt < 1$. Note that, to correctly approximate the continuous-time dynamics, we should use much smaller time steps such that $W(\sigma_i)dt \ll 1$. 
This is beyond what we can do numerically but we do not expect the use of these finite time steps to affect the qualitative behavior of the system.


\subsection{Microscopic simulations of the Vicsek Model}
The simulations of the microscopic Vicsek model rely on a parallel update of all particles.
At each time-step $dt$, the propulsion vectors $\bu_i$'s align according to \eref{eq:Vicsek_micro}.
After this alignment step, the position $\br_i$'s of the particles are simply advected by the $\bu_i$'s as
\begin{align}
  \br_i (t+dt)=\br_i(t) +v_0\bu_i dt \;,
\end{align}
where $dt=1$. As above, we used the CGAL package to determine the Voronoi tesselation.

\vspace{1cm}
\bibliographystyle{iopart-num}
\bibliography{Biblio}

\end{document}